\documentclass[aps,prb,reprint,letterpaper,superscriptaddress]{revtex4-2}

\usepackage{amsmath,amssymb,amsthm,bm}
\usepackage{graphicx}
\usepackage[colorlinks=true,linkcolor=blue,citecolor=blue,urlcolor=blue]{hyperref}
\usepackage{xcolor}
\usepackage{bbm}
\usepackage{upgreek}

\begin{document}

\title{Layer-selective proximity symmetry breaking enables anomalous and nonlinear Hall responses in $1H$-Nb$X_2$ ($X =$ S, Se, Te)}

\author{Yusuf Wicaksono}
\affiliation{International Center for Materials Nanoarchitectonics (WPI-MANA), National Institute for Materials Science, 1-1 Namiki, Tsukuba, Ibaraki, 305-0044 JAPAN}

\author{Toshikaze Kariyado}
\email{Contact author: KARIYADO.Toshikaze@nims.go.jp}
\affiliation{International Center for Materials Nanoarchitectonics (WPI-MANA), National Institute for Materials Science, 1-1 Namiki, Tsukuba, Ibaraki, 305-0044 JAPAN}

\date{\today}

\begin{abstract}
Nonlinear Hall responses provide an electrical probe of Berry-curvature dipoles, but they are symmetry forbidden in many pristine two-dimensional metals. We show that layer-selective magnetic proximity provides a symmetry-controlled route to induce and tune anomalous and nonlinear Hall responses in metallic monolayer $1H$-Nb$X_2$ ($X=\mathrm{S,Se,Te}$), where the nonmagnetic $D_{3h}$ crystal has vanishing anomalous Hall conductivity and Berry-curvature dipole. Fully relativistic density-functional theory combined with Wannier interpolation shows that an out-of-plane proximity exchange preserving $C_3$ generates a sizable sheet anomalous Hall conductivity, $\sigma^{\mathrm{sheet}}_{xy}\sim 10^{-2}(e^2/h)$ in representative active windows, while the Berry-curvature dipole remains zero. Breaking $C_3$ by introducing an in-plane exchange component, or by using an orthogonal two-sided exchange texture, produces a tunable Berry-curvature dipole and hence a nonlinear Hall response. Its exchange-odd part is linear in the in-plane exchange to leading order in the minimal model; the calculated spectra reach and can exceed $|D_y|\sim10^{-2}$~\AA, with the largest and sharpest features in NbTe$_2$. These trends are rationalized by symmetry analysis and an interface-induced $k$-linear Rashba--Zeeman minimal model. Within the idealized proximity model, an orthogonal dual-interface geometry further provides component-selective sign reversal of first- and second-harmonic Hall signals in the same Hall-bar configuration.
\end{abstract}

\maketitle


\emph{Introduction.}
Hall responses of spin-orbit-coupled metals provide direct access to band geometry and broken symmetries. With broken time-reversal symmetry (TRS), the intrinsic anomalous Hall effect (AHE) is governed by Berry curvature concentrated near spin-orbit-induced avoided crossings \cite{NagaosaRMP10}. Even with TRS, a transverse response can appear at second order in electric field when crystal symmetries permit a nonvanishing first moment of Berry curvature: the leading intrinsic nonlinear Hall conductivity (NLHC) is controlled by the Berry-curvature dipole (BCD) \cite{SodemannFu15} and has been observed in few-layer WTe$_2$ \cite{MaNature18,KangNatMat19}. These developments motivate symmetry-based routes to realize and control linear and nonlinear Hall responses within a single pristine two-dimensional metal.

Metallic $1H$-Nb$X_2$ ($X=\mathrm{S,Se,Te}$) provides a particularly clean setting for such a study. It is a pristine two-dimensional metal with strong spin-orbit coupling (SOC) and spin-valley locking, central to Ising superconductivity and related collective phenomena \cite{UgedaNatPhys15,XiNatPhys15,delaBarreraNatComm18}. Its van der Waals character also makes it compatible with magnetic-proximity heterostructures in which the induced exchange is short-ranged and predominantly interfacial. Experiments and calculations on transition-metal dichalcogenideTMDC/magnet stacks support strongly interface-dependent proximity effects \cite{SeylerNanoLett18,ZhongNatNano20,ZollnerPRB19}, although they do not uniquely determine the orbital partition assumed below inside Nb$X_2$. Related proximity-driven Hall phenomena have been discussed for semiconducting $2H$ TMDC monolayers \cite{HabeKoshinoPRB17}, NbSe$_2$-based spin/valley devices \cite{GuimaresNanoLett18,MatsuokaNatCommun22}, and interface-enabled nonlinear Hall responses \cite{DuNatCommun21,ZhongPRL24}. The distinctive question addressed here is whether the layer parity and orientation of the proximity exchange can be used as a symmetry knob to control both the intrinsic AHE and the BCD channel in a metallic $1H$ monolayer.

Here we show that layer-selective magnetic proximity in monolayer $1H$-Nb$X_2$ provides a minimal symmetry route to co-engineer intrinsic AHE and BCD-driven NLHC within a single two-dimensional metal. In pristine $1H$-Nb$X_2$, TRS forces the intrinsic AHE to vanish, while the threefold rotation $C_3$ contained in $D_{3h}$ forces the in-plane BCD to vanish. An interface-selective exchange acting asymmetrically on the two chalcogen sublayers lowers symmetry and enables proximity textures that separately affect Berry curvature (linear AHE) and its dipolar asymmetry (nonlinear Hall response). We formulate the corresponding selection rules, connect them to an interface-induced $k$-linear Rashba--Zeeman minimal model, and test them with fully relativistic first-principles Wannier calculations across the Nb$X_2$ series. We also discuss an ideal orthogonal dual-interface geometry that illustrates component-selective reversal of the two exchange-odd harmonic signals.

\begin{figure}[tb]
\includegraphics[width=\columnwidth]{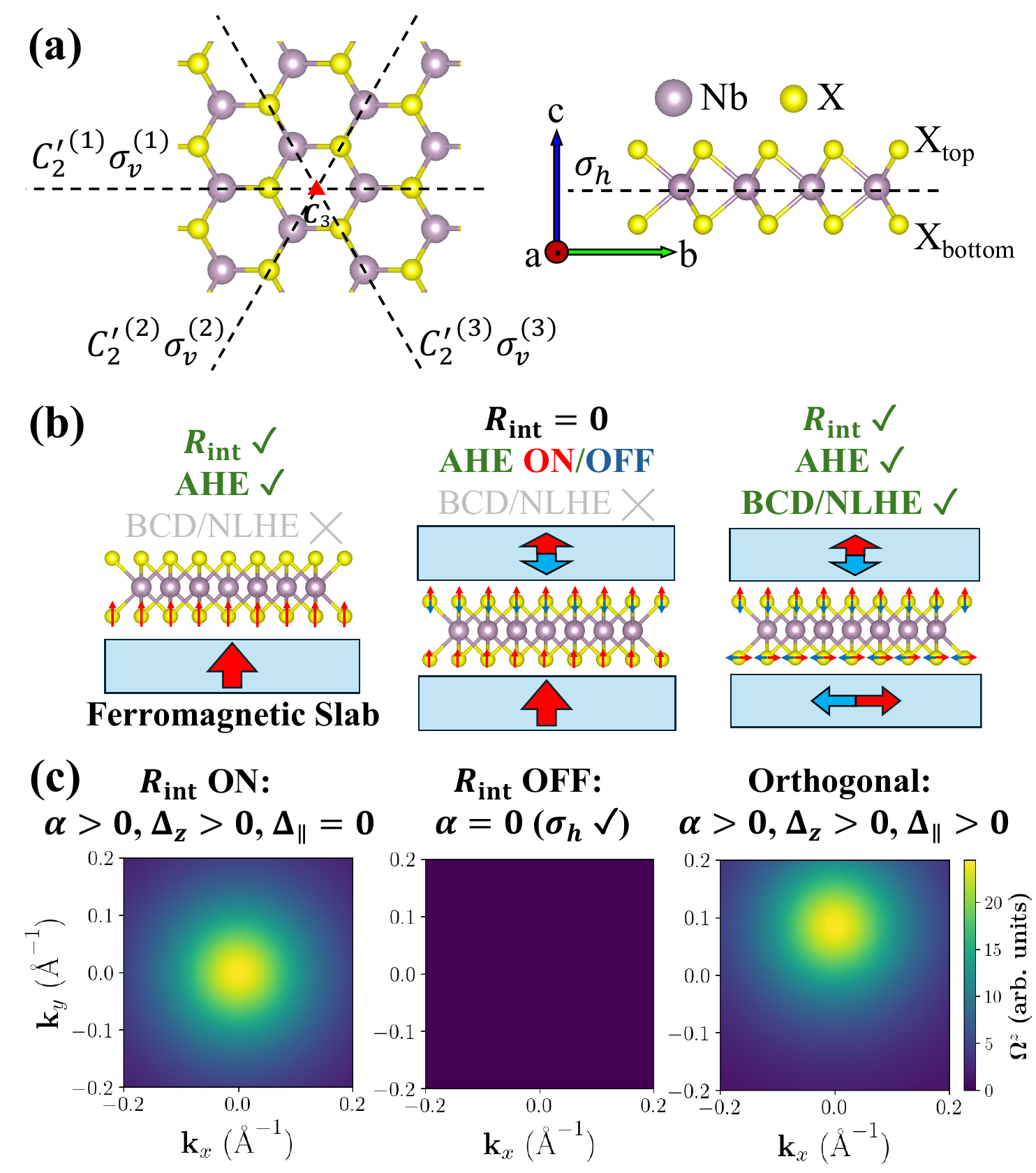}
\caption{ Symmetry route to Hall responses in monolayer $1H$-Nb$X_2$.
(a) Crystal structure (top/side views) with horizontal mirror $\sigma_h$ through the Nb plane and threefold axis $C_3$; chalcogen sublayers $X_{\mathrm{top}}$/$X_{\mathrm{bot}}$ host layer-selective exchange.
(b) Schematic proximity configurations: one-sided $m_z$ ($R_{\mathrm{int}}$ allowed; AHE allowed, BCD forbidden by $C_3$), two-sided Hall-valve geometry (parallel alignment ON, antiparallel alignment OFF), and orthogonal two-sided (AHE and BCD both allowed). Here $R_{\mathrm{int}}$ denotes the interface-induced Rashba term linear in the in-plane crystal momentum, $H_R=\alpha(\mathbf{k}\times\mathbf{s})_z=\alpha(k_xs_y-k_ys_x)$, and NLHE in the panel denotes the nonlinear Hall effect.
(c) Minimal-model Berry curvature $\Omega^z(\mathbf{k})$ from the interface-induced $k$-linear Rashba--Zeeman Hamiltonian, Eq.~(\ref{eq:RZ}). For $R_{\mathrm{int}}$ ON with $\Delta_z>0$ and $\boldsymbol{\Delta}_\parallel=\mathbf{0}$, the curvature is an isotropic hot spot and the BCD vanishes. For $R_{\mathrm{int}}$ OFF ($\alpha=0$, $\sigma_h$ preserved), $\Omega^z=0$ within the interface-induced $k$-linear Rashba-Zeeman minimal model; in the full material, however, a finite AHE can still arise in the $\sigma_h$-preserving two-sided $m_z$ geometry from pre-existing valley-textured Berry curvature (Supplemental Material Sec.~S3 \cite{supplemental_material}). In the orthogonal case ($\Delta_z\neq 0$ and $\boldsymbol{\Delta}_\parallel\neq\mathbf{0}$), the term $\hat{\mathbf z}\!\cdot(\mathbf{k}\times\boldsymbol{\Delta}_\parallel)$ shifts the avoided crossing in momentum space, creating a dipolar curvature asymmetry and hence a finite $\mathbf D\parallel \hat{\mathbf z}\times\boldsymbol{\Delta}_\parallel$.}
\label{fig:symmetry_device_concept}
\end{figure}

\emph{Symmetry principle and layer-selective proximity.}
We consider monolayer $1H$-Nb$X_2$ with point group $D_{3h}$ generated by the horizontal mirror $\sigma_h$ (through the Nb plane) and threefold rotation $C_3$ [Fig.~\ref{fig:symmetry_device_concept}(a)]. Magnetic proximity is modeled as an interfacial exchange $\Delta_{\rm ex}\,\mathbf{m}\!\cdot\!\mathbf{s}$ acting primarily on chalcogen $p$ orbitals. In the analytic minimal model, $s_i$ are dimensionless physical-spin Pauli matrices (eigenvalues $\pm1$). In the numerical Wannier implementation, analogous Pauli blocks act in projection-defined Wannier pseudospin pairs; because spin-orbit-entangled maximally localized Wannier
functions (MLWFs) need not be eigenstates of physical spin, this is an explicit effective-model convention rather than an exact representation of the projected physical-spin operator (Supplemental Material Sec.~S4.3). Thus $\Delta_{\rm ex}$ is the coefficient of the implemented pseudospin matrix, and a fully localized pure-spin orbital would have a splitting $2\Delta_{\rm ex}$. Because the exchange is short-ranged, we distinguish a \emph{one-sided} (layer-asymmetric) proximity pattern, acting predominantly on a single chalcogen sublayer, from \emph{two-sided} patterns acting on both sublayers. The residual horizontal-mirror symmetry depends on the detailed two-sided texture: Identical parallel out-of-plane exchange can preserve $\sigma_h$, whereas antiparallel and orthogonal textures obey different unitary or antiunitary mirror constraints. This distinction controls whether the leading interface-induced Rashba term, linear in the in-plane crystal momentum $\mathbf{k}$,
$
H_R=\alpha(\mathbf{k}\times\mathbf{s})_z=\alpha(k_xs_y-k_ys_x),
$
is symmetry allowed, and thereby determines the symmetry constraints on $\Omega^z(\mathbf{k})$ and its dipole [Fig.~\ref{fig:symmetry_device_concept}(b)].

The intrinsic AHC and BCD are
\begin{equation}
\sigma_{xy}=-\frac{e^2}{\hbar}\sum_n\!\int_{\rm BZ}\!\frac{d^2k}{(2\pi)^2}\; f_{n\mathbf{k}}\,\Omega^z_n(\mathbf{k}),
\label{eq:ahc-def}
\end{equation}
\begin{equation}
D_{az}=\sum_n\!\int_{\rm BZ}\!\frac{d^2k}{(2\pi)^2}\; f_{n\mathbf{k}}\,\partial_{k_a}\Omega^z_n(\mathbf{k}),
\label{eq:bcd-def}
\end{equation}
with $f_{n\mathbf{k}}$ the Fermi-Dirac distribution. Equation~\eqref{eq:bcd-def} defines the two-dimensional (sheet) BCD, whose components have units of length. We use the shorthand $D_a\equiv D_{az}^{\rm 2D}$, so that the component relevant below is $D_y\equiv D_{yz}^{\rm 2D}$. For the pristine nonmagnetic $D_{3h}$ monolayer, TRS enforces $\Omega^z_n(\mathbf{k})=-\Omega^z_n(-\mathbf{k})$, implying $\sigma_{xy}=0$ from Eq.~(\ref{eq:ahc-def}). The BCD $\mathbf{D}$ transforms as an in-plane polar vector, and the exact threefold rotation $C_3$ forbids any nonzero in-plane polar vector, so $\mathbf{D}=\mathbf{0}$ as long as $C_3$ is unbroken (see Supplemental Material Sec.~S1 \cite{supplemental_material}).

One-sided proximity breaks $\sigma_h$ and permits interface-induced $k$-linear Rashba SOC. With $\mathbf{m}\neq \mathbf{0}$ and SOC, $\Omega^z(\mathbf{k})$ and hence $\sigma_{xy}$ are generically allowed. For $\mathbf{m}\parallel\hat{\mathbf z}$, the residual $C_3$ symmetry forbids an in-plane polar vector, so $\mathbf{D}=\mathbf{0}$; introducing an in-plane component $\mathbf{m}_\parallel$ (or weak $C_3$ breaking) lifts this constraint and yields $\mathbf{D}\neq \mathbf{0}$, activating the intrinsic NLHC.

Two-sided proximity can preserve $\sigma_h$ for uniform out-of-plane exchange ($m_z$), thereby forbidding interface-induced $k$-linear Rashba terms. The residual $C_3$ symmetry, rather than $\sigma_h$ alone, enforces $\mathbf{D}=\mathbf{0}$, while TRS breaking still allows an AHE. In the ideal magnet/TMD/magnet model, parallel out-of-plane alignment yields a finite AHE, while antiparallel alignment nearly cancels the net exchange. If one interface supplies predominantly out-of-plane exchange and the other an in-plane component (orthogonal two-sided texture), then both $\sigma_{xy}$ and $\mathbf{D}$ are symmetry allowed without relying on fine control of a small canting angle.

\emph{Minimal interface-induced $k$-linear Rashba-Zeeman description.}
We denote the exchange induced on the top/bottom chalcogen sublayers by
$\Delta^{\rm top}_{\rm ex}\,\mathbf m_{\rm top}\!\cdot\!\mathbf s$ and
$\Delta^{\rm bot}_{\rm ex}\,\mathbf m_{\rm bot}\!\cdot\!\mathbf s$.
Projecting onto a Nb-derived Fermi sheet yields an effective Zeeman field
$\boldsymbol{\Delta}=w_{\rm top}\Delta^{\rm top}_{\rm ex}\mathbf m_{\rm top}+w_{\rm bot}\Delta^{\rm bot}_{\rm ex}\mathbf m_{\rm bot}$,
with $w_{\rm top/bot}$ the chalcogen weights.
Writing $\Delta_z=\boldsymbol{\Delta}\cdot\hat{\mathbf z}$ and $\boldsymbol{\Delta}_\parallel=\boldsymbol{\Delta}-\Delta_z\hat{\mathbf z}$, breaking $\sigma_h$ allows an interface-induced Rashba term linear in the in-plane crystal momentum, $H_R=\alpha(\mathbf{k}\times\mathbf{s})_z=\alpha(k_xs_y-k_ys_x)$, leading to the interface-induced $k$-linear Rashba--Zeeman Hamiltonian

\begin{equation}
H(\mathbf{k})=\varepsilon_0(k)\,\mathbbm{1}+\alpha(k_xs_y-k_ys_x)+\Delta_z s_z+\boldsymbol{\Delta}_\parallel\!\cdot\!\mathbf{s}.
\label{eq:RZ}
\end{equation}

One-sided proximity corresponds to $\Delta^{\rm top}_{\rm ex}\approx 0$; symmetric two-sided $m_z$ has $\mathbf m_{\rm top}=\mathbf m_{\rm bot}\parallel\hat{\mathbf z}$, yielding $\alpha=0$ while keeping $\Delta_z\neq 0$; an antiparallel configuration suppresses $\boldsymbol{\Delta}$ up to the weight imbalance $w_{\rm top}\Delta^{\rm top}_{\rm ex}-w_{\rm bot}\Delta^{\rm bot}_{\rm ex}$; and the orthogonal case produces simultaneous $\Delta_z\neq 0$ and $\boldsymbol{\Delta}_\parallel\neq\mathbf 0$.

Defining
$\mathbf{d}(\mathbf{k})=(-\alpha k_y+\Delta_\parallel^x,\alpha k_x+\Delta_\parallel^y,\Delta_z)$
and $E_\pm=\varepsilon_0(k)\pm|\mathbf{d}|$, the Berry curvature of the two spin-split bands is

\begin{multline}
\Omega_\pm^z(\mathbf{k})
=\mp\frac{1}{2}\frac{\mathbf{d}\cdot(\partial_{k_x}\mathbf{d}\times\partial_{k_y}\mathbf{d})}{|\mathbf{d}|^3} \\
=\mp\frac{\alpha^2\Delta_z}{2\!\left[\Delta_z^2+\alpha^2 k^2+\Delta_\parallel^2
+2\alpha\,\hat{\mathbf{z}}\!\cdot(\mathbf{k}\times\boldsymbol{\Delta}_\parallel)\right]^{3/2}}.
\label{eq:Omega}
\end{multline}

Equation~(\ref{eq:Omega}) captures the interface-induced $k$-linear Rashba-enabled Berry curvature generated when $\sigma_h$ is broken. In the $\sigma_h$-preserving two-sided $m_z$ limit ($\alpha=0$), this interface-induced $k$-linear Rashba--Zeeman mechanism is absent; nevertheless, Nb$X_2$ retains valley-textured Berry curvature originating from Ising SOC, and a finite AHE arises once TRS is broken, as described by a complementary valley $k\!\cdot\!p$ theory (Supplemental Material Sec.~S3 \cite{supplemental_material}). Three consequences of Eq.~(\ref{eq:Omega}) are immediate: (i) Within the interface-induced $k$-linear Rashba-Zeeman mechanism, if $\alpha=0$ or $\Delta_z=0$ then $\Omega_\pm^z\equiv 0$; (ii) for $\boldsymbol{\Delta}_\parallel=\mathbf 0$, $\Omega_\pm^z$ is isotropic in $k$ and hence the BCD vanishes; and (iii) a finite $\boldsymbol{\Delta}_\parallel$ enters through $\hat{\mathbf z}\!\cdot(\mathbf{k}\times\boldsymbol{\Delta}_\parallel)$, shifting the avoided crossing in momentum space and generating a dipolar curvature asymmetry with $\mathbf D\parallel \hat{\mathbf z}\times\boldsymbol{\Delta}_\parallel$ (i.e., $\mathbf D\perp\boldsymbol{\Delta}_\parallel$). A closed-form expression for $\sigma_{xy}(\mu)$ and a symmetry-based parametric scaling estimate for $\mathbf D(\mu)$ in the minimal model are given in Supplemental Material Sec.~S2 \cite{supplemental_material}.

\begin{figure}[tb]
\includegraphics[width=\columnwidth]{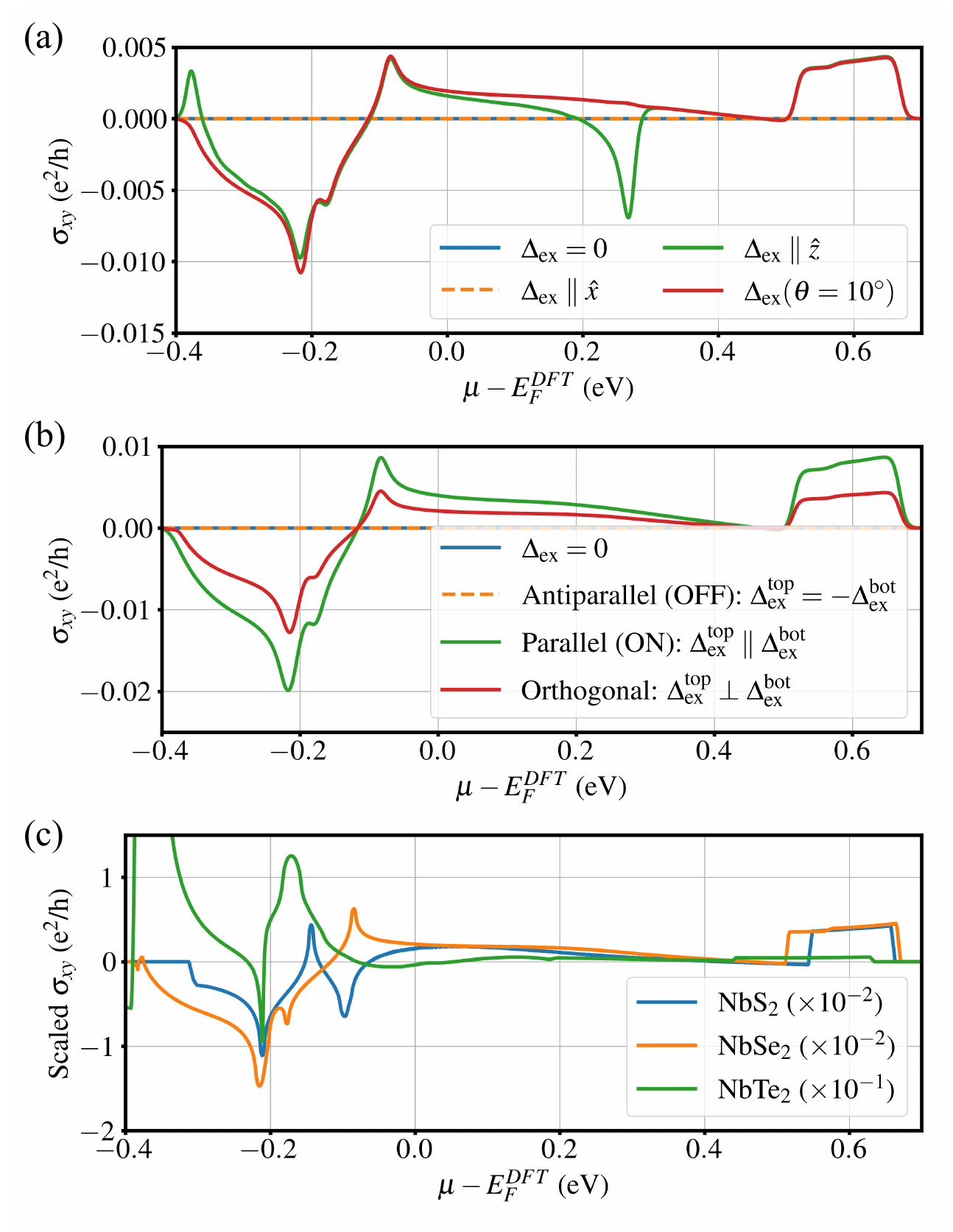}
\caption{ \textbf{First-principles anomalous Hall conductivity under layer-selective proximity exchange in $1H$-Nb$X_2$.}
(a) One-sided (layer-asymmetric) proximity exchange on a single chalcogen sublayer of NbSe$_2$: an out-of-plane exchange field $\boldsymbol{\Delta}_{\rm ex}\parallel\hat{z}$ generates a sizable intrinsic AHC with sign changes near SOC-induced avoided crossings, whereas a purely in-plane field $\boldsymbol{\Delta}_{\rm ex}\parallel\hat{x}$ yields $\sigma_{xy}\approx 0$; a small canting $\boldsymbol{\Delta}_{\rm ex}(\theta=10^\circ)$ with $\theta$ the canting angle between $\boldsymbol{\Delta}_{\rm ex}$ and $\hat{z}$ (tilt from $\hat{z}$ toward $\hat{x}$ in the $xz$ plane) interpolates between the two limits.
(b) Two-sided proximity exchange in a magnet/NbSe$_2$/magnet stack, characterized by the relative orientation of the exchange fields at the top and bottom interfaces, $\boldsymbol{\Delta}^{\rm top}_{\rm ex}$ and $\boldsymbol{\Delta}^{\rm bot}_{\rm ex}$: parallel alignment ($\boldsymbol{\Delta}^{\rm top}_{\rm ex}\parallel \boldsymbol{\Delta}^{\rm bot}_{\rm ex}$) yields a large AHC, antiparallel alignment ($\boldsymbol{\Delta}^{\rm top}_{\rm ex}=-\boldsymbol{\Delta}^{\rm bot}_{\rm ex}$) nearly cancels the net exchange, and the orthogonal configuration ($\boldsymbol{\Delta}^{\rm top}_{\rm ex}\perp \boldsymbol{\Delta}^{\rm bot}_{\rm ex}$) remains AHC-active.
(c) Chalcogen dependence of $\sigma_{xy}$ in the orthogonal configuration for NbS$_2$, NbSe$_2$, and NbTe$_2$, showing a pronounced increase from S$\rightarrow$Se$\rightarrow$Te. For every rescaled curve, the physical value equals the displayed ordinate multiplied by the factor printed in the legend. Energies are referenced to the DFT Fermi level; unless stated otherwise, $|\boldsymbol{\Delta}_{\rm ex}|=30~\mathrm{meV}$ is applied to the proximitized chalcogen $p$ orbitals.}
\label{fig:AHC_fig2}
\end{figure}

\emph{First-principles AHC trends.}
We test the symmetry analysis using fully relativistic density functional
theory (DFT) and Wannier interpolation (Supplemental Material Sec.~S4 \cite{supplemental_material}). In the Wannier Hamiltonian, magnetic proximity is introduced as a layer-resolved interfacial exchange acting on proximitized chalcogen $p$-orbital spinor Wannier functions. This construction is intended to capture symmetry-controlled proximity exchange at the effective-Hamiltonian level; explicit interface reconstruction, charge transfer, exchange on Nb and the remote chalcogen layer, and hopping renormalization from a specific magnetic substrate are not included. Thus, the model should be viewed as a proof-of-principle local spin-dependent component of the proximity self-energy after projecting out the magnetic substrate, rather than as a quantitatively validated material-specific magnet/Nb$X_2$ heterostructure calculation. The magnetic-point-group selection rules are more general than the calculated magnitudes and chemical-potential windows, which remain model dependent. We use $|\Delta_{\rm ex}|=30$~meV as a representative moderate local pseudospin-exchange scale, with its interpretation in the Wannier basis and benchmarking against reported proximity splittings discussed in Supplemental Material Sec.~S4.3 \cite{supplemental_material}. Unless stated otherwise, occupations are evaluated with a Fermi-Dirac distribution at $T=50$~K. The electronic structures themselves are those of the primitive $1\times1$ normal-state monolayers. In particular, the $3\times3$ charge-density-wave (CDW) reconstruction observed in low-temperature monolayer NbSe$_2$ \cite{UgedaNatPhys15} is not included; the NbSe$_2$ results should therefore be interpreted for a normal-state or CDW-suppressed regime, rather than as a calculation of the reconstructed low-temperature phase.

For one-sided NbSe$_2$ [Fig.~\ref{fig:AHC_fig2}(a)], out-of-plane exchange yields a sizable intrinsic AHC with dispersive features as $\mu$ crosses SOC-induced avoided crossings, while purely in-plane exchange gives $\sigma_{xy}\approx 0$; small canting interpolates continuously. In the two-sided geometry [Fig.~\ref{fig:AHC_fig2}(b)], $\sigma_{xy}$ realizes a Hall-valve response: Parallel alignment produces an ON state, whereas antiparallel alignment suppresses the net AHC through exchange cancellation. The orthogonal two-sided configuration remains AHC-active, providing a route to combine finite $\sigma_{xy}$ with the in-plane exchange component that activates a BCD. Across NbS$_2\!\rightarrow$NbSe$_2\!\rightarrow$NbTe$_2$ [Fig.~\ref{fig:AHC_fig2}(c)], the orthogonal AHC grows markedly. This cross-material trend correlates with the heavier chalcogen and larger SOC, but chemical substitution also changes lattice parameters, orbital hybridization, and Fermi-surface geometry; it should not be read as an isolated SOC-scaling calculation.

\begin{figure}[tb]
\includegraphics[width=\columnwidth]{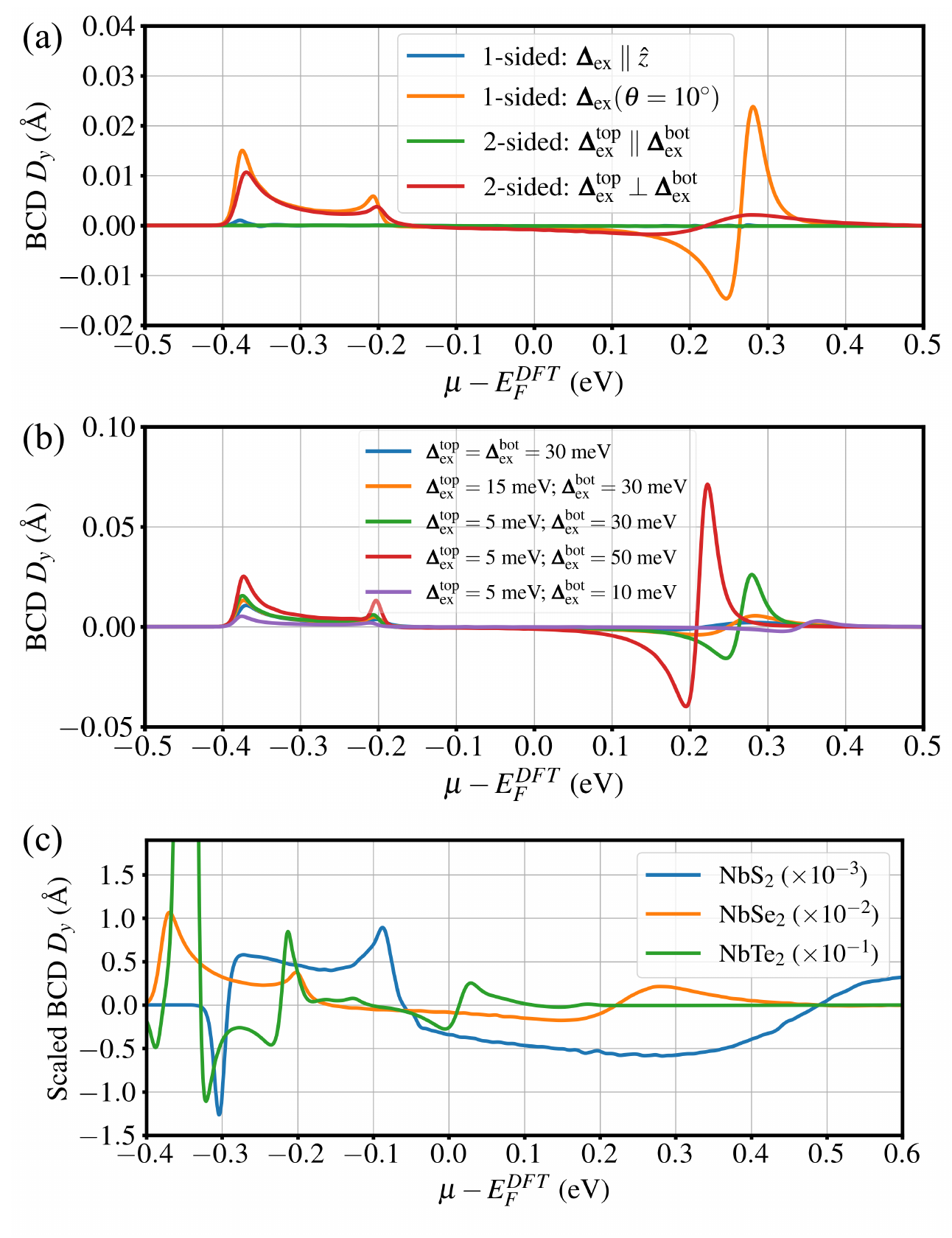}
\caption{ Berry-curvature dipole under layer-selective proximity exchange in $1H$-Nb$X_2$.
The leading intrinsic second-order Hall response is governed by the BCD component
$D_y \equiv D_{yz}^{\rm 2D}=\sum_n\!\int_{\rm BZ}\frac{d^2k}{(2\pi)^2}\, f_{n\mathbf{k}}\;\partial_{k_y}\Omega^z_n(\mathbf{k})$
(in units of \AA). With the quasistatic convention $\chi_{abc}^{\rm sheet}=-(e^3\tau/\hbar^2)\epsilon_{adc}D_{bd}^{\rm 2D}$ and $e>0$, this component yields $\chi_{xyy}^{\rm sheet}=(e^3\tau/\hbar^2)D_{yz}^{\rm 2D}$ for $\omega\tau\ll 1$.
(a) $D_y$ of NbSe$_2$ for representative exchange textures: one-sided $\boldsymbol{\Delta}_{\rm ex}\parallel\hat{z}$, one-sided canted $\boldsymbol{\Delta}_{\rm ex}(\theta=10^\circ)$ (tilted from $\hat{z}$ toward $\hat{x}$ in the $xz$ plane), two-sided parallel $\boldsymbol{\Delta}^{\rm top}_{\rm ex}\parallel\boldsymbol{\Delta}^{\rm bot}_{\rm ex}$, and two-sided orthogonal $\boldsymbol{\Delta}^{\rm top}_{\rm ex}\perp\boldsymbol{\Delta}^{\rm bot}_{\rm ex}$.
(b) Tunability of $D_y$ in the orthogonal geometry by independently varying the magnitudes of the top-interface in-plane exchange and the bottom-interface out-of-plane exchange (here $\boldsymbol{\Delta}^{\rm top}_{\rm ex}\parallel\hat{x}$ and $\boldsymbol{\Delta}^{\rm bot}_{\rm ex}\parallel\hat{z}$); parameter sets are indicated in the legend.
(c) Chalcogen dependence of $D_y$ in the orthogonal configuration for NbS$_2$, NbSe$_2$, and NbTe$_2$, showing a pronounced increase from S$\rightarrow$Se$\rightarrow$Te. For every rescaled curve, the physical value equals the displayed ordinate multiplied by the factor printed in the legend. Energies are referenced to the DFT Fermi level; unless stated otherwise, $|\boldsymbol{\Delta}_{\rm ex}|=30~\mathrm{meV}$ is applied to the proximitized chalcogen $p$ orbitals.}
\label{fig:BCD_fig3}
\end{figure}

\emph{Nonlinear Hall channel and BCD.}
In two dimensions, the intrinsic dc sheet nonlinear Hall tensor satisfies
$\chi_{abc}^{\rm sheet}=-(e^3\tau/\hbar^2)\epsilon_{adc}D_{bd}^{\rm 2D}$ in the quasistatic limit $\omega\tau\ll 1$ \cite{SodemannFu15}. Because the calculated component is $D_y\equiv D_{yz}^{\rm 2D}$, the corresponding Hall geometry uses a longitudinal field along $y$ and a transverse response along $x$:
\begin{equation}
j_x^{(2),\rm sheet}=\chi_{xyy}^{\rm sheet}E_y^2,\qquad
\chi_{xyy}^{\rm sheet}=\frac{e^3\tau}{\hbar^2}D_{y}.
\end{equation}
The transport time $\tau$ sets the overall scale, while symmetry and band geometry determine the sign structure and energy dependence of $D_y$.

Figure~\ref{fig:BCD_fig3} shows that proximity texture provides an independent symmetry knob for activating the intrinsic NLHC. For NbSe$_2$ [Fig.~\ref{fig:BCD_fig3}(a)], one-sided exchange strictly along $\hat{z}$ has $D_y=0$ by the exact residual $C_3$ symmetry, even though $\sigma_{xy}$ can be finite; the tiny nonzero values visible in the interpolated curve are numerical residuals and define a numerical floor rather than a physical BCD. Introducing an in-plane component by canting activates a finite, sharply dispersive $D_y$. In the two-sided geometry, parallel out-of-plane alignment likewise has $D_y=0$ by $C_3$ symmetry up to numerical residuals, whereas the orthogonal configuration produces a finite BCD without requiring fine control of a small canting angle: the out-of-plane component sustains $\Omega^z(\mathbf{k})$ while the in-plane component shifts and unbalances hot spots in $\mathbf{k}$ space, generating the required dipole. The magnitude and sign of $D_y$ vary with both exchange strengths [Fig.~\ref{fig:BCD_fig3}(b)], and the orthogonal BCD grows from NbS$_2$ to NbTe$_2$ [Fig.~\ref{fig:BCD_fig3}(c)]. Supplemental calculations show how the fixed-band spectra evolve under electronic Fermi-Dirac smearing up to $T=300$~K and under selected top-bottom exchange asymmetries \cite{supplemental_material}; these tests do not include temperature-dependent magnetic order, scattering, CDW reconstruction, or interface structure.

\begin{figure}[tb]
\includegraphics[width=\columnwidth]{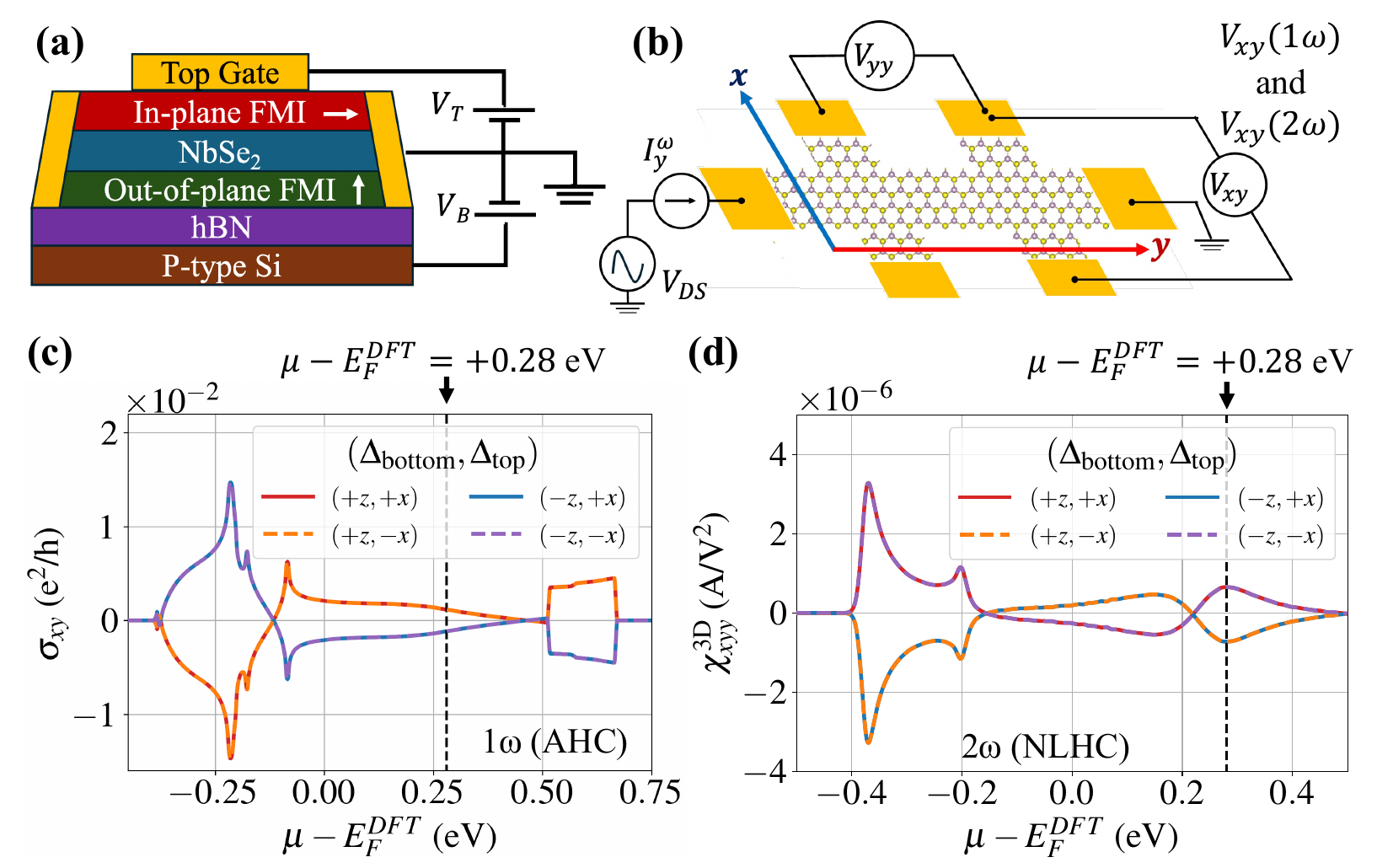}
\caption{ Harmonic Hall signatures of orthogonal two-sided magnetic proximity in monolayer NbSe$_2$.
(a) Dual-interface heterostructure: The bottom ferromagnetic insulator (FMI) provides predominantly out-of-plane proximity exchange, and the top FMI predominantly in-plane exchange; gating tunes the chemical potential $\mu$.
(b) Hall-bar geometry driven by an ac current $I_y(t)=I_0\cos\omega t$, with the transverse $x$-voltage $V_{xy}$ recorded at the first and second harmonics.
(c) Intrinsic AHC $\sigma_{xy}(\mu)$ for four magnetic configurations, showing sign control governed primarily by reversal of the out-of-plane component. The legend shorthand $\Delta_{\rm bottom}=\pm z$ and $\Delta_{\rm top}=\pm x$ denotes $\boldsymbol{\Delta}_{\rm ex}^{\rm bot}\parallel\pm\hat{\mathbf z}$ and $\boldsymbol{\Delta}_{\rm ex}^{\rm top}\parallel\pm\hat{\mathbf x}$, respectively.
(d) Bulk-normalized nonlinear Hall coefficient $\chi_{xyy}^{\rm 3D}(\mu)=\chi_{xyy}^{\rm sheet}(\mu)/c$ (in A/V$^2$) for the same configurations, where $c$ is the out-of-plane simulation-cell length. The device equations below use $\chi_{xyy}^{\rm sheet}=c\chi_{xyy}^{\rm 3D}$. At the representative operating point $\mu^*$, the exchange-odd part of the response follows the leading-order product $\Delta_z\Delta_x$ in the ideal model; it therefore changes under reversal of either component separately and is unchanged under simultaneous reversal of both. The vertical dashed line marks $\mu^*$.}
\label{fig:memory_fig4}
\end{figure}

\emph{Harmonic Hall signatures of orthogonal two-sided proximity.}
To connect the symmetry analysis with an experimentally relevant geometry, we consider the orthogonal two-sided structure of Fig.~\ref{fig:memory_fig4}(a,b), where the bottom interface provides predominantly out-of-plane exchange and the top interface predominantly in-plane exchange. Possible magnetic material choices and a vector-field switching protocol for independently reversing the out-of-plane and in-plane proximity components are discussed in Supplemental Material Sec.~S6.1~\mbox{\cite{supplemental_material}}. For a $y$-directed ac current, the open-circuit transverse field is along $x$. We denote the corresponding Hall voltage by $V_{xy}\equiv W E_x$ and decompose it as

\begin{equation}
V_{xy}(t)=V_{xy}^{0}+V_{xy}^{1\omega}\cos\omega t+V_{xy}^{2\omega}\cos2\omega t+\cdots .
\end{equation}
To leading order in the Hall angle ($|\sigma_{xy}^{\rm sheet}|\ll\sigma_{xx}^{\rm sheet}\simeq\sigma_{yy}^{\rm sheet}$) and retaining the intrinsic BCD contribution to $\chi_{xyy}^{\rm sheet}$, the harmonic amplitudes scale as
\begin{equation}
\begin{aligned}
V_{xy}^{1\omega}&\simeq
-\frac{\sigma_{xy}^{\rm sheet}(\mu)}{[\sigma_{xx}^{\rm sheet}(\mu)]^2}\,I_0,\\
V_{xy}^{2\omega}&\simeq
-\frac{\chi_{xyy}^{\rm sheet}(\mu)}{2[\sigma_{xx}^{\rm sheet}(\mu)]^3}\,
\frac{I_0^2}{W},
\end{aligned}
\label{eq:harmonic_scaling}
\end{equation}

where $W$ is the transverse Hall-bar width. Geometric factors and sign conventions are summarized in Supplemental Material Sec.~S6.2 \cite{supplemental_material}.

Using representative values
$\tau=25$~fs,
$|\sigma_{xy}^{\rm sheet}|\sim 10^{-2}(e^2/h)$,
$\sigma_{xx}^{\rm sheet}\sim 1$--$5$~mS,
$W=5~\upmu{\rm  m}$,
$I_0=0.1$--$1$~mA,
and
$|D_y|\sim 10^{-3}$--$3\times10^{-2}$~\AA\ as a conservative representative range rather than the global extrema of every calculated spectrum,
we estimate
$|V_{xy}^{1\omega}|\sim 2$--$400~\upmu{\rm V}$
and
$|V_{xy}^{2\omega}|\sim 10^{-10}$--$10^{-6}$~V for $I_0\sim 1$~mA, reaching $10^{-9}$--$10^{-6}$~V in the most active BCD windows
(Supplemental Material Sec.~S6.4 \cite{supplemental_material}).

A useful feature of the ideal orthogonal design is the component-selective sign response obtainable from two switching operations. At the operating point $\mu^*$, and up to a common overall sign fixed by the Hall-voltage lead convention, the calculated exchange-odd contributions satisfy $\mathrm{sgn}\,V_{xy}^{1\omega}\sim s_z$ and $\mathrm{sgn}\,V_{xy}^{2\omega}\sim s_zs_x$, consistent with the leading-order relation $D_y\propto\Delta_z\Delta_x$ (Supplemental Material Sec.~S6.3 \mbox{\cite{supplemental_material}}). With the explicit convention $V_{xy}=WE_x$ used in Eq.~\eqref{eq:harmonic_scaling}, both relations carry the same overall minus sign. Reversing the in-plane component at a fixed out-of-plane component flips the exchange-odd part of $V_{xy}^{2\omega}$ while leaving $V_{xy}^{1\omega}$ unchanged; the simultaneous reversal of both components flips $V_{xy}^{1\omega}$ while leaving that exchange-odd second-harmonic part unchanged. Reversing the out-of-plane component alone generally flips both. A real interface can also generate a structural $C_3$-breaking second-harmonic background that does not follow this product law; reversal combinations isolate the exchange-odd contribution from such an offset. Other extrinsic $2\omega$ backgrounds and separation strategies are discussed in Supplemental Material Sec.~S6.5 \cite{supplemental_material}. The four ideal configurations shown in Figs.~\ref{fig:memory_fig4}(c) and (d) therefore provide distinct relative sign combinations of the two exchange-odd Hall channels.

\emph{Conclusion.\textemdash}
Layer-selective magnetic proximity in metallic monolayer $1H$-Nb$X_2$ provides a symmetry-controlled route to co-engineer intrinsic linear and nonlinear Hall responses within a single two-dimensional metal. A one-sided exchange that breaks $\sigma_h$ and contains a finite out-of-plane component activates interface-induced $k$-linear Rashba-enabled Berry-curvature hot spots and yields an intrinsic AHC, while an independently oriented in-plane proximity component permits a dipolar curvature asymmetry and hence a finite BCD controlling the intrinsic NLHC. Fully relativistic Wannier calculations across Nb$X_2$ support the magnetic-point-group selection rules and Hall-valve behavior and show a cross-material increase correlated with heavier chalcogens and larger SOC, together with other band-structure changes. Within the idealized local-exchange model, the orthogonal two-sided geometry further illustrates component-selective reversal of the exchange-odd first- and second-harmonic Hall signals. Quantitative magnitudes and operating windows remain dependent on the interface model, normal-state electronic structure, and transport parameters.

\emph{Acknowledgments.}
The calculations in this study were performed on the Numerical Materials Simulator at the NIMS. This work is financially supported by JSPS KAKENHI Grant No. JP24K06968 from MEXT Japan. Y.W. thanks Ryo Kitaura and Daichi Kozawa for fruitful discussions.

\emph{Data availability.} There are no publicly available research data or software supporting this manuscript. Requests for fur- ther information or data should be sent to the authors. 
\nocite{*}

\bibliography{refs}

\clearpage
\onecolumngrid
\begin{center}
{\Large\bfseries Supplementary Materials for\par}
\vspace{0.4em}
{\large\bfseries "Layer-selective proximity symmetry breaking enables anomalous and nonlinear Hall responses in $1H$-Nb$X_2$ ($X =$ S, Se, Te)"\par}
\vspace{1.0em}
{\normalsize Yusuf Wicaksono\par}
\vspace{0.25em}
{\small International Center for Materials Nanoarchitectonics (WPI-MANA), National Institute for Materials Science, 1-1 Namiki, Tsukuba, Ibaraki, 305-0044 JAPAN\par}
\vspace{0.6em}
{\normalsize Toshikaze Kariyado\par}
\vspace{0.25em}
{\small International Center for Materials Nanoarchitectonics (WPI-MANA), National Institute for Materials Science, 1-1 Namiki, Tsukuba, Ibaraki, 305-0044 JAPAN\par}
\vspace{0.25em}
\vspace{0.8em}
{\small \today\par}
\end{center}
\vspace{0.8em}

\noindent\textbf{Abstract.} This Supplemental Material (SM) provides supporting analyses, model derivations, and additional first-principles results for the conclusions presented in the main text. The SM is organized into six sections. In Section S1 we present a magnetic-point-group symmetry analysis that establishes the selection rules for the intrinsic anomalous Hall conductivity (AHC) and the Berry-curvature dipole (BCD) across the proximity configurations considered in the main text, and clarify when interface-induced $k$-linear Rashba SOC is symmetry allowed. Section S2 collects analytic results for the interface-induced $k$-linear Rashba--Zeeman minimal model, including closed-form expressions for the Berry curvature and AHC, and a leading-order scaling estimate for the BCD. Section S3 provides a complementary valley $k\!\cdot\!p$ description appropriate to the $\sigma_h$-preserving two-sided out-of-plane limit, highlighting how AHC can arise from spin/valley filling of pre-existing valley Berry curvature while the BCD remains forbidden by $C_3$. Section S4 details the first-principles and Wannier workflows, the effective layer-resolved Wannier-pseudospin exchange model, and the evaluation and dimensional normalization of AHC, BCD, and NLHC. Section S5 presents additional numerical results and diagnostics, including representative band-structure fingerprints, $k$-resolved Berry-curvature/BCD integrands, electronic Fermi--Dirac smearing, exchange-asymmetry tests, and a PBEsol+$U$ check. Finally, Section S6 outlines harmonic Hall detection considerations and order-of-magnitude estimates relevant for experimental observables.
\vspace{1.0em}
\twocolumngrid

\setcounter{figure}{0}

\renewcommand{\thefigure}{S\arabic{figure}}

\setcounter{table}{0}

\renewcommand{\thetable}{S\arabic{table}}

\setcounter{equation}{0}

\renewcommand{\theequation}{S\arabic{equation}}
\renewcommand{\theHfigure}{S\arabic{figure}}
\renewcommand{\theHtable}{S\arabic{table}}
\renewcommand{\theHequation}{S\arabic{equation}}


\section*{S1. Symmetry analysis and magnetic point groups}

In this section we give the symmetry analysis underlying the selection rules for the anomalous Hall conductivity (AHC) and Berry-curvature dipole (BCD) quoted in the main text. We work at the level of magnetic point groups, focusing on the spatial operations that leave both the crystal and the magnetization pattern invariant (possibly combined with time reversal when appropriate).

Throughout, we consider a two-dimensional metal with Bloch eigenstates
$\lvert u_{n\mathbf{k}}\rangle$, band energies $\varepsilon_{n\mathbf{k}}$,
and out-of-plane Berry curvature
\begin{multline}
\Omega^{z}_{n}(\mathbf{k})
=
-2\,\operatorname{Im}   \\
\sum_{m\neq n}
\frac{
\langle u_{n\mathbf{k}} \vert
\partial_{k_x}H(\mathbf{k})
\vert u_{m\mathbf{k}}\rangle
\langle u_{m\mathbf{k}} \vert
\partial_{k_y}H(\mathbf{k})
\vert u_{n\mathbf{k}}\rangle
}{
\left(
\varepsilon_{n\mathbf{k}}
-
\varepsilon_{m\mathbf{k}}
\right)^{2}
}.
\end{multline}
Here the sum runs over all bands $m\neq n$, and we adopt the Berry-curvature
sign convention used in the WannierBerri calculations. The intrinsic anomalous
Hall conductivity and Berry-curvature dipole are
\begin{align}
\sigma_{xy} &= -\frac{e^2}{\hbar}\sum_n \int_{\mathrm{BZ}}\!\frac{d^2k}{(2\pi)^2}\,
f_{n\bm k}\,\Omega_n^z(\bm k), \label{eq:s1-ahc} \\
D_a &= \sum_n \int_{\mathrm{BZ}}\!\frac{d^2k}{(2\pi)^2}\,
f_{n\bm k}\,\partial_{k_a}\Omega_n^z(\bm k), \quad a\in\{x,y\}, \label{eq:s1-bcd}
\end{align}
with $f_{n\bm k}$ the Fermi–Dirac distribution.
In 2D only the out-of-plane curvature $\Omega^z$ enters. It is therefore convenient to retain the rank-2 notation for the two-dimensional (sheet) BCD,
\begin{equation}
D_{az}\equiv \sum_n\int_{\rm BZ}\frac{d^2k}{(2\pi)^2}\,
f_{n\mathbf{k}}\,\partial_{k_a}\Omega^z_n(\mathbf{k}),
\qquad a\in\{x,y\},
\end{equation}
and to use the shorthand $D_a\equiv D_{az}^{\rm 2D}$, in particular $D_y\equiv D_{yz}^{\rm 2D}$. These components have units of length.
For $e>0$, the semiclassical dc sheet nonlinear Hall tensor is
\begin{equation}
\chi_{abc}^{\rm sheet}=-\frac{e^3\tau}{\hbar^2}\,\epsilon_{adc}D_{bd}^{\rm 2D},
\qquad \omega\tau\ll1.
\label{eq:s1-chi-general}
\end{equation}
Consequently, the calculated component $D_y$ controls
$\chi_{xyy}^{\rm sheet}=(e^3\tau/\hbar^2)D_y$, whereas
$\chi_{yxx}^{\rm sheet}=-(e^3\tau/\hbar^2)D_x$.

Because the Berry curvature is an axial vector, under a unitary spatial operation represented by the orthogonal matrix $R$,
\begin{align}
\Omega_b(\mathbf{k}) &\to \det(R)R_{bb'}\,
\Omega_{b'}(R^{-1}\mathbf{k}),\\
D_{ab} &\to \det(R)R_{aa'}R_{bb'}D_{a'b'}.
\label{eq:s1-D-transform}
\end{align}
For an antiunitary operation $R\mathcal T$, the Berry curvature acquires an additional minus sign and its argument becomes $-R^{-1}\mathbf{k}$; the derivative in $D_{ab}$ supplies a second minus sign, so the BCD obeys the same spatial pseudotensor transformation in Eq.~\eqref{eq:s1-D-transform}. In two dimensions,
\begin{equation}
D_a'\equiv D_{az}'=
\det(R)R_{zz}\sum_{a'=x,y}R_{aa'}D_{a'}.
\label{eq:s1-D-2d-transform}
\end{equation}
Thus $\mathbf D$ transforms as an ordinary in-plane polar vector under proper rotations such as $C_3$, but its transformation under mirrors must retain the factor $\det(R)R_{zz}$.

Time reversal acts as
\begin{equation}
\Omega^z_n(\mathbf{k})\xrightarrow{\mathcal T}
-\Omega^z_n(-\mathbf{k}),\qquad
f_{n\mathbf{k}}\xrightarrow{\mathcal T}f_{n,-\mathbf{k}}.
\end{equation}
Time reversal therefore forces the intrinsic AHC to vanish, but it does not by itself force the BCD to vanish. Indeed, $\Omega_n^z(\mathbf{k})$ is odd under $\mathbf{k}\to-\mathbf{k}$, whereas $\partial_{k_a}\Omega_n^z(\mathbf{k})$ is even. Consequently, the BCD integrand is compatible with time-reversal symmetry, and additional crystalline symmetries, such as $C_3$, are required to enforce $\mathbf D=\mathbf0$.

\vspace{0.5em}
\subsection*{S1.1. Pristine monolayer $1H$-Nb$X_2$}

The pristine $1H$-Nb$X_2$ monolayer has point group $D_{3h}$
(see Fig.~1(a) of the main text), generated by a threefold rotation $C_3$ around $z$ and the horizontal mirror $\sigma_h$ through the Nb plane, together with three vertical mirrors. The non-magnetic system also has time reversal $\mathcal T$.

Time reversal enforces
\begin{equation}
\Omega^z_n(\mathbf{k})=-\Omega^z_n(-\mathbf{k}),
\end{equation}
and hence $\sigma_{xy}=0$ from Eq.~\eqref{eq:s1-ahc}. The BCD is not forbidden by time reversal alone. Instead, under the proper rotation $C_3$, Eq.~\eqref{eq:s1-D-2d-transform} reduces to the ordinary in-plane vector transformation
\begin{equation}
\mathbf D=R_{C_3}\mathbf D.
\end{equation}
The only in-plane vector invariant under a $120^\circ$ rotation is $\mathbf D=\mathbf0$. Thus the pristine monolayer has vanishing intrinsic AHC because of time reversal and vanishing BCD because of $C_3$.

\vspace{0.5em}
\subsection*{S1.2. Horizontal mirror and interface-induced $k$-linear Rashba spin–orbit coupling}

The horizontal mirror $\sigma_h$ acts as $(x,y,z)\to(x,y,-z)$.
Under $\sigma_h$, polar vectors transform as
\begin{equation}
(k_x,k_y,k_z) \xrightarrow{\sigma_h} (k_x,k_y,-k_z),
\end{equation}
while axial vectors transform with the opposite parity of the in-plane and out-of-plane components. For the spin operator $\textbf{s}$ (an axial vector),
\begin{equation}
(s_x,s_y,s_z) \xrightarrow{\sigma_h} (-s_x,-s_y,s_z).
\end{equation}
The interface-induced $k$-linear Rashba term in the Hamiltonian has the form
\begin{equation}
H_R = \alpha(\mathbf{k}\times\mathbf{s})_z
= \alpha(k_xs_y-k_ys_x).
\end{equation}
Under $\sigma_h$,
\begin{equation}
H_R \xrightarrow{\sigma_h}
\alpha\big[k_x(-s_y)-k_y(-s_x)\big] = -H_R.
\end{equation}
If $\sigma_h$ is an exact symmetry of the magnetic point group, the Hamiltonian must satisfy $H_R = \sigma_h H_R \sigma_h^{-1}$, which implies $\alpha=0$. Thus a nonzero interface-induced $k$-linear Rashba coefficient $\alpha$ is only allowed when the horizontal mirror is broken by the environment and/or the magnetic configuration.

This simple observation will be used below to distinguish mirror-preserving two-sided configurations (no interface-induced $k$-linear Rashba) from mirror-breaking one-sided and canted configurations (interface-induced $k$-linear Rashba allowed).

\vspace{0.5em}
\subsection*{S1.3. One-sided proximity}

We now consider magnetic proximity exchange acting primarily on a single chalcogen sublayer (a ``one-sided'' or layer-asymmetric configuration). The one-sided environment breaks $\sigma_h$ but may preserve $C_3$ and the three vertical mirror planes before magnetization is introduced. We distinguish two limiting magnetization directions.

\paragraph*{(i) One-sided, out-of-plane magnetization.}

For $\mathbf m\parallel\hat{\mathbf z}$, the unitary threefold rotation $C_3$ remains a symmetry. A vertical mirror $\sigma_v$ reverses the axial component $m_z$ and therefore is not a unitary symmetry of the magnetized state; instead, the antiunitary operation $\sigma_v\mathcal T$ is preserved. The corresponding magnetic point group is the type-III group customarily denoted $3m'$, with unitary subgroup $C_3$.

Because $\sigma_h$ is broken, an interface-induced $k$-linear Rashba term is symmetry allowed. The antiunitary vertical mirrors $\sigma_v\mathcal T$ do not forbid an out-of-plane AHC, so a finite $\sigma_{xy}$ is allowed. The exact $C_3$ rotation nevertheless imposes
\begin{equation}
\mathbf D=R_{C_3}\mathbf D,
\qquad
R_{C_3}=
\begin{pmatrix}
-\frac12 & -\frac{\sqrt3}{2}\\
\frac{\sqrt3}{2} & -\frac12
\end{pmatrix},
\end{equation}
whose only solution is $\mathbf D=\mathbf0$. Thus one-sided out-of-plane exchange allows an intrinsic AHC but forbids the BCD through $C_3$.

\paragraph*{(ii) One-sided, canted or in-plane magnetization.}

A generic in-plane component $\mathbf m_\parallel$ breaks $C_3$ and all mirror-related operations, leaving the trivial magnetic point group $C_1$; both $\sigma_{xy}$ and $\mathbf D$ are then symmetry allowed. If the canted magnetization lies in a vertical mirror plane, the antiunitary operation $\sigma_v\mathcal T$ can remain. For a mirror plane chosen as $xz$, Eq.~\eqref{eq:s1-D-2d-transform} gives
\begin{equation}
(D_x,D_y)\xrightarrow{\sigma_v\mathcal T}(-D_x,D_y),
\end{equation}
so invariance requires $D_x=0$ while $D_y$ is allowed. Hence the surviving mirror fixes the allowed BCD component perpendicular to the in-plane mirror line; it is not a microscopic choice between parallel and perpendicular orientations. This is the symmetry origin of the canting-activated BCD used in the main text.

\begin{table*}[tb]
\centering
\caption{Summary of symmetry-allowed intrinsic responses for the proximity configurations considered in the main text. Here $C_3$ indicates an exact threefold rotation around $z$, $\sigma_h$ the unitary horizontal mirror, ``interface-induced $k$-linear Rashba'' the symmetry allowance of an interface-induced $k$-linear Rashba SOC term, and the last column indicates whether a finite intrinsic AHC $\sigma_{xy}$ and BCD $\mathbf D$ are symmetry allowed (``$\checkmark$'') or forbidden (``0'') in the clean limit.}
\label{tab:s1-symmetry-summary}

\small
\setlength{\tabcolsep}{5pt}
\renewcommand{\arraystretch}{1.15}
\begin{tabular}{l c c c c}
\hline\hline
Configuration & $C_3$ & $\sigma_h$ & Interface-induced $k$-linear Rashba & $(\sigma_{xy},\mathbf D)$ \\
\hline
Pristine (no exchange) & yes & yes & 0 & $(0,0)$ \\
One-sided, $m_z$ & yes & broken & allowed & $(\checkmark,0)$ \\
One-sided, canted & no & broken & allowed & $(\checkmark,\checkmark)$ \\
Two-sided, parallel $m_z$ & yes & yes & 0 & $(\checkmark,0)$ \\
Two-sided, antiparallel $m_z$ & yes & broken, but $\sigma_h\mathcal T$ preserved & forbidden by $\sigma_h\mathcal T$ & $(0^\dagger,0^\dagger)$ \\
Orthogonal two-sided & no & broken & allowed & $(\checkmark,\checkmark)$ \\
\hline\hline
\end{tabular}

\vspace{0.25em}
\noindent
$^\dagger$ In the ideal symmetric antiparallel stack, the antiunitary symmetry $\sigma_h\mathcal T$ forbids both the interface-induced $k$-linear Rashba term and the intrinsic AHC, while $C_3$ forbids the BCD. Unequal top/bottom coupling or gating asymmetry can break $\sigma_h\mathcal T$ and generate a weak residual AHC. The BCD remains zero as long as $C_3$ is retained and becomes allowed only if an additional in-plane anisotropy also breaks $C_3$.
\end{table*}

\vspace{0.5em}
\subsection*{S1.4. Two-sided proximity: parallel and antiparallel $m_z$}

We next consider exchange acting on both chalcogen sublayers (``two-sided'' configurations). Their parity under top--bottom layer exchange and their residual horizontal-mirror symmetry depend on the relative orientations of the two interfacial exchange fields. We distinguish several magnetization patterns.

\paragraph*{(i) Two-sided, parallel $m_z$.}

When both interfaces induce the same out-of-plane magnetization,
$\mathbf m_{\rm top}=\mathbf m_{\rm bot}\parallel\hat{\mathbf z}$,
the horizontal mirror $\sigma_h$ and the threefold rotation $C_3$ remain unitary symmetries. The vertical mirrors and the in-plane twofold rotations reverse the axial component $m_z$ and therefore survive only when combined with time reversal. The magnetized state is consequently described by a type-III magnetic point group with unitary subgroup $C_{3h}$, rather than by the ordinary type-I group $D_{3h}$.

The unitary horizontal mirror enforces $\alpha=0$ and forbids the interface-induced $k$-linear Rashba term. The remaining Ising- and valley-coupled SOC can nevertheless support valley-textured Berry curvature, and the broken time-reversal symmetry permits a finite AHC. Independently, the exact $C_3$ rotation forces $\mathbf D=\mathbf0$. Thus the parallel two-sided configuration allows a finite, generally modest AHC but has a symmetry-forbidden BCD.

\paragraph*{(ii) Two-sided, antiparallel $m_z$.}

For $\mathbf m_{\rm top}=-\mathbf m_{\rm bot}\parallel\hat{\mathbf z}$, the layer-odd exchange changes sign under $\sigma_h$, but the combined antiunitary operation $\sigma_h\mathcal T$ leaves the structure invariant. This antiunitary symmetry forbids both the interface-induced $k$-linear Rashba term and the intrinsic AHC. The exact $C_3$ rotation independently forbids the BCD.

The effective exchange field on a Fermi sheet is
$\boldsymbol{\Delta}\propto
w_{\rm top}\Delta_{\rm ex}^{\rm top}\mathbf m_{\rm top}
+w_{\rm bot}\Delta_{\rm ex}^{\rm bot}\mathbf m_{\rm bot}$.
In the ideal symmetric limit it cancels exactly, consistent with the symmetry result. Interface inequivalence can break $\sigma_h\mathcal T$ and generate a weak residual AHC, whereas the BCD remains zero unless $C_3$ is also broken.

\vspace{0.5em}
\subsection*{S1.5. Orthogonal two-sided configuration}

Finally, we consider the orthogonal two-sided configuration used in the experimental geometry discussed in the main text:
the bottom interface induces predominantly out-of-plane exchange, while the top interface induces a predominantly in-plane component, e.g.
\begin{equation}
\bm m_{\rm bot} \parallel \hat{\bm z}, \qquad
\bm m_{\rm top} \parallel \hat{\bm x}.
\end{equation}
The resulting effective field
$\boldsymbol{\Delta}= (\Delta_\parallel^x,0,\Delta_z)$
has both out-of-plane and in-plane components.
This configuration breaks $\sigma_h$ and $C_3$. For a generic in-plane orientation, the remaining magnetic point group is $C_1$ and imposes no constraint on $\sigma_{xy}$ or $\mathbf D$. For the orientation used in the calculations, $\mathbf m_{\rm top}\parallel\hat{\mathbf x}$ is chosen in a vertical mirror plane. The unitary mirror reverses both axial components $m_x$ and $m_z$, so the surviving symmetry is the antiunitary operation $\sigma_v\mathcal T$. Taking the mirror plane to be $xz$, Eq.~\eqref{eq:s1-D-2d-transform} enforces $D_x=0$ and allows $D_y$, consistent with the numerical response. In the interface-induced $k$-linear Rashba–Zeeman minimal model discussed in the main text, the curvature hot spot is shifted by
\begin{equation}
\bm k^\ast \propto \hat{\bm z}\times \boldsymbol{\Delta}_\parallel,
\end{equation}
so that the AHC is controlled by the out-of-plane exchange component (via $m_z$) while the in-plane component shifts the hot spot by $\mathbf{k}^\ast\propto \hat{\mathbf z}\times\boldsymbol{\Delta}_\parallel$ and generates a BCD aligned with this shift direction,
\begin{equation}
\mathbf{D}\propto \hat{\mathbf z}\times\boldsymbol{\Delta}_\parallel .
\label{eq:S17}
\end{equation}
consistent with the numerical trends.

\vspace{0.5em}
\subsection*{S1.6. Summary of symmetry-allowed responses}

Table~\ref{tab:s1-symmetry-summary} summarizes the residual spatial symmetries, interface-induced $k$-linear Rashba allowance, and symmetry-allowed intrinsic responses for the configurations discussed above.

This group-theoretical analysis underpins the selection rules and qualitative trends for $\sigma_{xy}$ and the BCD reported in the main text and provides the symmetry basis for the interface-induced $k$-linear Rashba–Zeeman minimal model of Sec.~S2.

\section*{S2. Interface-induced $k$-linear Rashba--Zeeman minimal model: Berry curvature, AHC, and BCD}
\label{sec:S2.Rashba-Zeeman_minimal_model}

In this section we collect the analytic results for the interface-induced $k$-linear Rashba--Zeeman minimal model used in the main text. The goal is to make explicit how the layer-selective proximity configuration---encoded in an effective exchange field $\boldsymbol{\Delta}$ and interface-induced $k$-linear Rashba coefficient $\alpha$---controls the Berry curvature, anomalous Hall conductivity (AHC), and Berry--curvature dipole (BCD).

\subsection*{S2.1. Model and eigenvalues}

We consider a single spinful band with interface-induced $k$-linear Rashba spin--orbit coupling and an effective Zeeman field $\boldsymbol{\Delta}$. The Hamiltonian reads
\begin{equation}
H(\bm k) = \varepsilon_0(k)\,\mathbb{1}
+ \alpha (k_x s_y - k_y s_x)
+ \Delta_z s_z
+ \boldsymbol{\Delta}_\parallel\!\cdot\!\bm s,
\label{eq:s2-HRZ}
\end{equation}
where $\varepsilon_0(k)$ is an isotropic scalar dispersion (e.g.\ parabolic), $k=|\bm k|$, and $s_i$ are dimensionless Pauli matrices in spin space, normalized to eigenvalues $\pm1$. With this convention the two branches are split by $2|\mathbf d|$. The parameter $\alpha$ is the interface-induced $k$-linear Rashba coefficient, and we decompose the effective Zeeman field as
\begin{equation}
\boldsymbol{\Delta} = \Delta_z \hat{\bm z} + \boldsymbol{\Delta}_\parallel,
\qquad
\boldsymbol{\Delta}_\parallel = (\Delta_\parallel^x,\Delta_\parallel^y,0).
\end{equation}
As discussed in Sec.~S1, $\alpha\neq 0$ is symmetry-allowed only when the horizontal mirror $\sigma_h$ is broken by the environment and/or magnetization pattern, while $\Delta_z$ and $\boldsymbol{\Delta}_\parallel$ encode the projections of the layer-resolved proximity fields onto a given Fermi sheet.

Equation~\eqref{eq:s2-HRZ} may be written in the standard $2\times 2$ form
\begin{equation}
H(\bm k) = \varepsilon_0(k)\,\mathbb{1} + \bm d(\bm k)\cdot\bm s,
\end{equation}
with
\begin{equation}
\bm d(\bm k) = \big(-\alpha k_y+\Delta_\parallel^x,\;\alpha k_x+\Delta_\parallel^y,\;\Delta_z\big).
\label{eq:s2-dvector}
\end{equation}
The eigenvalues are
\begin{equation}
E_\pm(\bm k) = \varepsilon_0(k) \pm |\bm d(\bm k)|,
\label{eq:s2-Epm}
\end{equation}
with
\begin{equation}
|\bm d(\bm k)|^2 = \Delta_z^2 + \alpha^2 k^2 + \Delta_\parallel^2
+ 2\alpha\,\hat{\bm z}\!\cdot(\bm k\times\boldsymbol{\Delta}_\parallel),
\label{eq:s2-d2}
\end{equation}
and $\Delta_\parallel^2=(\Delta_\parallel^x)^2+(\Delta_\parallel^y)^2$.

Equivalently, Eq.~\eqref{eq:s2-d2} can be rewritten by completing the square as
$|\bm d(\bm k)|^2=\Delta_z^2+\alpha^2|\bm k-\bm k^\ast|^2$ with
$\bm k^\ast=(\hat{\bm z}\times\boldsymbol{\Delta}_\parallel)/\alpha$
[Eq.~\eqref{eq:s2-kstar}]. Thus the in-plane field enters the spectrum through an
interface-induced $k$-linear Rashba-controlled momentum shift $\bm k^\ast$ (or inversely
$\boldsymbol{\Delta}_\parallel=\alpha\,\bm k^\ast\times\hat{\bm z}$), making the
$\alpha$--$\boldsymbol{\Delta}_\parallel$ relation explicit.

\subsection*{S2.2. Berry curvature}

For a two-level Hamiltonian $H=\varepsilon_0\mathbb{1}+\bm d\cdot\bm s$ the Berry curvature of band $\nu=\pm$ is
\begin{equation}
\Omega_\nu^z(\bm k)
= \nu\,\Omega^z(\bm k)
= \mp\frac{1}{2}
\frac{\bm d\cdot(\partial_{k_x}\bm d\times\partial_{k_y}\bm d)}{|\bm d|^3},
\label{eq:s2-Om-general}
\end{equation}
where the sign convention is such that $\nu=+$ ($-$) corresponds to the upper (lower) Rashba band.

Using Eq.~\eqref{eq:s2-dvector}, we have
\begin{align}
\partial_{k_x}\bm d &= (0,\alpha,0), \\
\partial_{k_y}\bm d &= (-\alpha,0,0),
\end{align}
so that
\begin{equation}
\partial_{k_x}\bm d\times\partial_{k_y}\bm d
= (0,0,\alpha^2).
\end{equation}
The numerator in Eq.~\eqref{eq:s2-Om-general} is therefore
\begin{equation}
\bm d\cdot(\partial_{k_x}\bm d\times\partial_{k_y}\bm d)
= d_z \alpha^2 = \Delta_z \alpha^2,
\end{equation}
independent of $\bm k$ and $\boldsymbol{\Delta}_\parallel$. The Berry curvature becomes
\begin{equation}
\Omega_\pm^z(\bm k)
=\mp\frac{1}{2}
\frac{\alpha^2\Delta_z}{|\bm d(\bm k)|^3},
\label{eq:s2-Om-compact}
\end{equation}
with $|\bm d(\bm k)|$ given by Eq.~\eqref{eq:s2-d2}. Explicitly,
\begin{equation}
\Omega_\pm^z(\bm k)
=\mp\frac{\alpha^2\Delta_z/2}{
\big[\Delta_z^2+\alpha^2 k^2+\Delta_\parallel^2
+2\alpha\,\hat{\bm z}\!\cdot(\bm k\times\boldsymbol{\Delta}_\parallel)\big]^{3/2}}.
\label{eq:s2-Om-explicit}
\end{equation}
Several limiting cases are immediate:

\begin{itemize}
\item If $\alpha=0$ (mirror intact in this minimal model), then $\Omega_\pm^z(\bm k)\equiv 0$ regardless of $\boldsymbol{\Delta}$. The interface-induced $k$-linear Rashba-Zeeman mechanism for intrinsic Berry curvature is absent.

\item If $\Delta_z=0$, then $\Omega_\pm^z(\mathbf k)\equiv0$ because the $\mathbf d$ vector is coplanar and $\mathbf d\cdot(\partial_{k_x}\mathbf d\times\partial_{k_y}\mathbf d)=0$. An in-plane exchange field can still break time reversal; the vanishing curvature here is therefore a property of this minimal two-level Hamiltonian, not a restoration of TRS.

\item If $\boldsymbol{\Delta}_\parallel=\mathbf 0$, then $|\bm d|^2=\Delta_z^2+\alpha^2 k^2$ and
\begin{equation}
\Omega_\pm^z(\bm k)
=\mp\frac{\alpha^2\Delta_z/2}{
\big(\Delta_z^2+\alpha^2 k^2\big)^{3/2}},
\end{equation}
which is isotropic in $\bm k$; this isotropy implies a vanishing BCD, as discussed below.
\end{itemize}

Equation~\eqref{eq:s2-Om-explicit} reproduces the expression quoted in the main text and underlies the curvature maps of Fig.~1(c).

\subsection*{S2.3. Anomalous Hall conductivity for ${\Delta}_\parallel = 0$}

For $\boldsymbol{\Delta}_\parallel=\mathbf 0$, the problem is rotationally symmetric and the AHC can be evaluated in closed form.
At zero temperature, the AHC is
\begin{equation}
\sigma_{xy}(\mu)
=-\frac{e^2}{\hbar}\sum_{\nu=\pm}
\int\!\frac{d^2k}{(2\pi)^2}\,
\Theta\big(\mu-E_\nu(\bm k)\big)\,\Omega_\nu^z(\bm k),
\label{eq:s2-sxy-def}
\end{equation}
where $\mu$ is the chemical potential, $\Theta$ is the Heaviside step function, and $E_\nu(\bm k)$ are given by Eq.~\eqref{eq:s2-Epm} with $\boldsymbol{\Delta}_\parallel=0$.

Using Eq.~\eqref{eq:s2-Om-compact} with $|\bm d|^2=\Delta_z^2+\alpha^2 k^2$ and going to polar coordinates, we obtain
\begin{equation}
\sigma_{xy}(\mu)
=\frac{e^2}{2\hbar}\sum_{\nu=\pm}\nu
\int_0^{k_{F,\nu}}\!\frac{k\,dk}{2\pi}\,
\frac{\alpha^2\Delta_z}{\big(\Delta_z^2+\alpha^2 k^2\big)^{3/2}},
\end{equation}
where $k_{F,\nu}$ is the Fermi momentum in band $\nu$, defined implicitly by $\mu= \varepsilon_0(k_{F,\nu})+\nu|\bm d(k_{F,\nu})|$, and the integral is present only if $\mu$ lies above the band minimum of band $\nu$.

The radial integral can be performed analytically. Writing $u=\alpha^2 k^2+\Delta_z^2$ and $du=2\alpha^2 k\,dk$, we find
\begin{multline}
\int_0^{k_{F,\nu}}\frac{k\,dk}{\big(\Delta_z^2+\alpha^2 k^2\big)^{3/2}}
=\frac{1}{\alpha^2}
\Bigg[ -\frac{1}{\sqrt{\Delta_z^2+\alpha^2 k^2}} \Bigg]_{0}^{k_{F,\nu}} \\
= \frac{1}{\alpha^2}\left(
\frac{1}{|\Delta_z|} - \frac{1}{\sqrt{\Delta_z^2+\alpha^2 k_{F,\nu}^2}}
\right).
\end{multline}
Substituting back, we obtain
\begin{multline}
\sigma_{xy}(\mu)
=\frac{e^2}{4\pi\hbar}\sum_{\nu=\pm}\nu\,\Delta_z \times \\
\left(
\frac{1}{|\Delta_z|} - \frac{1}{\sqrt{\Delta_z^2+\alpha^2 k_{F,\nu}^2}}
\right)
\Theta\big(\mu-E_\nu^{\mathrm{min}}\big),
\label{eq:s2-sxy-SFS}
\end{multline}
where $E_\nu^{\mathrm{min}}$ is the band minimum for band $\nu$.
The first term in parenthesis yields a quantized contribution proportional to $\mathrm{sgn}(\Delta_z)$ that cancels between the two bands when both are partially occupied. A convenient way to write the result, emphasizing the dependence on the interface-induced $k$-linear Rashba--Zeeman anticrossings, is
\begin{equation}
\sigma_{xy}(\mu)
=-\frac{e^2}{4\pi\hbar}\sum_{\nu=\pm}\nu\,
\frac{\Delta_z}{\sqrt{\Delta_z^2+\alpha^2 k_{F,\nu}^2}}\,
\Theta\big(\mu-E_\nu^{\mathrm{min}}\big),
\label{eq:s2-sxy-final}
\end{equation}
Equation~\eqref{eq:s2-sxy-final} is the closed-form expression referenced in the main text. It applies in the two-band regime where both Rashba-split branches are partially occupied; in the single-Fermi-surface regime, one should use Equation~\eqref{eq:s2-sxy-SFS}.

Equation~\eqref{eq:s2-sxy-final} makes several features transparent:

\begin{itemize}
\item The AHC changes sign under $\Delta_z\to -\Delta_z$ (reversal of the out-of-plane magnetization), as expected for an axial response.

\item For fixed $\Delta_z$, the magnitude of $\sigma_{xy}$ is enhanced when $\mu$ lies near the interface-induced $k$-linear Rashba--Zeeman avoided crossings, where one of the $k_{F,\nu}$ becomes small and the corresponding denominator $\sqrt{\Delta_z^2+\alpha^2 k_{F,\nu}^2}$ decreases.

\item In the limit $|\mu-\varepsilon_0(0)| \gg \sqrt{\Delta_z^2+\alpha^2 k_{F,\nu}^2}$, the contribution of each band tends to zero and $\sigma_{xy}$ decays, as the Fermi surface moves far from the curvature hot spots.
\end{itemize}

These features are directly reflected in the energy dependence of the first-principles AHC curves in Fig.~2 of the main text.

\subsection*{S2.4. Berry--curvature dipole for small in-plane field}

We now consider the effect of a small in-plane component $\boldsymbol{\Delta}_\parallel$ on the BCD.
The BCD is defined by
\begin{equation}
D_a
= \sum_\nu\int_{\mathrm{BZ}}\!\frac{d^2k}{(2\pi)^2}\,
f_\nu(\bm k)\,\partial_{k_a}\Omega_\nu^z(\bm k),
\label{eq:s2-Da-def}
\end{equation}
with $f_\nu(\bm k)$ the Fermi--Dirac distribution in band $\nu$.
For an ideal two-dimensional continuum model, the momentum integral may be taken over all $\bm k$.

\subsubsection*{(i) Shifted hot spot and direction of $\mathbf D$}

A useful way to understand the effect of $\boldsymbol{\Delta}_\parallel$ is to rewrite the denominator of Eq.~\eqref{eq:s2-Om-explicit} in terms of a shifted momentum.
Using
\begin{equation}
\hat{\bm z}\!\cdot(\bm k\times\boldsymbol{\Delta}_\parallel)
= -\bm k\!\cdot(\hat{\bm z}\times\boldsymbol{\Delta}_\parallel),
\end{equation}
we can complete the square as
\begin{multline} \label{eq:s2-kstar}
\Delta_z^2+\alpha^2 k^2+\Delta_\parallel^2
+2\alpha\,\hat{\bm z}\!\cdot(\bm k\times\boldsymbol{\Delta}_\parallel)
= \Delta_z^2 + \alpha^2\big|\bm k - \bm k^\ast\big|^2,
\\
\bm k^\ast=\frac{\hat{\bm z}\times\boldsymbol{\Delta}_\parallel}{\alpha},
\qquad
|\bm k^\ast|=\frac{|\boldsymbol{\Delta}_\parallel|}{|\alpha|}\ll k_F .
\end{multline}
Thus the Berry curvature can be written as
\begin{equation}
\Omega_\pm^z(\bm k)
=\mp\frac{\alpha^2\Delta_z/2}{
\big[\Delta_z^2+\alpha^2|\bm k-\bm k^\ast|^2\big]^{3/2}},
\label{eq:s2-Om-shifted}
\end{equation}
so that it is radially symmetric around the shifted point $\bm k^\ast$ in momentum space.
Within the isotropic continuum description, a nonzero $\bm k^\ast$ shifts the Berry-curvature hot spot away from the expansion point and produces a dipolar imbalance across the Fermi contour, which is precisely what the BCD measures.

For $|\mathbf k^\ast|\ll k_F$, both the Berry curvature and the occupation function acquire first-order corrections. Writing
$f_\nu=f_{\nu,0}+\delta f_\nu$ and
$\Omega_\nu^z=\Omega_{\nu,0}^z+\delta\Omega_\nu^z$, the BCD to linear order is
\begin{equation}
D_a^{(1)}=
\sum_\nu\int\frac{d^2k}{(2\pi)^2}
\left[
f_{\nu,0}\,\partial_{k_a}\delta\Omega_\nu^z
+\delta f_\nu\,\partial_{k_a}\Omega_{\nu,0}^z
\right].
\label{eq:s2-Da-linear}
\end{equation}
The separate curvature-shift and Fermi-surface-shift terms are model dependent, but rotational covariance fixes their sum. In an isotropic continuum the only available in-plane vector at first order is $\mathbf k^\ast$, and therefore
\begin{equation}
D_a^{(1)}=\mathcal M(\mu,T)\,(\mathbf k^\ast)_a,
\qquad
\mathbf D=\mathcal M(\mu,T)\,\mathbf k^\ast
\propto\hat{\mathbf z}\times\boldsymbol{\Delta}_\parallel ,
\label{eq:s2-D-direction}
\end{equation}
where $\mathcal M(\mu,T)$ contains both contributions in Eq.~\eqref{eq:s2-Da-linear}. Reversing $\boldsymbol{\Delta}_\parallel$ reverses $\mathbf k^\ast$ and hence reverses $\mathbf D$. For the orthogonal geometry used in the main text, $\boldsymbol{\Delta}_\parallel\parallel\hat{\mathbf x}$ gives $\mathbf D\parallel\hat{\mathbf y}$ and thus $D_y\neq0$.

\subsubsection*{(ii) Parametric scaling}

We next extract the parametric dependence of $|\mathbf D|$ on the microscopic parameters.
For concreteness, consider zero temperature and an isotropic parabolic dispersion
$\varepsilon_0(k)=\hbar^2 k^2/2m$,
for which the Fermi contours of the Rashba-split branches are approximately circular with radii $k_{F,\nu}$.
Equations~\eqref{eq:s2-Da-linear} and \eqref{eq:s2-D-direction} show that the leading BCD is linear in the momentum shift
\(
\mathbf k^\ast=(\hat{\mathbf z}\times\boldsymbol{\Delta}_\parallel)/\alpha
\)
together with the radial variation of the unshifted Berry curvature on the Fermi contour.
A convenient estimate is therefore
\begin{equation}
\mathbf D \approx \gamma(\mu,T)\,k_F
\left.\frac{\partial \Omega_{z,0}(k)}{\partial k}\right|_{k_F}\mathbf k^\ast,
\qquad
\mathbf k^\ast=\frac{\hat{\mathbf z}\times\boldsymbol{\Delta}_\parallel}{\alpha},
\label{eq:s2-D-scaling}
\end{equation}
where $\Omega_{z,0}(k)\equiv \Omega_z(k)\big|_{\boldsymbol{\Delta}_\parallel=\mathbf 0}$,
$k_F$ denotes a characteristic Fermi momentum of the occupied branch(es), and
$\gamma(\mu,T)$ is a dimensionless factor of order unity that depends on the detailed Fermi-surface weighting and band occupancy.
Equation~\eqref{eq:s2-D-scaling} should be regarded as a parametric estimate rather than a universal closed form.

Using Eq.~\eqref{eq:s2-Om-compact} for $\boldsymbol{\Delta}_\parallel=\mathbf 0$,
\begin{equation}
\Omega_{z,0}(k)
=\mp \frac{\alpha^2\Delta_z}{2\big(\Delta_z^2+\alpha^2 k^2\big)^{3/2}},
\end{equation}
we obtain
\begin{equation}
\frac{\partial \Omega_{z,0}(k)}{\partial k}
=
\pm \frac{3\alpha^4\Delta_z\,k}
{2\big(\Delta_z^2+\alpha^2 k^2\big)^{5/2}}.
\end{equation}
Substituting this into Eq.~\eqref{eq:s2-D-scaling} gives the parametric estimate
\begin{equation}
\mathbf D \sim
\gamma(\mu,T)\,
\frac{\alpha^3\Delta_z\,k_F^2}
{\big(\Delta_z^2+\alpha^2k_F^2\big)^{5/2}}
\left(\hat{\mathbf z}\times\boldsymbol{\Delta}_\parallel\right).
\label{eq:s2-D-scaling-explicit}
\end{equation}
The precise prefactor is not universal: it depends on which Rashba-split branches are occupied and on the detailed Fermi-surface weighting encoded in Eq.~\eqref{eq:s2-Da-linear}. The important points for the discussion in the main text are the following.

\begin{itemize}
\item $\mathbf D$ is linear in $\boldsymbol{\Delta}_\parallel$ for small in-plane field, as required by symmetry.

\item $\mathbf D$ vanishes when either $\alpha=0$ (no interface-induced $k$-linear Rashba term) or $\Delta_z=0$ (zero Berry curvature in this minimal coplanar-$\mathbf d$ limit).

\item The direction of $\mathbf D$ is fixed by $\hat{\mathbf z}\times\boldsymbol{\Delta}_\parallel$.

\item The magnitude of $\mathbf D$ is enhanced when the Fermi surface cuts through the interface-induced Rashba--Zeeman anticrossings, where the Berry curvature varies most rapidly with momentum.
\end{itemize}

The curvature maps in Fig.~1(c) of the main text are obtained by evaluating Eq.~\eqref{eq:s2-Om-shifted} for representative parameter sets.
In the one-sided out-of-plane configuration ($\boldsymbol{\Delta}_\parallel=\mathbf 0$), the curvature is a rotationally symmetric hot spot centered at $\bm k=0$, leading to a finite AHC but vanishing BCD.
In the orthogonal two-sided configuration ($\Delta_z\neq 0$ and $\boldsymbol{\Delta}_\parallel\neq\mathbf 0$), the curvature profile is centered at $\bm k^\ast=\hat{\bm z}\times\boldsymbol{\Delta}_\parallel/\alpha$, generating a dipolar asymmetry and hence a finite BCD proportional to $\boldsymbol{\Delta}_\parallel$ as summarized by Eq.~\eqref{eq:s2-D-scaling-explicit}.

\section*{S3. Valley $k\cdot p$ description for the $\sigma_h$-preserving two-sided $m_z$ limit}

In this section we provide a valley $k\cdot p$ description appropriate for the
$\sigma_h$-preserving two-sided out-of-plane proximity configuration
(two-sided, parallel $m_z$) discussed in the main text.
In this geometry the horizontal mirror $\sigma_h$ is preserved, so interface-induced
$k$–linear Rashba spin–orbit coupling is forbidden (Sec.~S1.2),
but strong Ising- and valley-coupled SOC are present. The AHC arises
from spin-dependent occupation of pre-existing valley-textured Berry curvature, while
the Berry–curvature dipole (BCD) remains symmetry-forbidden by $C_3$ unless additional
in-plane anisotropies are introduced.

\subsection*{S3.1. Valley Hamiltonian and eigenvalues}

Near a valley center $\bm K_\tau$ ($\tau=\pm 1$ labeling $K$ and $K'$),
we consider a minimal two-band massive Dirac Hamiltonian
for a given spin $s=\pm 1$ (spin quantized along $z$),
\begin{multline}
H_{\tau,s}(\bm q)
= \varepsilon_\tau(\bm q)\,\mathbb{1}
+ v_\tau(\tau q_x \sigma_x + q_y \sigma_y)
\\
+ \big[m_0(\tau)+\lambda_I \tau s\big]\sigma_z
+ \Delta_z s,
\label{eq:s3-valley-H}
\end{multline}
where $\bm q=\bm k-\bm K_\tau$ is the momentum measured from valley $\tau$, 
$\varepsilon_\tau(\bm q)$ is a smooth valley-dependent scalar dispersion
(e.g.\ incorporating particle–hole asymmetry), $v_\tau$ is the Dirac velocity,
$m_0(\tau)$ is a valley-dependent mass term (e.g.\ from remote-band coupling),
$\lambda_I$ is the Ising SOC, and $\Delta_z s$ is an out-of-plane exchange term
that is \emph{spin-diagonal} and proportional to the identity in the pseudospin space.
The Pauli matrices $\bm \sigma$ act in the two-band (orbital/sublattice) subspace.

For fixed $(\tau,s)$, $H_{\tau,s}$ has the standard two-level structure
\begin{equation}
H_{\tau,s}(\bm q)
= \varepsilon_\tau(\bm q)\,\mathbb{1}
+ \bm d_{\tau,s}(\bm q)\cdot\bm \sigma
+ \Delta_z s,
\end{equation}
with
\begin{equation}
\bm d_{\tau,s}(\bm q)
= \big(v_\tau\tau q_x,\ v_\tau q_y,\ M_{\tau,s}\big),
~
M_{\tau,s} = m_0(\tau)+\lambda_I \tau s.
\label{eq:s3-dvec}
\end{equation}
The eigenvalues are
\begin{equation}
E_{\tau,s,\nu}(\bm q)
= \varepsilon_\tau(\bm q) + \Delta_z s
+ \nu \sqrt{v_\tau^2 q^2 + M_{\tau,s}^2},
~
\nu=\pm 1,
\label{eq:s3-E}
\end{equation}
corresponding to conduction ($\nu=+$) and valence ($\nu=-$) branches.
Importantly, the exchange term $\Delta_z s$ is proportional to the identity in
pseudospin space and therefore does \emph{not} modify the eigenvectors or the
Berry curvature; it only shifts the energies and hence the occupations.

\subsection*{S3.2. Berry curvature and independence from $\Delta_z$}

For a two-band Hamiltonian of the form
$H=\varepsilon\mathbb{1}+\bm d\cdot\bm \sigma$,
the Berry curvature of band $\nu=\pm$ is given by
\begin{multline}
\Omega_{\tau,s,\nu}^z(\bm q)
= \nu\,\Omega_{\tau,s}^z(\bm q) = \\
-\nu\,
\frac{1}{2}
\frac{
\mathbf{d}_{\tau,s}(\mathbf{q})\cdot
\left[
\partial_{q_x}\mathbf{d}_{\tau,s}(\mathbf{q})
\times
\partial_{q_y}\mathbf{d}_{\tau,s}(\mathbf{q})
\right]
}{
\left|\mathbf{d}_{\tau,s}(\mathbf{q})\right|^{3}
}.
\label{eq:s3-Om-general}
\end{multline}
Using Eq.~\eqref{eq:s3-dvec}, we have
\begin{align}
\partial_{q_x}\bm d_{\tau,s} &= (v_\tau\tau, 0, 0), \\
\partial_{q_y}\bm d_{\tau,s} &= (0, v_\tau, 0),
\end{align}
hence
\begin{equation}
\partial_{q_x}\bm d_{\tau,s}\times\partial_{q_y}\bm d_{\tau,s}
= (0,0,\tau v_\tau^2),
\end{equation}
and therefore
\begin{equation}
\bm d_{\tau,s}\cdot(\partial_{q_x}\bm d_{\tau,s}\times\partial_{q_y}\bm d_{\tau,s})
= M_{\tau,s}\,\tau v_\tau^2.
\end{equation}
Substituting into Eq.~\eqref{eq:s3-Om-general} yields
\begin{equation}
\Omega_{\tau,s,\nu}^z(\bm q)
= -\nu\,
\frac{\tau v_\tau^2 M_{\tau,s}}{2\big(v_\tau^2 q^2+M_{\tau,s}^2\big)^{3/2}}.
\label{eq:s3-Om-explicit}
\end{equation}
Crucially, the Berry curvature depends on $M_{\tau,s}$ (and hence on $\lambda_I$)
but \emph{does not} depend on the exchange term $\Delta_z$, which only shifts the energies through Eq.~\eqref{eq:s3-E}. In other words, in this mirror-preserving valley picture the out-of-plane exchange acts purely as a valley- and spin-dependent ``filling knob'' for Berry curvature that is already present.

\subsection*{S3.3. AHC from valley filling}

The intrinsic AHC at zero temperature is
\begin{equation}
\sigma_{xy}
= -\frac{e^2}{\hbar}\sum_{\tau,s,\nu}
\int\!\frac{d^2q}{(2\pi)^2}\,
\Theta\big(\mu - E_{\tau,s,\nu}(\bm q)\big)\,
\Omega_{\tau,s,\nu}^z(\bm q),
\label{eq:s3-sxy}
\end{equation}
where the sum runs over valleys $\tau$, spins $s$, and bands $\nu$, and $\mu$ is the chemical potential. In the absence of exchange ($\Delta_z=0$) and for a time-reversal-symmetric crystal, each $(\tau,s,\nu)$ contribution is canceled by the corresponding $(-\tau,-s,\nu)$ partner, yielding $\sigma_{xy}=0$.

When $\Delta_z\neq 0$, the energies $E_{\tau,s,\nu}(\bm q)$ are shifted by $\pm\Delta_z$ depending on $s$, while the Berry curvature
$\Omega_{\tau,s,\nu}^z(\bm q)$ remains unchanged. This produces a spin- (and hence valley-) dependent redistribution of occupied states in momentum space and a net AHC. For small $\Delta_z$ one can linearize Eq.~\eqref{eq:s3-sxy} in $\Delta_z$ to obtain
\begin{equation}
\sigma_{xy} \approx \frac{e^2}{\hbar}\Delta_z
\sum_{\tau,s,\nu} s
\int\!\frac{d^2q}{(2\pi)^2}\,
\delta\big(\mu - E_{\tau,s,\nu}^{(0)}(\bm q)\big)\,
\Omega_{\tau,s,\nu}^z(\bm q),
\label{eq:s3-sxy-linear}
\end{equation}
where $E_{\tau,s,\nu}^{(0)}$ are the band energies at $\Delta_z=0$.
Equation~\eqref{eq:s3-sxy-linear} makes explicit that in this
$\sigma_h$-preserving ``valley-filling'' regime the AHC is controlled by the
\emph{pre-existing} Berry curvature encoded in Eq.~\eqref{eq:s3-Om-explicit},
with $\Delta_z$ acting only as a prefactor and as a Fermi-surface selector.

Several qualitative features follow:

\begin{itemize}
\item The AHC changes sign under $\Delta_z\to -\Delta_z$ (spin reversal), as expected for an axial response.

\item The magnitude of $\sigma_{xy}$ is material and filling dependent. In this schematic mechanism, the Berry-curvature texture is set by the Ising mass $M_{\tau,s}$ and the valley structure, whereas $\Delta_z$ shifts the occupations without changing the eigenvectors of Eq.~\eqref{eq:s3-valley-H}.

\item The dominant contribution can be traced to asymmetric filling of $(\tau,s)$ sectors with favorable momentum and a small but nonzero $|M_{\tau,s}|$. At $q=0$, Eq.~\eqref{eq:s3-Om-explicit} gives $|\Omega|\propto |M_{\tau,s}|^{-2}$, whereas at finite $q$ the mass dependence is nonmonotonic.
\end{itemize}

These features provide a schematic filling-based interpretation of the finite AHC without a concomitant BCD in the two-sided parallel $m_z$ configuration of Fig.~2(b). Because the implemented Wannier exchange is orbital projected rather than proportional to the identity of this two-band model, the minimal valley description is not a quantitative magnitude comparison.

\subsection*{S3.4. Vanishing BCD under $C_3$}

The BCD is defined by
\begin{equation}
D_a
= \sum_{\tau,s,\nu}
\int\!\frac{d^2q}{(2\pi)^2}\,
f_{\tau,s,\nu}(\bm q)\,\partial_{q_a}\Omega_{\tau,s,\nu}^z(\bm q),
~
a\in\{x,y\},
\label{eq:s3-Da}
\end{equation}
with $f_{\tau,s,\nu}(\bm q)=\Theta[\mu-E_{\tau,s,\nu}(\bm q)]$ at $T=0$.
Integrating by parts, one can recast Eq.~\eqref{eq:s3-Da} as
\begin{equation}
D_a
= -\sum_{\tau,s,\nu}
\int\!\frac{d^2q}{(2\pi)^2}\,
\big(\partial_{q_a}f_{\tau,s,\nu}(\bm q)\big)\,
\Omega_{\tau,s,\nu}^z(\bm q),
\end{equation}
making it clear that $\mathbf D$ transforms as an in-plane polar vector under the point group.

In the two-sided parallel $m_z$ configuration, the magnetic point group is a type-III group with unitary subgroup $C_{3h}$, rather than the ordinary type-I group $D_{3h}$ (Sec.~S1.4). In particular, the threefold rotation $C_3$ about $\hat{\mathbf z}$ remains an exact unitary symmetry. Under $C_3$, the in-plane polar vector $\mathbf D$ must satisfy
\begin{equation}
\mathbf D = R_{C_3}\,\mathbf D.
\end{equation}
As in Sec.~S1.3, the only in-plane vector invariant under $120^\circ$ rotation is $\mathbf D=\mathbf 0$. Thus the BCD is \emph{exactly zero} in this limit, independent of the detailed values of $M_{\tau,s}$, $\Delta_z$, and the Fermi level, provided that $C_3$ is unbroken.

Equivalently, Eq.~\eqref{eq:s3-Om-explicit} shows that the Berry curvature is rotationally symmetric about each valley center. The exact $C_3$ symmetry therefore forces the total in-plane vector $\mathbf D$ to vanish after summing over all symmetry-related momenta. At $\Delta_z=0$, time reversal relates the sectors $(\tau,s)$ and $(-\tau,-s)$ and gives opposite Berry curvatures for the corresponding time-reversed states. The vanishing BCD in the parallel $m_z$ configuration, however, follows directly from the exact $C_3$ symmetry and does not require a fixed-spin cancellation between the two valleys.

\subsection*{S3.5. Weak $C_3$ breaking and connection to the interface-induced $k$-linear Rashba regime}

If $C_3$ is weakly broken while $\sigma_h$ remains intact (for example by in-plane uniaxial strain or substrate anisotropy that lowers the symmetry to $C_{2v}$ or $C_s$), the above argument for $\mathbf D=\mathbf 0$ no longer holds strictly. In such a case the valley $k\cdot p$ model may be augmented by additional symmetry-allowed terms that tilt the Dirac cones or introduce anisotropic velocities, leading to a small but finite BCD even without a genuine interface-induced $k$-linear Rashba term.

However, in realistic $1H$–Nb$X_2$ heterostructures, the more natural route to large and tunable BCD is via \emph{breaking} $\sigma_h$ through layer-selective proximity and/or in-plane magnetization components, as emphasized in the main text. In that regime, interface-induced $k$–linear Rashba SOC becomes allowed and the interface-induced $k$-linear Rashba–Zeeman minimal model of Sec.~S2 provides a more appropriate description. The valley $k\cdot p$ picture then smoothly connects to the interface-induced $k$-linear Rashba regime as $\alpha$ is turned on and the curvature hot spots sharpen near interface-induced $k$-linear Rashba–exchange anticrossings, consistent with the evolution seen in the first-principles calculations across the different proximity configurations.

\section*{S4. First-principles, Wannier, and transport-calculation details}

\subsection*{S4.1 DFT setup}

All first-principles calculations were performed within density-functional theory using
\textsc{Quantum ESPRESSO} \cite{qespresso, qe2}, with the PBEsol generalized-gradient functional
\cite{PBEsol} and fully relativistic projector-augmented-wave pseudopotentials
from the \textsc{pslibrary} family \cite{paw,fr,pslibrary}. Spin--orbit coupling was treated
self-consistently. 
We modeled the primitive $1\times1$ normal-state monolayers of
$1H$--NbX$_2$ in a slab geometry with an out-of-plane
simulation-cell length of approximately $30$~\AA, providing a
large vacuum region along $z$ to suppress spurious interlayer
interactions.
The $1H$ stacking (space group $P\bar{6}m2$, point group $D_{3h}$) is realized
in a primitive cell containing one Nb and two chalcogen atoms. The $3\times3$ charge-density-wave reconstruction observed in low-temperature monolayer NbSe$_2$ \cite{UgedaNatPhys15} was not included. Accordingly, the NbSe$_2$ spectra represent a normal-state or CDW-suppressed electronic structure; changing the Fermi--Dirac smearing temperature does not incorporate CDW band folding or gap opening.

For NbSe$_2$, the in-plane lattice constant was set to the relaxed value
$a=3.42$~\AA.
The atomic positions were taken in the ideal $1H$ structural motif, with Nb at the
center of the trigonal prism and the Se sublayers displaced symmetrically above and below
the Nb plane. For NbS$_2$ and NbTe$_2$, we used the same structural motif, with relaxed
in-plane lattice constants $a=3.30$~\AA and $a=3.62$~\AA, respectively,
and the corresponding relaxed chalcogen $z$ coordinates.

Plane-wave cutoffs of $90$~Ry for the wave functions and $720$~Ry for the charge density
were used for all three compounds. Metallic occupations were treated with Marzari--Vanderbilt cold smearing \cite{mv_smearing} with a width of $0.01$~Ry. Self-consistent field
calculations employed a dense $42\times 42\times 1$ Monkhorst–Pack $\mathbf{k}$ mesh and
tight convergence criteria for the charge density and total energy
(energy threshold $10^{-10}$~Ry). These parameters define the calculations reported here, but the response-function convergence test in Sec.~S4.4 varies only the Wannier-interpolation integration grid. Separate convergence with respect to the DFT smearing, DFT $\mathbf{k}$ mesh, plane-wave cutoffs, and Wannier subspace is not established by that test.

\subsection*{S4.2 Wannierization and tight-binding models}

\begin{figure}[tb]
\centering
\includegraphics[width=\columnwidth]{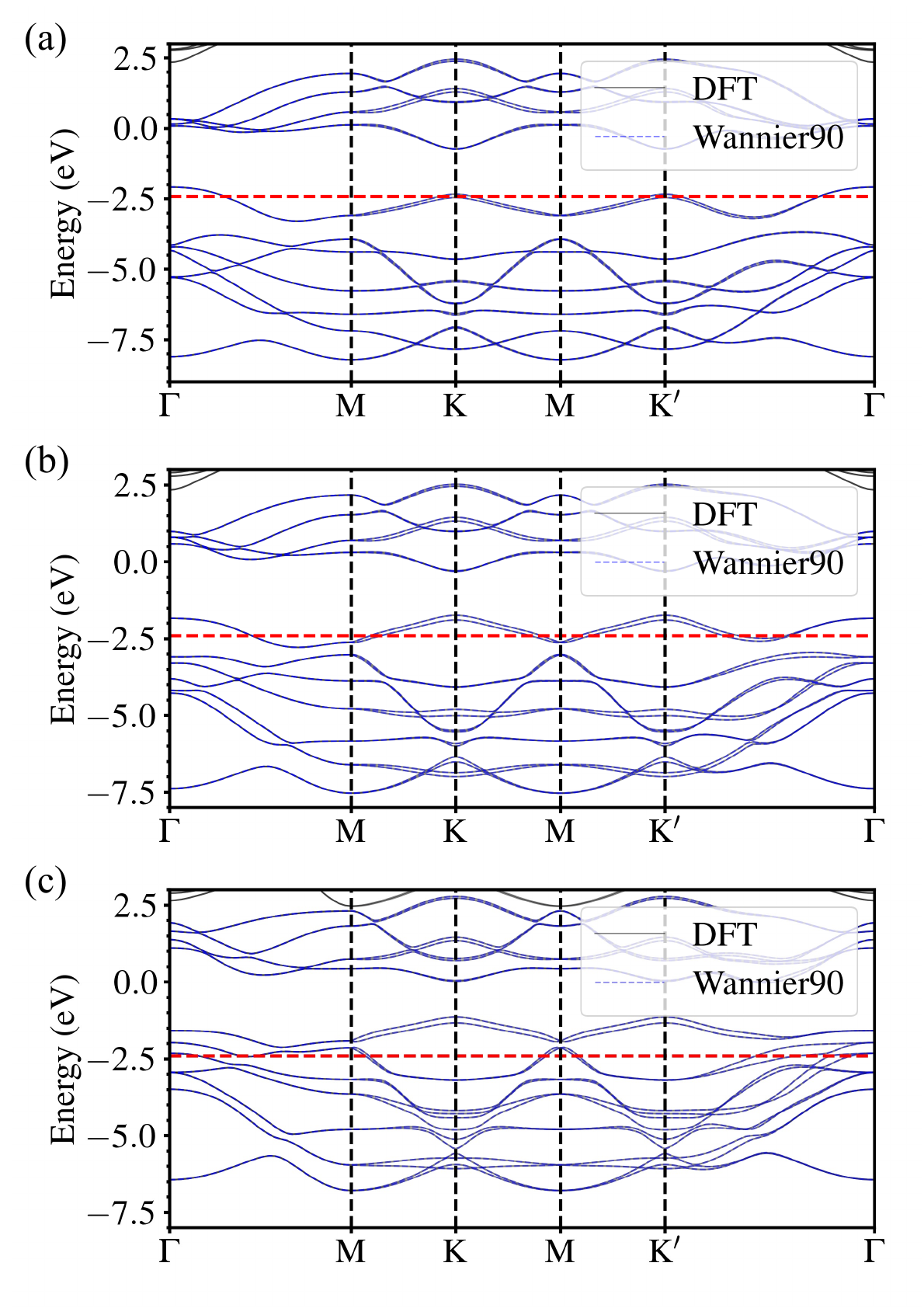}
\caption{
\textbf{Quality of Wannier interpolation for monolayer $1H$--Nb$X_2$.}
Comparison of fully relativistic DFT bands (gray) and Wannier-interpolated bands from
\textsc{wannier90} (blue) along the high-symmetry path
$\Gamma$–M–K–M–K$'$–$\Gamma$ for
\textbf{(a)} NbS$_2$, \textbf{(b)} NbSe$_2$, and \textbf{(c)} NbTe$_2$.
The horizontal red dashed line marks a representative chemical potential used in the Hall
response analysis in the main text.
All Wannier Hamiltonians are constructed from $50$ Bloch bands and $22$ spinor MLWFs
(5 Nb $d$-like and 6 chalcogen $p$-like orbitals per spin) using the energy windows
described in Sec.~S4.2.
Over the energy range shown, and in particular near the Fermi level that controls the
Hall responses discussed in the main text, the Wannier bands track the DFT dispersions
to within $\mathcal{O}(10)$~meV, faithfully capturing the spin–orbit-induced splittings
and valley structure.}
\label{fig:Wannier_DFT_S1}
\end{figure}

Maximally localized spinor Wannier functions (MLWFs) were constructed using \textsc{wannier90}
\cite{Mustofi2012,Pizzi2020,Mostofi2008,Mostofi2014}, starting from the fully relativistic
Bloch states. For NbSe$_2$ we included $50$ Kohn–Sham bands in the disentanglement window and
targeted $22$ spinor Wannier functions per unit cell. The corresponding spatial orbitals
comprise five Nb $d$-like and six Se $p$-like Wannier functions (per spin), chosen to capture
the Nb-derived Fermi sheet and the nearby Se $p$ bands that control the Berry curvature.

The outer (disentanglement) energy window was set to $[-8,3]$~eV relative to the DFT Fermi
level, and the frozen window to $[-8,2]$~eV. Disentanglement was iterated for up to $500$
steps with a mixing ratio of $0.5$, followed by $200$ spread-minimization iterations.
Initial projections were chosen as Nb-centered $d$ orbitals and Se-centered $p$ orbitals,
with spinors aligned along the $z$ axis (spin quantization axis along $\hat{\mathbf z}$).
As shown in Fig.~\ref{fig:Wannier_DFT_S1}(b), the MLWF band structure reproduces the fully relativistic DFT bands
to within a few meV along the plotted high-symmetry lines, and to better than
$\sim 10$~meV in the vicinity of the Fermi level that governs the anomalous and nonlinear
Hall responses, including the spin–orbit-induced splittings and avoided crossings.

The same Wannierization protocol was applied to NbS$_2$ and NbTe$_2$, with the same number of Kohn-Sham bands, number of wannier orbitals, and energy windows, using their respective
lattice constants and atomic positions. For all three compounds the Brillouin-zone sampling
for the Wannierization step used a $60\times60\times1$ uniform $\mathbf{k}$ grid, ensuring
smooth interpolation of Berry curvature and related quantities; the corresponding Wannier–DFT
agreement for NbS$_2$ and NbTe$_2$ is illustrated in Figs.~\ref{fig:Wannier_DFT_S1}(a) and \ref{fig:Wannier_DFT_S1}(c), respectively.

\subsection*{S4.3 Modeling layer-selective proximity exchange at the Wannier level}

Magnetic proximity is introduced at the level of the Wannier-interpolated tight-binding
Hamiltonian by adding a layer-resolved on-site exchange term acting on the chalcogen-like
spinor Wannier functions. Starting from the spinor Hamiltonian
$H_0(\mathbf{k})$ obtained from \textsc{Wannier90} (Sec.~S4.2), we modify the
Wannier-space Hamiltonian as
\begin{equation}
H(\mathbf{k}) = H_0(\mathbf{k}) + H_{\rm ex},
\end{equation}
with
\begin{equation}
H_{\rm ex}
= \sum_{\ell \in \mathcal{L}_{\rm prox}}
\Delta_{\rm ex}^{(\ell)} \mathbf{m}_\ell\cdot\mathbf{s}_\ell,
\end{equation}
where $\mathcal{L}_{\rm prox}$ denotes the set of proximitized chalcogen Wannier orbitals,
$\Delta_{\rm ex}^{(\ell)}$ is the effective exchange strength on orbital $\ell$,
$\mathbf{m}_\ell$ is a unit vector specifying its local exchange orientation, and
$\mathbf{s}_\ell$ denotes a dimensionless Pauli matrix acting within the corresponding projection-defined Wannier pair, with eigenvalues $\pm1$. In the actual implementation, the two spinor-projection-derived MLWFs associated with each spatial orbital are ordered as a pair and standard $2\times2$ Pauli blocks are inserted in that chosen Wannier pseudospin subspace. The complete physical-spin operator in an SOC-entangled Wannier gauge would instead contain matrix elements
\begin{equation}
\mathbf S_{ij}=\langle W_i|\mathbf{s}|W_j\rangle
\end{equation}
over the full Wannier subspace. We did not replace the Pauli-block approximation by this full projected-spin matrix. Thus $\Delta_{\rm ex}^{(\ell)}$ is the coefficient of an implemented Wannier-pseudospin matrix, not a basis-independent physical-spin splitting; a fully localized pure-spin orbital would have a splitting $2\Delta_{\rm ex}^{(\ell)}$, while the shifts of Bloch bands are reduced by their layer, orbital, and pseudospin weights. In practice
we identify projection-derived Wannier pairs corresponding to the top and bottom chalcogen $p$ orbitals and
apply either (i) a \emph{one-sided} (layer-asymmetric) exchange acting only on one chalcogen
sublayer, or (ii) a \emph{two-sided} exchange acting on both sublayers, with
parallel, antiparallel, or orthogonal orientations for $\mathbf{m}_{\rm top}$ and
$\mathbf{m}_{\rm bot}$ as in the main text. All other Wannier matrix elements, including
Nb-centered $d$ orbitals and inter-site hoppings, are left unchanged, so that the only
control parameter is the layer-resolved exchange texture. 
This local form should be understood as the leading static proximity self-energy
obtained after integrating out the magnetic substrate and projecting onto the
NbX$_2$ Wannier subspace. In an explicit magnet/NbX$_2$ heterostructure,
spin-dependent hybridization with the magnetic layer can in general generate
both local exchange-like terms and nonlocal corrections, including modified
hoppings, interface-induced electrostatic potentials, and substrate-dependent
orbital mixing. The present model retains only the dominant local spin-dependent
component because it provides a minimal proof-of-principle way to isolate how the layer parity and
orientation of the proximity exchange constrain $\sigma_{xy}$ and the Berry-curvature
dipole by symmetry. Quantitative results can depend on the chosen Wannier gauge and orbital partition until the full projected-spin matrix or a material-specific interface downfolding is supplied.

The assumption of short-ranged, layer-selective exchange is motivated by the microscopic origin of magnetic proximity in van der Waals heterostructures.
The induced exchange arises from spin-dependent hybridization and virtual hopping between the magnetic layer and the adjacent TMD orbitals.
It is therefore controlled by wave-function overlap across the van der Waals gap and is expected to decay rapidly away from the interface.
A useful phenomenological representation is
\begin{equation}
\Delta_{\rm ex}^{(\ell)}\sim \Delta_0 \exp[-d_\ell/\xi],
\end{equation}
where $d_\ell$ is the distance of layer $\ell$ from the magnetic interface and $\xi$ is an interfacial decay length set by the evanescent wave-function extent.
Our one-sided exchange model corresponds to the limiting assumption in which the adjacent chalcogen sublayer receives the dominant exchange self-energy, while the opposite chalcogen sublayer and Nb orbitals are neglected. Interfacial decay across a van der Waals gap motivates locality but does not by itself validate this detailed orbital partition inside NbX$_2$.
This short-range picture is consistent with layer-resolved WSe$_2$/CrI$_3$ experiments, where monolayer WSe$_2$ interfaced with bilayer or trilayer CrI$_3$ shows that the spin-dependent charge-transfer response is dominated by the interfacial CrI$_3$ layer~\cite{ZhongNatNano20}.
It is also supported by first-principles calculations for monolayer MoSe$_2$/CrI$_3$ and WSe$_2$/CrI$_3$, which find that the proximity exchange is mainly confined to the TMD layer adjacent to CrI$_3$ and that inserting an hBN spacer strongly suppresses the proximity effect~\cite{ZollnerPRB19}.

Throughout the main figures we choose a representative magnitude
$\Delta_{\rm ex}^{(\ell)} \equiv \Delta_{\rm ex} = 30~\mathrm{meV}$ on the proximitized
chalcogen orbitals, independent of $\ell$ and of the chalcogen species
($X=\mathrm{S,Se,Te}$), unless stated otherwise.
This value should be interpreted as a local exchange parameter in the proximitized chalcogen-like Wannier subspace, not as a rigid 30~meV splitting of all low-energy Nb-derived bands.
The effective exchange seen by a Bloch state is reduced by its layer- and orbital-resolved chalcogen weight.
At the level of first-order perturbation theory, one may write the exchange-induced band-energy shift schematically as
\begin{equation}
\delta E_{n\mathbf{k}}
\simeq
\sum_{\ell\in\mathcal{L}_{\rm prox}}
\Delta_{\rm ex}^{(\ell)}
\langle u_{n\mathbf{k}}|P_\ell \mathbf{m}_\ell\cdot\mathbf{s}_\ell P_\ell|u_{n\mathbf{k}}\rangle,
\label{eq:S4_Deff_projected}
\end{equation}
where $P_\ell$ projects onto the proximitized chalcogen Wannier subspace.
The observable splitting of two low-energy branches is therefore controlled by the difference between their projected exchange shifts, rather than directly by the input local value $\Delta_{\rm ex}$.
A diagnostic comparison for one-sided NbSe$_2$ with $\Delta_{\rm ex}=30$~meV applied to the bottom Se sublayer gives shifts of the near-$E_F$ Wannier bands on the scale of several meV per branch. We therefore use 30~meV only as an order-of-magnitude local model parameter; a quantitative branch-splitting claim requires band- and momentum-resolved data together with projected orbital weights.

To place this value in the context of known van der Waals magnetic-proximity systems, it is useful to distinguish optical exciton valley splittings from local band exchange parameters.
For an optically measured exciton valley splitting $\Delta E_{\rm X}$, one often quotes an effective exchange field
\begin{equation}
B_{\rm eff}^{\rm exc}\simeq \frac{\Delta E_{\rm X}}{g_{\rm X}\mu_B},
\end{equation}
or equivalently compares $\Delta E_{\rm X}$ with the bare monolayer-TMD valley-Zeeman slope, typically of order $0.2$--$0.27$~meV/T.
In monolayer WSe$_2$ on approximately $10$~nm CrI$_3$ encapsulated by hBN, optical photoluminescence measurements reported an effective magnetic exchange field of order $13$~T, and optical tuning of the CrI$_3$ magnetization varied this exchange field over a range of about $20$~T~\cite{SeylerNanoLett18}.
Using the above optical conversion, these fields correspond to valley exciton splittings of order a few meV.
Layer-resolved measurements on monolayer WSe$_2$ interfaced with bilayer or trilayer CrI$_3$ further showed that the spin-dependent charge-transfer response is dominated by the interfacial CrI$_3$ layer, while the induced valley Zeeman splitting remains sensitive to the full layered magnetic configuration~\cite{ZhongNatNano20}.

First-principles calculations for monolayer MoSe$_2$/CrI$_3$ and WSe$_2$/CrI$_3$ give proximity-induced conduction- and valence-band exchange splittings of roughly $1$--$5$~meV, with exciton valley splittings depending strongly on twist angle and electrostatic gating~\cite{ZollnerPRB19}.
For example, calculated exciton valley splittings are about $1.13$~meV and $3.89$~meV for MoSe$_2$/CrI$_3$ at $0^\circ$ and $30^\circ$ twist, respectively, and about $1.43$~meV and $6.35$~meV for WSe$_2$/CrI$_3$ at $0^\circ$ and $30^\circ$ twist, respectively~\cite{ZollnerPRB19}.

Larger proximity scales have been reported for stronger magnetic substrates.
In monolayer WS$_2$ on ferromagnetic EuS, magneto-reflectance measurements found a giant valley exciton splitting, with an initial slope of about $-16$~meV/T for small magnetic fields and a saturated splitting of order $10$--$20$~meV~\cite{NordenNC19}.
First-principles analysis of EuS/TMD interfaces also gives band splittings in the tens-of-meV range, depending on the assumed EuS surface termination and band alignment~\cite{NordenNC19}.
Ferromagnetic Fe$_3$GeTe$_2$ is another experimentally established van der Waals magnetic platform, with atomically thin ferromagnetism and strong out-of-plane anisotropy~\cite{DengNature2018}.
We do not use that reference as a quantitative benchmark for an Fe$_3$GeTe$_2$/NbX$_2$ proximity splitting. Because Fe$_3$GeTe$_2$ is metallic, substrate-band hybridization, charge transfer, and current shunting can be important; such a platform requires an explicit heterostructure calculation for quantitative predictions.

Most directly relevant to the present metallic Nb-based system, NbSe$_2$/V$_5$Se$_8$ heterostructures with four-layer NbSe$_2$ and variable-thickness V$_5$Se$_8$ exhibit anomalous-Hall signatures attributed to ferromagnetically proximitized NbSe$_2$~\cite{MatsuokaNatCommun22}.
In that system, comparison between experiment and theory indicates an interfacial exchange scale of a few tens of meV, with representative model calculations using an exchange field of order $|M|=40$~meV~\cite{MatsuokaNatCommun22}.
Relative to these benchmarks, our choice $\Delta_{\rm ex}=30$~meV corresponds to a moderate local Wannier-basis proximity scale: it is larger than the few-meV optical splittings typical of CrI$_3$-based semiconducting TMD heterostructures, but comparable to the tens-of-meV exchange scales inferred for stronger magnetic substrates and metallic Nb-based interfaces.
The several-meV near-$E_F$ band shifts in our NbSe$_2$ diagnostic comparison are consistent at the order-of-magnitude level with such proximity scales, but they are not a material-specific prediction for a selected magnetic substrate.

{In this weak-to-intermediate regime, the induced exchange is small compared with the intrinsic Ising spin--orbit coupling and bandwidth of Nb$X_2$, but large enough to produce clearly resolved Berry-curvature hot spots and associated Hall responses.}
{Changing $\Delta_{\rm ex}$ does not modify the symmetry-based selection rules. Quantitatively, however, it can shift proximity-induced avoided crossings and Berry-curvature hot spots relative to the Fermi level, thereby changing response magnitudes, peak positions, and even the local sign at a fixed chemical potential. The comparative trends discussed in the main text should therefore be understood within the tested weak-to-intermediate exchange regime rather than as a universal pure-rescaling relation.}

Importantly, only the additional on-site pseudospin-dependent proximity block is modified; the intrinsic fully relativistic spinor Wannier Hamiltonian obtained from the nonmagnetic first-principles calculation, including spin--orbit coupling and all original hopping matrix elements, is kept fixed. The proximity term should therefore be regarded as a static, local, layer-resolved effective self-energy rather than as a full explicit magnet/Nb$X_2$ interface calculation. No sensitivity test in the present work redistributes exchange onto Nb or the remote chalcogen layer, adds spin-dependent hopping, or varies an interface electrostatic potential. The approximation can therefore be used to test ideal magnetic-point-group selection rules, but its response magnitudes, peak positions, and fixed-$\mu$ signs are not validated against those omitted terms. Strongly hybridized metallic interfaces, charge transfer, structural reconstruction, moir\'e potentials, disorder, or structural $C_3$ breaking can change the quantitative results and can add a background BCD. A material-specific prediction requires an explicit heterostructure calculation followed by interface-resolved Wannier downfolding and a documented physical-spin exchange matrix.

\subsection*{S4.4 WannierBerri evaluation of AHC, BCD, and NLHC}

\begin{figure}[b]
    \centering
    \includegraphics[width=\columnwidth]{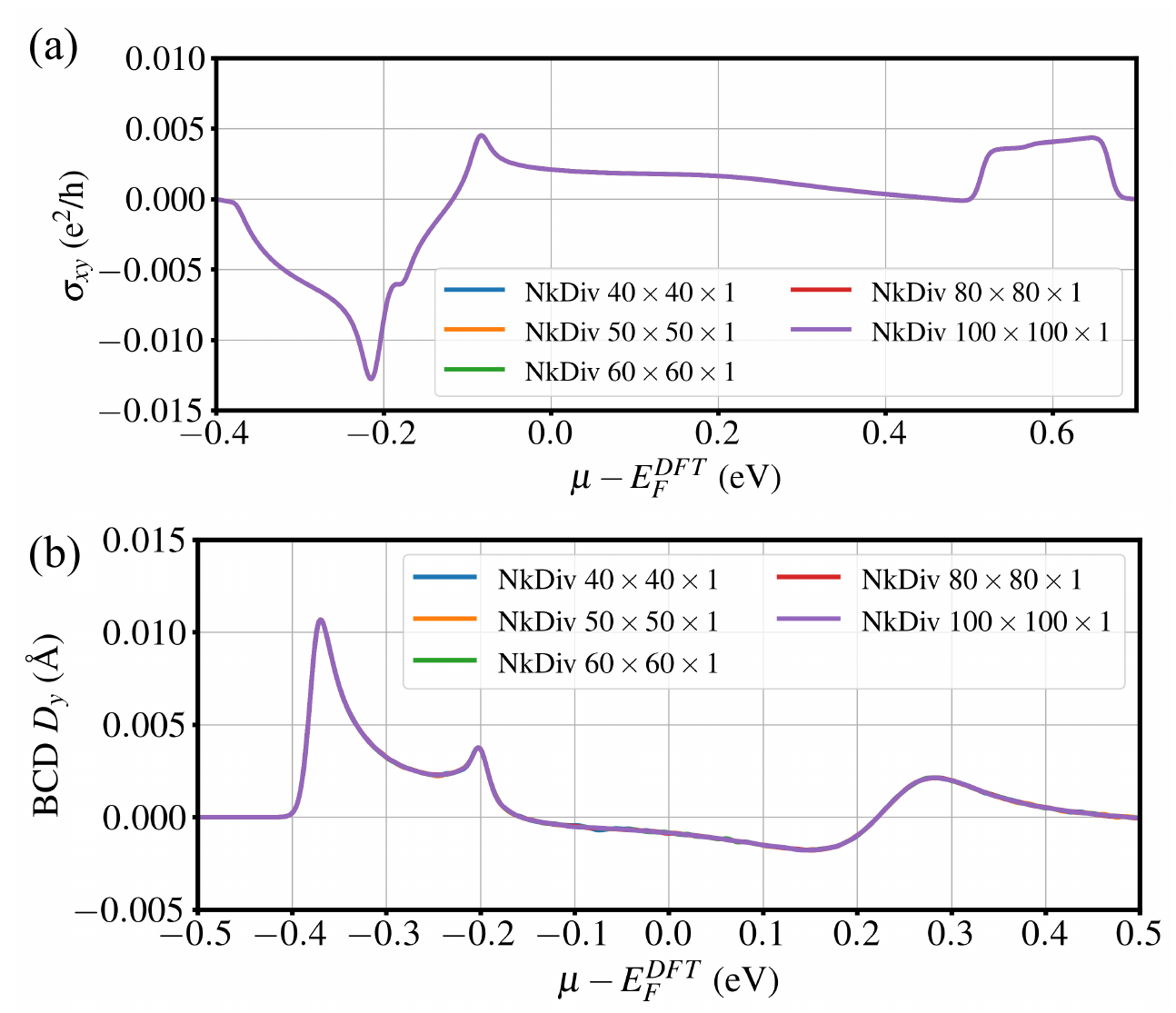}
    \caption{\textbf{Convergence of AHC and BCD with respect to the interpolation grid
    \texttt{NkDiv} in the orthogonal two-sided proximity configuration.}
    Representative convergence test for monolayer NbSe$_2$ in the orthogonal two-sided
    proximity geometry used in the main text (bottom interface predominantly out-of-plane,
    top interface predominantly in-plane). Panels show (a) sheet anomalous Hall conductivity
    $\sigma_{xy}^{\rm sheet}$ (in units of $e^2/h$) and (b) Berry-curvature dipole $D_y$
    (in \AA) as functions of the chemical potential, computed with \textsc{WannierBerri}
    using a fixed internal FFT grid $\texttt{NkFFT}=80\times80\times1$ and varying
    interpolation grids $\texttt{NkDiv}=40\times40\times1$, $50\times50\times1$,
    $60\times60\times1$, $80\times80\times1$, and $100\times100\times1$. The nearly
    indistinguishable curves for $\texttt{NkDiv}\ge 60\times60\times1$ demonstrate integration-grid convergence for this NbSe$_2$ configuration; they do not constitute convergence tests of the DFT or Wannier construction. In the
    main calculations we adopt $\texttt{NkDiv}=60\times60\times1$.}
    \label{fig:NkDiv_convergence}
\end{figure}

Berry curvature, anomalous Hall conductivity (AHC), Berry-curvature dipole (BCD), and
intrinsic nonlinear Hall conductivity (NLHC) were computed from the exchange-modified
Wannier Hamiltonians using the \textsc{WannierBerri} code \cite{wanberri}. For each
tight-binding model \texttt{<seed>\_tb.dat}, we constructed a \texttt{System\_tb} object with
Berry-curvature evaluation enabled and performed interpolation on dense reciprocal-space
grids optimized for a two-dimensional Brillouin zone. In the calculations underlying
Figs.~2--4 we used $\texttt{NkFFT}=80\times80\times1$ and $\texttt{NkDiv}=60\times60\times1$. In the \textsc{WannierBerri} grid convention used here, these are multiplicative subdivisions, so the effective dense mesh is their elementwise product, $N_k^{\rm eff}=4800\times4800\times1$; the convergence sweep below spans $3200\times3200\times1$ to $8000\times8000\times1$. Refinement of $\texttt{NkDiv}$ changes the plotted $\sigma_{xy}$ and $D_{yz}$ curves only at the level of a few percent for the tested orthogonal NbSe$_2$ case.
In particular, Fig.~\ref{fig:NkDiv_convergence} shows that, for the orthogonal two-sided
proximity configuration in NbSe$_2$ (bottom interface predominantly out-of-plane, top
interface predominantly in-plane), increasing the interpolated $k$-mesh (\texttt{NkDiv})
from $40\times40\times1$ up to $100\times100\times1$ at fixed
$\texttt{NkFFT}=80\times80\times1$ leaves both the energy-dependent AHC and BCD essentially
unchanged; we therefore adopt $\texttt{NkDiv}=60\times60\times1$ as the default integration setting. This single test does not establish convergence of the sharper NbTe$_2$ extrema or of the symmetry-forbidden numerical floor.
For each proximity configuration we also inspected the symmetry operations seen by
\textsc{WannierBerri} (including the presence or absence of $C_3$) to confirm that the
intended magnetic point group was correctly encoded.

The AHC returned by \textsc{WannierBerri} is a three-dimensional conductivity tensor in SI
units (S/m). To convert this to a sheet conductance appropriate for a monolayer, we extract
the real-space lattice vectors $\{\mathbf{a}_1,\mathbf{a}_2,\mathbf{a}_3\}$ directly from
the corresponding \texttt{<seed>\_tb.dat} file. The in-plane area and out-of-plane
simulation-cell length are defined as
\[
A=\|\mathbf{a}_1\times\mathbf{a}_2\|,\qquad
c=\|\mathbf{a}_3\|.
\]
Here $\mathbf{a}_3$ is along the slab normal. The same out-of-plane simulation-cell
length, $c=30.00$~\AA, is used for NbS$_2$, NbSe$_2$, and NbTe$_2$, as obtained
directly from the respective \texttt{tb.dat} lattice vectors. The sheet AHC in units
of the conductance quantum $e^2/h$ is then obtained as
\begin{equation}
\sigma_{xy}^{\rm sheet}(E_{\rm F})
= \frac{\sigma_{xy}^{\rm 3D}(E_{\rm F})\,c}{e^2/h},
\end{equation}
which is the quantity plotted in Fig.~2 of the main text. Here
$\sigma_{xy}^{\rm 3D}(E_{\rm F})$ is the $z$-directed Hall component (in S/m) evaluated at
Fermi energy $E_{\rm F}$.

The same supercell normalization must be made explicit for the BCD and nonlinear coefficient. We distinguish bulk-normalized code outputs from intrinsic monolayer sheet quantities as
\begin{equation}
D_{az}^{\rm 2D}=cD_{az}^{\rm 3D},\qquad
\chi_{abc}^{\rm sheet}=c\chi_{abc}^{\rm 3D}.
\label{eq:S4_sheet_conversion}
\end{equation}
Here $D^{\rm 2D}$ has units of length and $\chi^{\rm sheet}$ has units
A\,m/V$^2$, whereas $\chi^{\rm 3D}$ has units A/V$^2$ and depends on the
simulation-cell length $c$. Figure~3 and all harmonic-voltage equations use
$D^{\rm 2D}$ and $\chi^{\rm sheet}$. Figure~4(d) displays the corresponding
bulk-normalized coefficient
$\chi_{xyy}^{\rm 3D}=\chi_{xyy}^{\rm sheet}/c$.
The NLHC data plotted in Fig.~4(d) are the bulk-normalized
\textsc{WannierBerri} outputs in A/V$^2$; no additional factor of $c$ is
applied to those plotted data. Conversion to the sheet coefficient used in
the device equations is performed through
$\chi_{xyy}^{\rm sheet}=c\chi_{xyy}^{\rm 3D}$.

The energy dependence of all response functions was obtained by scanning the chemical
potential over a window that covers the relevant interface-induced $k$-linear Rashba–exchange anticrossings. For
NbSe$_2$, we used $E_{\rm F}\in[-2.95,-1.65]$~eV (relative to the \textsc{wannier90}
reference) with $N=1001$ equally spaced points; analogous windows were used for NbS$_2$
and NbTe$_2$. This chemical-potential scan is a rigid-band construction. Mapping it to a gate voltage requires the carrier-density change $\Delta n(\mu)$ and the geometric and quantum capacitances, which are not calculated here. Finite temperature was incorporated via a Fermi--Dirac smoother corresponding
to $T=50$~K, which regularizes sharp features near van Hove singularities without obscuring
the trends associated with proximity exchange and interface-induced $k$-linear Rashba–exchange anticrossings.

The Berry-curvature dipole $D_{ab}(E_{\rm F})$ was computed using the Fermi-surface
formulation implemented in \textsc{WannierBerri}. In 2D, and for band index $n$, this may
be written schematically as
\begin{equation}
D^{\rm surf}_{ab}(E_{\rm F})
= -\sum_n \int_{\rm BZ} \frac{d^2k}{(2\pi)^2}\,
\left[\frac{\partial}{\partial k_a} f_0\!\left(\varepsilon_{n\mathbf{k}}-E_{\rm F}\right)\right]
\Omega^b_{n\mathbf{k}},
\end{equation}
where $f_0$ is the equilibrium Fermi function and $\Omega^b_{n\mathbf{k}}$ is the $b$-th
component of the Berry curvature of band $n$. This expression is most directly connected
to the semiclassical nonlinear Hall response and is the one underlying the quantitative
results shown in Fig.~3; we also evaluated the corresponding Fermi-sea contribution as a
consistency check, finding good agreement in the regimes of interest.

The intrinsic nonlinear Hall conductivity was related to the two-dimensional BCD through Eq.~\eqref{eq:s1-chi-general}. For the component calculated here,
\begin{equation}
\chi_{xyy}^{\rm sheet}(E_{\rm F})
= \frac{e^3\tau}{\hbar^2}\,D_{yz}^{\rm 2D}(E_{\rm F}),
\end{equation}
corresponding to a longitudinal field along $y$ and a transverse current along $x$. We use a phenomenological relaxation time $\tau = 25$~fs that is constant in energy.
This choice fixes the overall scale of $\chi_{xyy}^{\rm sheet}$ but does not affect the symmetry
trends, sign patterns, or relative enhancements across proximity configurations and across
the NbS$_2\!\rightarrow$NbSe$_2\!\rightarrow$NbTe$_2$ series. The energy-dependent AHC,
BCD, and NLHC shown in Figs.~2–4 are obtained by combining these \textsc{WannierBerri}
outputs with the conversion procedure described above. Quantitative device estimates require $\sigma_{xx}^{\rm sheet}$, $D_y$, and $\sigma_{xy}^{\rm sheet}$ at the same chemical potential; the illustrative estimates in Sec.~S6.4 instead span independent representative ranges and are not a self-consistent transport calculation for one gated state.

\section*{S5. $k$-space origin of the anomalous and nonlinear Hall responses}

\subsection*{S5.1 Band structures (without/with proximity exchange) for NbS$_2$, NbSe$_2$, and NbTe$_2$}

\begin{figure*}[tb]
    \centering
    \includegraphics[width=\textwidth]{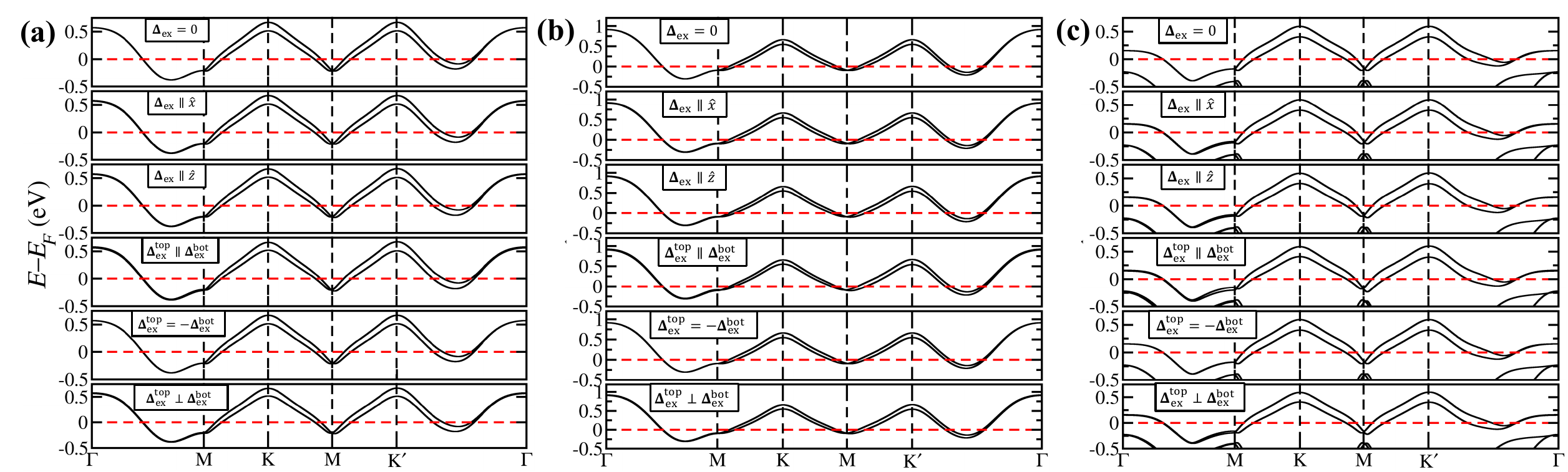}
    \caption{\textbf{Band-structure fingerprints of layer-selective proximity exchange
    in Nb$X_2$.}
    Wannier-interpolated band structures along $\Gamma$–M–K–M–K$'$–$\Gamma$ for
    (a) NbSe$_2$, (b) NbS$_2$, and (c) NbTe$_2$ under six representative proximity
    textures (top to bottom): no exchange, $\Delta_{\rm ex}=0$; one-sided in-plane
    exchange, $\Delta_{\rm ex}\!\parallel\!\hat{\mathbf x}$, acting on the chalcogen
    sublayer adjacent to the magnetic interface; one-sided out-of-plane exchange,
    $\Delta_{\rm ex}\!\parallel\!\hat{\mathbf z}$; two-sided parallel alignment,
    $\Delta_{\rm ex}^{\rm top}\!\parallel\!\Delta_{\rm ex}^{\rm bot}$; two-sided
    antiparallel alignment, $\Delta_{\rm ex}^{\rm top}=-\Delta_{\rm ex}^{\rm bot}$;
    and orthogonal two-sided alignment,
    $\Delta_{\rm ex}^{\rm top}\!\perp\!\Delta_{\rm ex}^{\rm bot}$. In all panels the
    exchange magnitude on the proximitized chalcogen orbitals is
    $\Delta_{\rm ex}=30$~meV, and the red dashed line marks a Fermi energy. The overall dispersions remain
    close to the zero-exchange case, but small spin splittings and the reshaping of
    interface-induced $k$-linear Rashba–exchange anticrossings, which become more pronounced from NbS$_2$ to
    NbTe$_2$, provide the $k$-space structure that gives rise to the anomalous and
    nonlinear Hall responses discussed in the main text.}
    \label{fig:S3_bands}
\end{figure*}

In this section we illustrate how the anomalous Hall conductivity (AHC) and
Berry-curvature dipole (BCD) in the main text arise from the band structure
and $k$-resolved Berry curvature of NbX$_2$ in different proximity
configurations. Because the exchange scale used in the calculations
($\Delta_{\rm ex}=30$~meV) is small compared to the overall bandwidth, the
changes in the dispersion and in $|\Omega_z(\mathbf k)|$ are modest; the key
point is the \emph{symmetry and weighting} of the Berry curvature on the
Fermi surface, rather than the appearance of dramatically new hot spots.

Fig.~\ref{fig:S3_bands} summarizes how the different layer-selective proximity textures affect the
low-energy band structure of the Nb$X_2$ series. Panels (a), (b), and (c) show
Wannier-interpolated bands along $\Gamma$–M–K–M–K$'$–$\Gamma$ for
NbSe$_2$, NbS$_2$, and NbTe$_2$, respectively. For each material we display six
configurations: (top to bottom) (i) no exchange, $\Delta_{\rm ex}=0$; (ii) one-sided
in-plane exchange, $\Delta_{\rm ex}\!\parallel\!\hat{\mathbf x}$, applied to the chalcogen
sublayer adjacent to the magnetic interface; (iii) one-sided out-of-plane exchange,
$\Delta_{\rm ex}\!\parallel\!\hat{\mathbf z}$; (iv) two-sided parallel alignment
$\Delta_{\rm ex}^{\rm top}\!\parallel\!\Delta_{\rm ex}^{\rm bot}$; (v) two-sided
antiparallel alignment $\Delta_{\rm ex}^{\rm top}=-\Delta_{\rm ex}^{\rm bot}$; and
(vi) the orthogonal two-sided configuration
$\Delta_{\rm ex}^{\rm top}\!\perp\!\Delta_{\rm ex}^{\rm bot}$ used in the device
proposal. In all cases the exchange magnitude on the proximitized chalcogen orbitals is
$\Delta_{\rm ex}=30$~meV.

Because $\Delta_{\rm ex}$ is small compared to the overall bandwidth, the dispersions in
Fig.~S3 remain very similar to the zero-exchange case: the Fermi surface pockets around
K/K$'$ are preserved, and the overall metallic character is unchanged. The proximity
primarily induces (i) a small spin splitting of the K/K$'$-centered bands and (ii) a
modest reshaping of interface-induced $k$-linear Rashba–exchange anticrossings near the Fermi level. These changes
are most visible when comparing the $\Delta_{\rm ex}=0$ and $\Delta_{\rm ex}\!\parallel\!\hat{\mathbf z}$
rows, and they become more pronounced from NbS$_2$ to NbTe$_2$, tracking the increase in
intrinsic spin–orbit coupling. In the parallel two-sided configuration the exchange
splittings are enhanced relative to the one-sided case, while in the antiparallel
configuration they are strongly reduced, consistent with the Hall-valve behavior of
$\sigma_{xy}$ in Fig.~2. The orthogonal two-sided geometry retains sizable splittings
while lowering the crystalline symmetry, thereby enabling both a finite anomalous Hall
conductivity and a Berry-curvature dipole. Thus, even though the band structures in
Fig.~\ref{fig:S3_bands} appear only weakly perturbed, they encode the symmetry breaking and valley
selectivity that underlie the AHC and BCD trends reported in the main text.

\subsection*{S5.2 $k$-resolved Berry curvature maps $\Omega_z(\mathbf{k})$}

\subsubsection*{(i) $k$-resolved Berry curvature and the one-sided anomalous Hall effect}

\begin{figure}[t]
    \centering
    \includegraphics[width=\columnwidth]{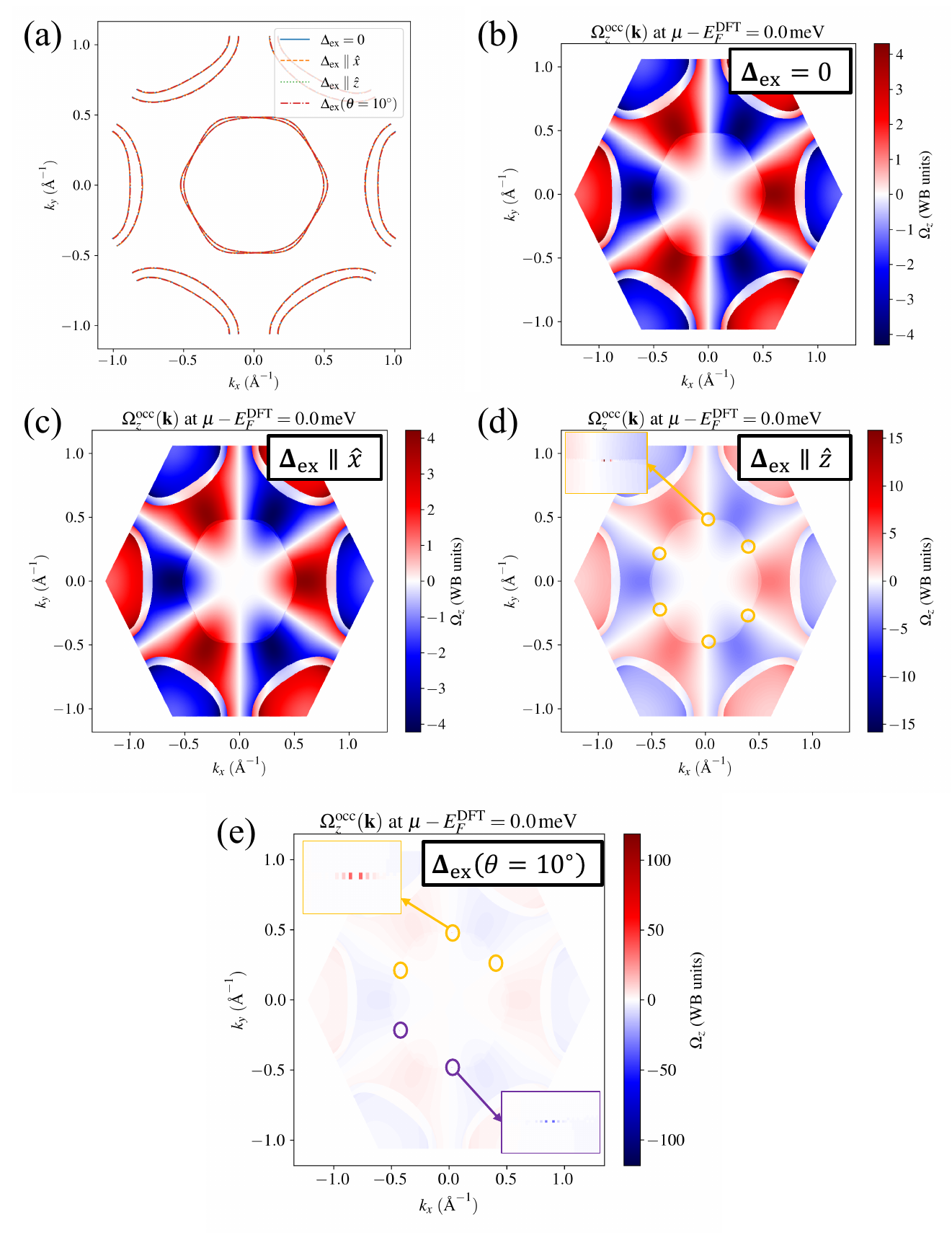}
    \caption{\textbf{$k$-resolved occupied-band Berry curvature and the origin of the
    one-sided anomalous Hall effect in NbSe$_2$.}
    (a) Fermi surface at $\mu-E_F^{\rm DFT}=0$ overlaid for four configurations:
    no exchange ($\Delta_{\rm ex}=0$), $\Delta_{\rm ex}\!\parallel\!\hat{\mathbf x}$,
    $\Delta_{\rm ex}\!\parallel\!\hat{\mathbf z}$, and
    $\Delta_{\rm ex}(\theta=10^\circ)$ with a small in-plane component. The
    $\Gamma$-centered pocket is almost unchanged, indicating that proximity mainly
    acts through Berry curvature. (b)–(e) Occupied-band Berry curvature
    $\Omega_z^{\rm occ}(\mathbf{k};\mu=E_F^{\rm DFT})$ in the Brillouin zone for
    $\Delta_{\rm ex}=0$ (b), $\Delta_{\rm ex}\!\parallel\!\hat{\mathbf x}$ (c),
    $\Delta_{\rm ex}\!\parallel\!\hat{\mathbf z}$ (d), and
    $\Delta_{\rm ex}(\theta=10^\circ)$ (e). Time-reversal symmetry enforces cancellation of the Brillouin-zone integral in the zero-exchange case. The purely in-plane
    exchange retains finite local curvature but produces a negligible net AHC because the positive and negative contributions nearly cancel.
    A finite out-of-plane component generates uncompensated curvature patches
    along the $\Gamma$-centered contour and produces the sizable $\sigma_{xy}$
    seen in Fig.~2(a), while a small canting angle leaves both the Berry-curvature
    pattern and the AHC almost unchanged (see insets).}
    \label{fig:S4_BC_AHC}
\end{figure}

To connect the energy-dependent anomalous Hall conductivity in Fig.~2(a) to the
underlying band geometry, Fig.~\ref{fig:S4_BC_AHC} shows the $k$-resolved Fermi surface
and occupied-band Berry curvature for monolayer NbSe$_2$ in the one-sided proximity
configuration. In all panels we consider the Berry-curvature density that directly
enters the intrinsic AHC,
\begin{equation}
\Omega_z^{\rm occ}(\mathbf{k};\mu)
= \sum_n f_0\!\left(\varepsilon_{n\mathbf{k}}-\mu\right)\,\Omega_{n,z}(\mathbf{k}),
\end{equation}
so that $\sigma_{xy}(\mu)\propto\int_{\rm BZ}d^2k\,\Omega_z^{\rm occ}(\mathbf{k};\mu)$
(up to the usual prefactor $-e^2/\hbar$). At the finite temperature used in the
main WannierBerri calculations ($T=50$~K), the Fermi function is sufficiently sharp
that $\Omega_z^{\rm occ}(\mathbf{k};\mu)$ may be interpreted as the Berry curvature
summed over the occupied bands at the chosen chemical potential.

Fig.~\ref{fig:S4_BC_AHC}(a) compares the Fermi surface at $\mu-E_F^{\rm DFT}=0$ for four
cases: no exchange ($\Delta_{\rm ex}=0$), purely in-plane exchange
$\Delta_{\rm ex}\!\parallel\!\hat{\mathbf x}$, purely out-of-plane exchange
$\Delta_{\rm ex}\!\parallel\!\hat{\mathbf z}$, and a slightly tilted exchange
$\Delta_{\rm ex}(\theta=10^\circ)$ with a small in-plane component. The contours
nearly coincide, reflecting that the chosen proximity scale
$\Delta_{\rm ex}=30$~meV only weakly perturbs the dispersion; the one-sided AHC in
Fig.~2(a) is therefore not driven by large Fermi-surface distortions.

The anomalous Hall response is instead controlled by the redistribution of Berry
curvature in $k$-space. Figs.~\ref{fig:S4_BC_AHC}(b)–(e) display
$\Omega_z^{\rm occ}(\mathbf{k};\mu=E_F^{\rm DFT})$ for the same four configurations.
In the TRS-symmetric case $\Delta_{\rm ex}=0$ [Fig.~\ref{fig:S4_BC_AHC}(b)],
$\Omega_z^{\rm occ}(\mathbf{k})$ exhibits the familiar valley-contrasting pattern of
Ising NbSe$_2$. However, the contributions from $\mathbf{k}$ and $-\mathbf{k}$ pockets
cancel in the Brillouin-zone integral, and the net AHC vanishes. For a purely
in-plane proximity field $\Delta_{\rm ex}\!\parallel\!\hat{\mathbf x}$
[Fig.~\ref{fig:S4_BC_AHC}(c)], the interface-induced $k$-linear Rashba–Zeeman mechanism is effectively inactive
($\Delta_z\simeq 0$ in the minimal model), so the Berry-curvature texture on the
$\Gamma$-centered pocket remains almost unchanged and the resulting
$\sigma_{xy}(\mu)$ is negligibly small, consistent with the orange curve in
Fig.~2(a).

By contrast, a purely out-of-plane exchange
$\Delta_{\rm ex}\!\parallel\!\hat{\mathbf z}$ [Fig.~\ref{fig:S4_BC_AHC}(d)] breaks
time-reversal symmetry while preserving threefold lattice rotation, and generates
sizable regions of uncompensated curvature along the $\Gamma$-centered Fermi
contour. These appear as red/blue lobes whose imbalance yields a finite integral
\begin{equation}
\sigma_{xy}(\mu)\propto \int_{\rm BZ} d^2k\;\Omega_z^{\rm occ}(\mathbf{k};\mu),
\end{equation}
and produces the pronounced peaks and sign changes as $\mu$ passes through
interface-induced $k$-linear Rashba–exchange anticrossings in Fig.~2(a). Introducing a small canting angle
$\theta=10^\circ$ [Fig.~\ref{fig:S4_BC_AHC}(e)] leaves the overall pattern of
$\Omega_z^{\rm occ}(\mathbf{k})$ essentially unchanged, as highlighted by the insets
that zoom into representative segments of the $\Gamma$-centered pocket; the local
curvature patches are only weakly modified, and the AHC curve for the tilted
configuration therefore almost coincides with the
$\Delta_{\rm ex}\!\parallel\!\hat{\mathbf z}$ result.

Together, these $k$-space maps clarify that the one-sided AHC in the main text is
controlled mainly by the presence of a finite out-of-plane exchange component,
which unbalances pre-existing SOC-induced valley-textured curvature hotspots, rather than by large
changes in the Fermi surface or by a qualitative reshaping of the Berry-curvature
profile.

\subsubsection*{(ii) Two-sided Hall valve and orthogonal proximity}

\begin{figure}[t]
    \centering
    \includegraphics[width=\columnwidth]{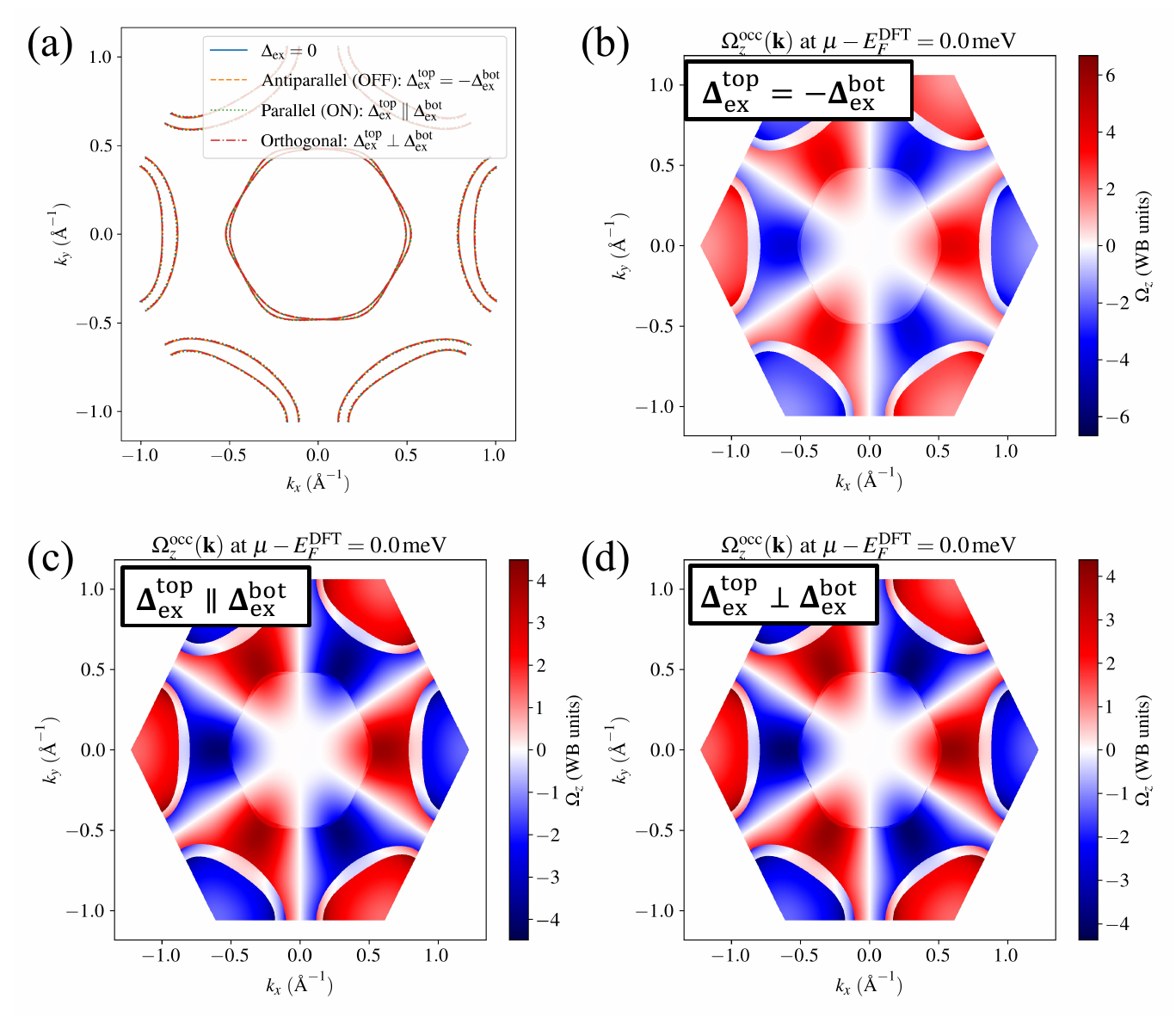}
    \caption{\textbf{$k$-resolved occupied-band Berry curvature for the two-sided Hall
    valve and orthogonal configurations in NbSe$_2$.}
    (a) Fermi surface at $\mu-E_F^{\rm DFT}=0$ overlaid for four configurations:
    no exchange ($\Delta_{\rm ex}=0$), antiparallel two-sided proximity
    $\boldsymbol{\Delta}_{\rm ex}^{\rm top}=-\boldsymbol{\Delta}_{\rm ex}^{\rm bot}$
    (valve OFF), parallel two-sided proximity
    $\boldsymbol{\Delta}_{\rm ex}^{\rm top}\parallel\boldsymbol{\Delta}_{\rm ex}^{\rm bot}$
    (valve ON), and orthogonal two-sided proximity
    $\boldsymbol{\Delta}_{\rm ex}^{\rm top}\perp\boldsymbol{\Delta}_{\rm ex}^{\rm bot}$.
    As in Fig.~\ref{fig:S4_BC_AHC}, the $\Gamma$-centered pocket is only weakly modified by
    the proximity exchange. (b)–(d) Occupied-band Berry curvature
    $\Omega_z^{\rm occ}(\mathbf{k};\mu=E_F^{\rm DFT})$ for the antiparallel (b),
    parallel (c), and orthogonal (d) configurations. In the antiparallel state, the
    effective Zeeman field from the two interfaces largely cancels, yielding a negligible
    net AHC after Brillouin-zone integration (Hall valve OFF). The parallel state generates finite curvature patches and a nonzero AHC while retaining the symmetry constraints of a
    $\sigma_h$-preserving geometry (Hall valve ON). The orthogonal state also gives a nonzero, anisotropic curvature pattern and permits a BCD. Because panels (b)--(d) use different color scales, their local amplitudes should not be compared quantitatively from the artwork alone.}
    \label{fig:S5_BC_valve}
\end{figure}

We now turn to the two-sided proximity geometries that underlie the Hall valve
and orthogonal configurations in Fig.~2(b,c) of the main text. Fig.~\ref{fig:S5_BC_valve}(a)
compares the Fermi surface at $\mu-E_F^{\rm DFT}=0$ for four cases:
no exchange ($\Delta_{\rm ex}=0$), antiparallel two-sided proximity
$\boldsymbol{\Delta}_{\rm ex}^{\rm top}=-\boldsymbol{\Delta}_{\rm ex}^{\rm bot}$ (valve OFF),
parallel two-sided proximity
$\boldsymbol{\Delta}_{\rm ex}^{\rm top}\parallel\boldsymbol{\Delta}_{\rm ex}^{\rm bot}$
(valve ON), and the orthogonal configuration
$\boldsymbol{\Delta}_{\rm ex}^{\rm top}\perp\boldsymbol{\Delta}_{\rm ex}^{\rm bot}$.
As in the one-sided case, the $\Gamma$-centered contours are nearly indistinguishable,
indicating that the chosen exchange scale $\Delta_{\rm ex}=30$~meV primarily perturbs
the Berry curvature rather than the Fermi surface itself.

The corresponding occupied-band Berry curvatures are shown in
Figs.~\ref{fig:S5_BC_valve}(b)–(d). For the antiparallel configuration
$\boldsymbol{\Delta}_{\rm ex}^{\rm top}=-\boldsymbol{\Delta}_{\rm ex}^{\rm bot}$
[Fig.~\ref{fig:S5_BC_valve}(b)], contributions from the two interfaces largely cancel in the effective
Zeeman field, so that the Brillouin-zone integral of $\Omega_z^{\rm occ}(\mathbf{k})$ yields a negligible
net AHC, consistent with the OFF state in Fig.~2(b).

In the parallel two-sided case
$\boldsymbol{\Delta}_{\rm ex}^{\rm top}
\parallel
\boldsymbol{\Delta}_{\rm ex}^{\rm bot}$
[Fig.~\ref{fig:S5_BC_valve}(c)], the effective out-of-plane exchange is enhanced,
while the horizontal mirror $\sigma_h$ and the threefold rotation
$C_3$ remain exact symmetries of the ideal symmetric model.
The resulting
$\Omega_z^{\rm occ}(\mathbf{k})$ develops finite patches along the $\Gamma$-centered
contour, but with a pattern that remains constrained by the underlying threefold
rotation. The associated AHC realizes the ON state of the Hall valve in
Fig.~2(b). Its finite AHC is consistent with the schematic ``valley-filling''
mechanism described by the Ising SOC model of Sec.~S3.2; the minimal model is not used to predict a universal magnitude relative to the one-sided geometry.

Finally, the orthogonal two-sided configuration
$\boldsymbol{\Delta}_{\rm ex}^{\rm top}\perp\boldsymbol{\Delta}_{\rm ex}^{\rm bot}$
[Fig.~\ref{fig:S5_BC_valve}(d)] produces an effective Zeeman field with both
out-of-plane and in-plane components, thereby breaking $\sigma_h$ and activating
interface-induced $k$-linear Rashba terms in the same spirit as the one-sided case. The resulting
$\Omega_z^{\rm occ}(\mathbf{k})$ on the $\Gamma$-centered pocket exhibits an anisotropic pattern whose integral lacks the near cancellation of the antiparallel state. This explains why the orthogonal configuration remains
robustly AHC-active in Fig.~2(b) while simultaneously supporting a finite
Berry-curvature dipole and nonlinear Hall response (discussed in Sec.~S5.3 and
Fig.~3 of the main text). Together with Fig.~\ref{fig:S4_BC_AHC}, these maps make
explicit how different layer-resolved exchange textures redistribute
$\Omega_z^{\rm occ}(\mathbf{k})$ on an almost unchanged Fermi surface, thereby
realizing the Hall-valve and orthogonal-proximity principles central to our proposal.

\subsubsection*{(iii) Chalcogen dependence in the orthogonal configuration}

\begin{figure}[t]
    \centering
    \includegraphics[width=\columnwidth]{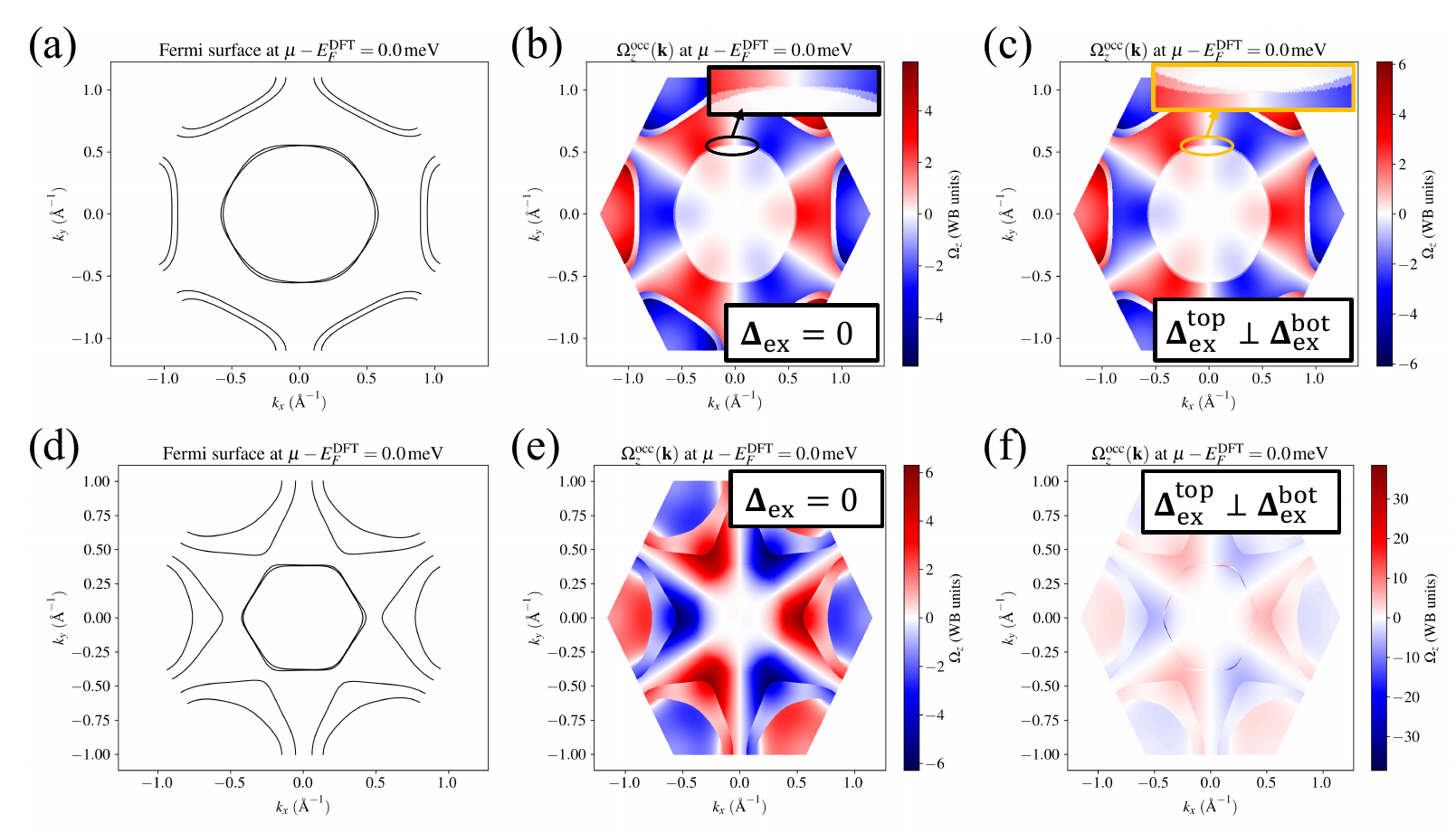}
    \caption{\textbf{Chalcogen dependence of $k$-resolved occupied-band Berry curvature
    in the orthogonal two-sided configuration.}
    (a) Fermi surface of NbS$_2$ at $\mu-E_F^{\rm DFT}=0$; (b) occupied-band Berry
    curvature $\Omega_z^{\rm occ}(\mathbf{k};\mu=E_F^{\rm DFT})$ for NbS$_2$ with
    $\Delta_{\rm ex}=0$; (c) $\Omega_z^{\rm occ}(\mathbf{k};\mu=E_F^{\rm DFT})$ for
    NbS$_2$ in the orthogonal two-sided proximity configuration
    $\boldsymbol{\Delta}_{\rm ex}^{\rm top}\perp\boldsymbol{\Delta}_{\rm ex}^{\rm bot}$.
    Insets in (b,c) zoom the curvature along a representative segment of the
    $\Gamma$-centered pocket, highlighting the extremely small amplitude and the subtle
    change induced by orthogonal exchange. (d)–(f) Analogous plots for NbTe$_2$.
    Note the different color scales between the top and bottom rows: the displayed curvature
    amplitude on the $\Gamma$-centered pocket is larger for NbTe$_2$ than for NbS$_2$, especially once the orthogonal
    exchange is applied. This cross-material change in
    $\Omega_z^{\rm occ}(\mathbf{k})$ accompanies the increase of the anomalous
    Hall conductivity and Berry-curvature dipole across the Nb$X_2$ series reported
    in the main text.}
    \label{fig:S6_BC_chalcogen}
\end{figure}

Finally, Fig.~\ref{fig:S6_BC_chalcogen} illustrates how the $k$-resolved Berry curvature
responds to orthogonal two-sided proximity across the Nb$X_2$ series. Panels (a)–(c)
show NbS$_2$, while panels (d)–(f) show NbTe$_2$, all at $\mu-E_F^{\rm DFT}=0$ and in the
orthogonal geometry used in Figs.~2(c) and 3(c) of the main text.

For NbS$_2$ [Fig.~\ref{fig:S6_BC_chalcogen}(a)–(c)], the $\Gamma$-centered Fermi contour has a
nearly circular shape and the occupied-band Berry curvature
$\Omega_z^{\rm occ}(\mathbf{k})$ is extremely small, both in the TRS-symmetric case
$\Delta_{\rm ex}=0$ [Fig.~\ref{fig:S6_BC_chalcogen}(b)] and in the orthogonal two-sided
configuration $\boldsymbol{\Delta}_{\rm ex}^{\rm top}\perp\boldsymbol{\Delta}_{\rm ex}^{\rm bot}$
[Fig.~\ref{fig:S6_BC_chalcogen}(c)]. The insets in panels (b,c), plotted with an expanded
color scale, underscore this point: although a finite curvature texture is present
and is slightly reorganized by the orthogonal exchange, its amplitude along the
$\Gamma$-centered pocket remains so small that the Brillouin-zone integral yields only
a tiny AHC, consistent with the weak NbS$_2$ signal in Fig.~2(c).

In NbTe$_2$ [Fig.~\ref{fig:S6_BC_chalcogen}(d)–(f)], the Fermi surface exhibits a more
hexagonally warped $\Gamma$-centered pocket, concurrent with the heavier chalcogen, stronger spin--orbit coupling, and other material-dependent band-structure changes. Even without proximity
exchange [Fig.~\ref{fig:S6_BC_chalcogen}(e)], $\Omega_z^{\rm occ}(\mathbf{k})$ is already
much larger than in NbS$_2$, though TRS still enforces cancellation of the
Brillouin-zone integral. When the orthogonal two-sided exchange is turned on
[Fig.~\ref{fig:S6_BC_chalcogen}(f)], the curvature on the $\Gamma$-centered contour is
larger on the displayed scale and develops an anisotropic pattern.
Importantly, the Fermi-surface geometry itself changes only modestly between
NbS$_2$ and NbTe$_2$; the concurrent change in the magnitude and anisotropy of
$\Omega_z^{\rm occ}(\mathbf{k})$ contributes to the growth of both AHC and
Berry-curvature dipole across the NbS$_2\!\rightarrow$NbSe$_2\!\rightarrow$NbTe$_2$
series seen in Figs.~2(c) and 3(c).

Overall, Fig.~\ref{fig:S6_BC_chalcogen} makes explicit that the chalcogen dependence of
the linear and nonlinear Hall responses in the orthogonal geometry is governed
by the evolution of the occupied-band Berry curvature and Fermi-surface geometry across the three distinct compounds. Chemical substitution does not isolate SOC alone; a controlled SOC-scaling calculation would be required for a causal attribution.

\subsection*{S5.3 $k$-resolved Berry-curvature dipole}

\begin{figure}[t]
    \centering
    \includegraphics[width=\columnwidth]{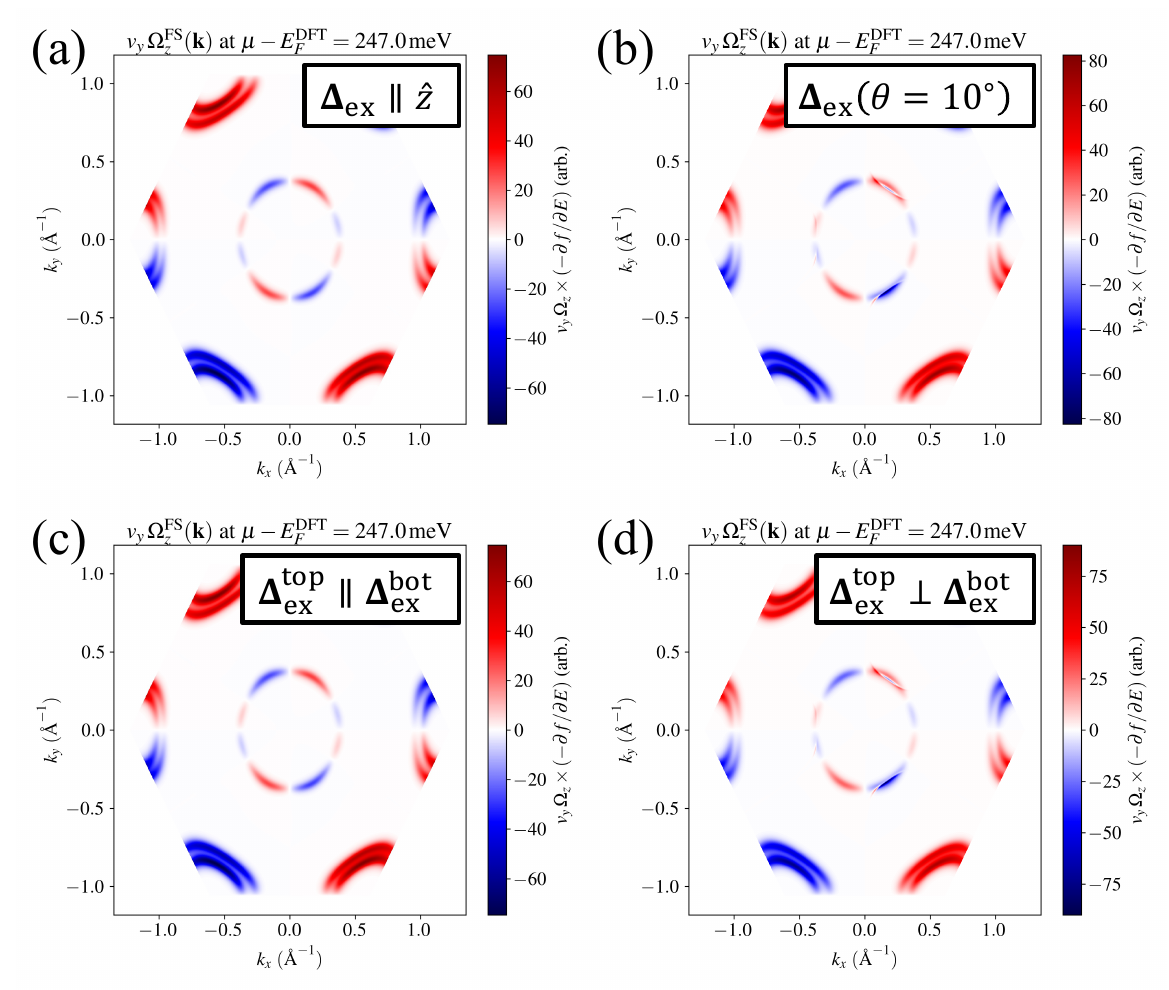}
    \caption{\textbf{Geometry dependence of the BCD kernel in NbSe$_2$.}
    Plotted quantity is the Berry-curvature-dipole kernel
    $\sum_n M_{yz,n}(\mathbf{k})=\sum_n v_{n,y}(\mathbf{k})\Omega_{n,z}(\mathbf{k})(-\partial f_0/\partial\varepsilon)$
    at $\mu-E_F^{\rm DFT}=247$~meV, close to the prominent BCD peak in
    Fig.~3(a) of the main text.
    (a) One-sided proximity with purely out-of-plane exchange
    $\Delta_{\rm ex}\!\parallel\!\hat{\mathbf z}$;
    (b) one-sided canted exchange $\Delta_{\rm ex}(\theta=10^\circ)$ with a
    small in-plane component;
    (c) parallel two-sided proximity
    $\boldsymbol{\Delta}_{\rm ex}^{\rm top}\parallel\boldsymbol{\Delta}_{\rm ex}^{\rm bot}$;
    (d) orthogonal two-sided proximity
    $\boldsymbol{\Delta}_{\rm ex}^{\rm top}\perp\boldsymbol{\Delta}_{\rm ex}^{\rm bot}$.
    The redistribution and asymmetry of $M_{yz}(\mathbf{k})$ between these
    four cases explains both the geometry dependence of the BCD in
    Fig.~3(a) and its sensitivity to the relative orientation of the two
    exchange fields. The panels use the plotting normalization of the supplied data; quantitative comparison requires a common physical normalization and matched color scale.}
    \label{fig:S7_BCD_maps}
\end{figure}

\begin{figure}[t]
    \centering
    \includegraphics[width=\columnwidth]{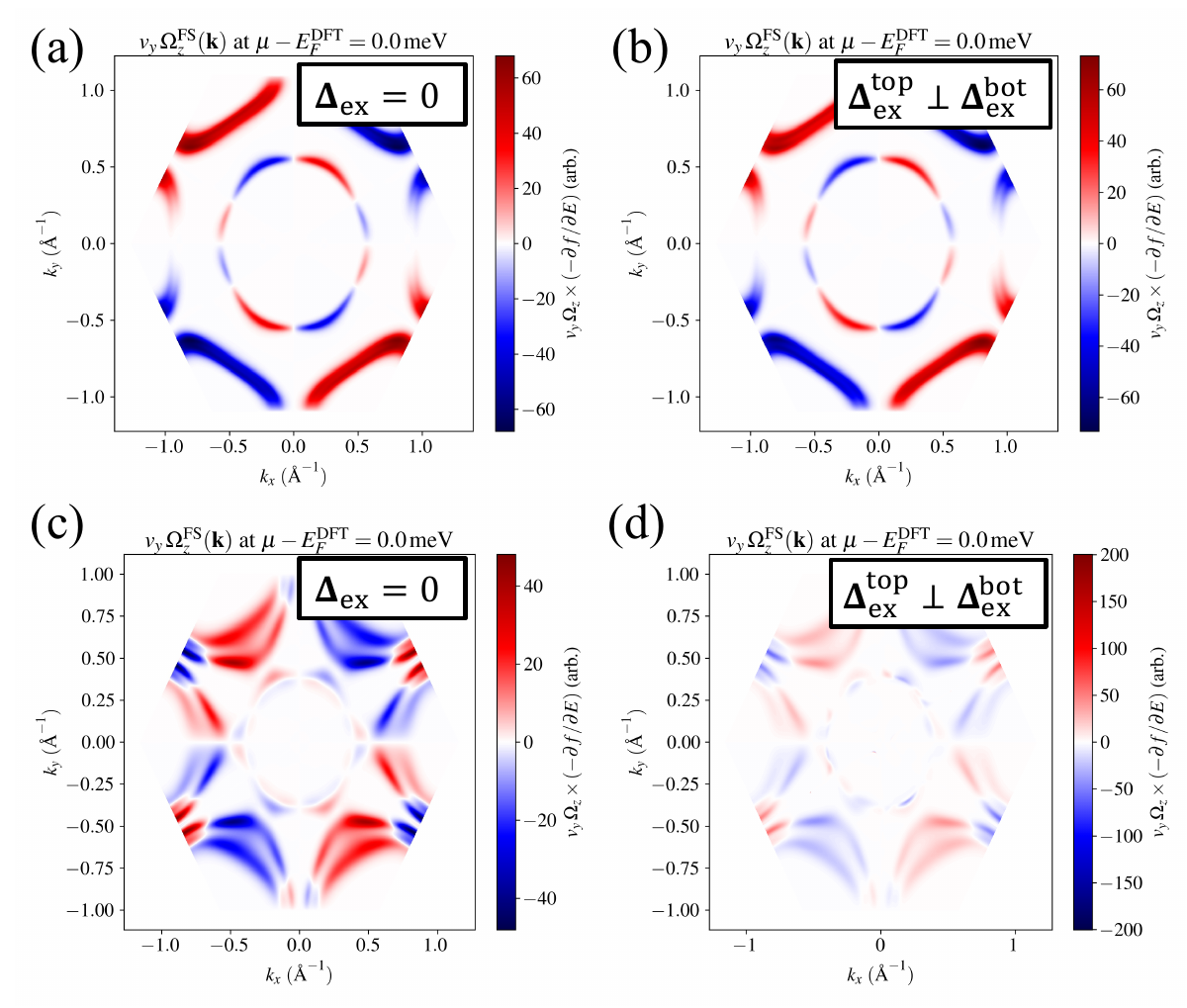}
    \caption{\textbf{Chalcogen dependence of the BCD kernel in the orthogonal
    two-sided configuration.}
    All panels show the band-summed BCD kernel
    $M_{yz}(\mathbf{k};E_F)=\sum_n M_{yz,n}(\mathbf{k};E_F)$ defined above at
    $\mu-E_F^{\rm DFT}=0$ in the orthogonal geometry used in
    Figs.~2(c) and 3(c) of the main text.
    (a),(b) NbS$_2$ with zero exchange ($\Delta_{\rm ex}=0$) (a) and with
    orthogonal two-sided proximity
    $\boldsymbol{\Delta}_{\rm ex}^{\rm top}\perp\boldsymbol{\Delta}_{\rm ex}^{\rm bot}$ (b).
    (c),(d) Corresponding maps for NbTe$_2$.
    Note the different color-bar scales between the top and bottom rows. The displayed kernel is larger for NbTe$_2$ than for NbS$_2$, in line with the growth of the BCD across the Nb$X_2$ series in Fig.~3(c), but the arbitrary plotting normalization and unmatched scales do not support a quantitative orders-of-magnitude ratio.}
    \label{fig:S8_BCD_maps}
\end{figure}

Within the Fermi-surface formulation used in the main text, the Berry-curvature
dipole (BCD) entering the nonlinear Hall conductivity is
\begin{equation}
D_{ab}(E_{\rm F})
= -\sum_n \int_{\rm BZ} \frac{d^2k}{(2\pi)^2}\,
\left[\frac{\partial}{\partial k_a}
f_0\!\left(\varepsilon_{n\mathbf{k}}-E_{\rm F}\right)\right]
\Omega^b_{n\mathbf{k}},
\end{equation}
and, for the Hall geometry used here, $D_y\equiv D_{yz}^{\rm 2D}$ controls $\chi_{xyy}^{\rm sheet}$. Using $\partial_{k_y}\varepsilon_{n\mathbf k}=\hbar v_{n,y}$ gives the exact Fermi-surface form in terms of the band-resolved ``BCD kernel''
\begin{equation}
M_{yz,n}(\mathbf{k};E_{\rm F})
\equiv v_{n,y}(\mathbf{k})\,\Omega_{n,z}(\mathbf{k})
\left[-\frac{\partial f_0(\varepsilon_{n\mathbf k}-E_{\rm F})}{\partial\varepsilon}\right],
\end{equation}
so that
\begin{equation}
D_y(E_{\rm F})
=\hbar\sum_n\int_{\rm BZ}\frac{d^2k}{(2\pi)^2}\,
M_{yz,n}(\mathbf{k};E_{\rm F}).
\end{equation}
Below we use $M_{yz}\equiv\sum_n M_{yz,n}$ for the band-summed kernel.
Figures~\ref{fig:S7_BCD_maps} and \ref{fig:S8_BCD_maps} show the band-summed kernel at selected
energies and proximity configurations, making explicit how the BCD peaks and
sign changes in Fig.~3 are encoded in the $k$-resolved structure of
$v_y\Omega_z(-\partial f_0/\partial\varepsilon)$. The supplied maps use a plotting normalization and panel-dependent color scales; they establish the sign texture but do not by themselves provide quantitative cross-panel amplitude ratios.

\subsubsection*{(i) Geometry dependence near the BCD maxima in NbSe$_2$}

To connect the BCD peaks of Fig.~3(a) to the underlying band geometry in
NbSe$_2$, Fig.~\ref{fig:S7_BCD_maps} shows $M_{yz}(\mathbf{k})$ at
$\mu-E_F^{\rm DFT}=247$~meV for the four proximity configurations considered in
the main text.

In the one-sided configuration with purely out-of-plane exchange
$\Delta_{\rm ex}\!\parallel\!\hat{\mathbf z}$ [Fig.~\ref{fig:S7_BCD_maps}(a)], the Berry curvature
on the interface-induced $k$-linear Rashba-split pockets is finite, but the exact
$C_3$ symmetry forces the Brillouin-zone integral of the in-plane BCD vector to vanish.
Accordingly, the positive and negative patches of the velocity-weighted kernel
$M_{yz}(\mathbf{k})$ cancel after integration. The tiny nonzero values visible in the
$\Delta_{\rm ex}\!\parallel\!\hat{\mathbf z}$ curve of Fig.~3(a) should therefore be
interpreted as finite-grid or interpolation residuals rather than as a physical BCD.

When the exchange is slightly canted by $\theta=10^\circ$
[Fig.~\ref{fig:S7_BCD_maps}(b)], the same interface-induced $k$-linear Rashba–exchange anticrossings generate a much
more asymmetric pattern of $M_{yz}(\mathbf{k})$: bright lobes of one sign
dominate one side of the $\Gamma$-centered contour, while the opposite side is
substantially weakened. This imbalance removes the cancellations present in
panel (a) and produces a large net $D_y$, corresponding to the pronounced BCD
peak of the orange curve in Fig.~3(a).

The symmetric parallel two-sided configuration
$\boldsymbol{\Delta}_{\rm ex}^{\rm top}\parallel\boldsymbol{\Delta}_{\rm ex}^{\rm bot}
\parallel\hat{\mathbf z}$ [Fig.~\ref{fig:S7_BCD_maps}(c)] preserves both $\sigma_h$
and $C_3$ in the ideal model. Although $M_{yz}(\mathbf{k})$ contains finite positive
and negative lobes, their Brillouin-zone integral vanishes exactly by $C_3$ symmetry.
The tiny residual values visible in the nearly vanishing green curve of Fig.~3(a)
should therefore be regarded as finite-grid or interpolation residuals rather than
as a physical BCD.

By contrast, in the orthogonal two-sided configuration
$\boldsymbol{\Delta}_{\rm ex}^{\rm top}\perp\boldsymbol{\Delta}_{\rm ex}^{\rm bot}$
[Fig.~\ref{fig:S7_BCD_maps}(d)], the effective Zeeman field acquires both out-of-plane
and in-plane components and simultaneously breaks $\sigma_h$ and certain in-plane
mirrors. This produces a strongly anisotropic $M_{yz}(\mathbf{k})$ with
enhanced weight on selected segments of the $\Gamma$-centered Fermi contour and
reduced weight on the opposite side. The resulting Brillouin-zone integral
yields a sizable BCD with a sign and magnitude that match the red curve in
Fig.~3(a); the fact that the pattern differs qualitatively from
Fig.~\ref{fig:S7_BCD_maps}(b) also explains why the orthogonal configuration can generate
a BCD comparable in size to the canted one-sided case, but with a distinct
energy dependence and, in certain windows, an opposite sign.

Finally, Fig.~3(b) shows that changing the exchange magnitude shifts and reshapes the energy-dependent BCD features. The minimal model predicts an exchange-odd contribution linear in a small in-plane field at fixed band structure and sufficiently close to a specified operating window. The multiband curves over the full 5--50~meV range need not scale globally or monotonically because the hot spots move relative to the chemical potential; the nonmonotonic examples in Sec.~S5.5 make this limitation explicit.

\subsubsection*{(ii) Chalcogen dependence at the Fermi level}

Fig.~\ref{fig:S8_BCD_maps} illustrates how the BCD kernel
$M_{yz}(\mathbf{k})$ responds to orthogonal two-sided proximity across the
Nb$X_2$ series at $\mu-E_F^{\rm DFT}=0$, complementing the energy-dependent BCD
shown in Fig.~3(c). Panels (a) and (b) correspond to NbS$_2$ in the
TRS-symmetric case ($\Delta_{\rm ex}=0$) and in the orthogonal two-sided
geometry $\boldsymbol{\Delta}_{\rm ex}^{\rm top}\perp\boldsymbol{\Delta}_{\rm ex}^{\rm bot}$,
respectively. In both situations the $\Gamma$-centered Fermi contour is nearly
circular and $M_{yz}(\mathbf{k})$ is small on the displayed scale. In the pristine case, positive and negative
lobes constrained by the exact $D_{3h}$ symmetry cancel in
the Brillouin-zone integral, yielding a very small $D_y$ consistent with the
weak NbS$_2$ signal in Fig.~3(c).

Panels (c) and (d) show the corresponding maps for NbTe$_2$. The
$\Gamma$-centered pocket is more hexagonally warped, concurrent with the heavier chalcogen, stronger spin--orbit coupling, and other band-structure changes. Even without proximity exchange [Fig.~\ref{fig:S8_BCD_maps}(c)], the local magnitude of
$M_{yz}(\mathbf{k})$ is larger on the displayed NbTe$_2$ scale, although the exact $C_3$ symmetry of pristine $D_{3h}$ NbTe$_2$ forces the in-plane BCD vector to vanish. When the orthogonal two-sided exchange
is switched on [Fig.~\ref{fig:S8_BCD_maps}(d)], the BCD kernel on the
$\Gamma$-centered contour becomes larger and
develops an anisotropic pattern that lacks the nearly cancelling
structure of NbS$_2$. This enhancement, together with the modest change in
Fermi-surface shape, directly explains the strong increase of the BCD
across the NbS$_2\!\rightarrow$NbSe$_2\!\rightarrow$NbTe$_2$ series shown in
Fig.~3(c).

Overall, Figs.~\ref{fig:S7_BCD_maps} and \ref{fig:S8_BCD_maps} illustrate that the
geometry- and chalcogen-dependent BCD reported in Fig.~3 is governed by the
exchange-dependent reshaping of the BCD kernel
$M_{yz}(\mathbf{k})=\sum_n v_{n,y}\Omega_{n,z}(-\partial f_0/\partial\varepsilon)$ on an almost
unchanged $\Gamma$-centered Fermi surface. This provides a qualitative microscopic
$k$-space picture for the Hall-valve and orthogonal-proximity mechanisms
central to the main text.

\subsection*{S5.4 Temperature smearing effects on $\sigma_{xy}$ and $D_y$}

\begin{figure}[t]
    \centering
    \includegraphics[width=\columnwidth]{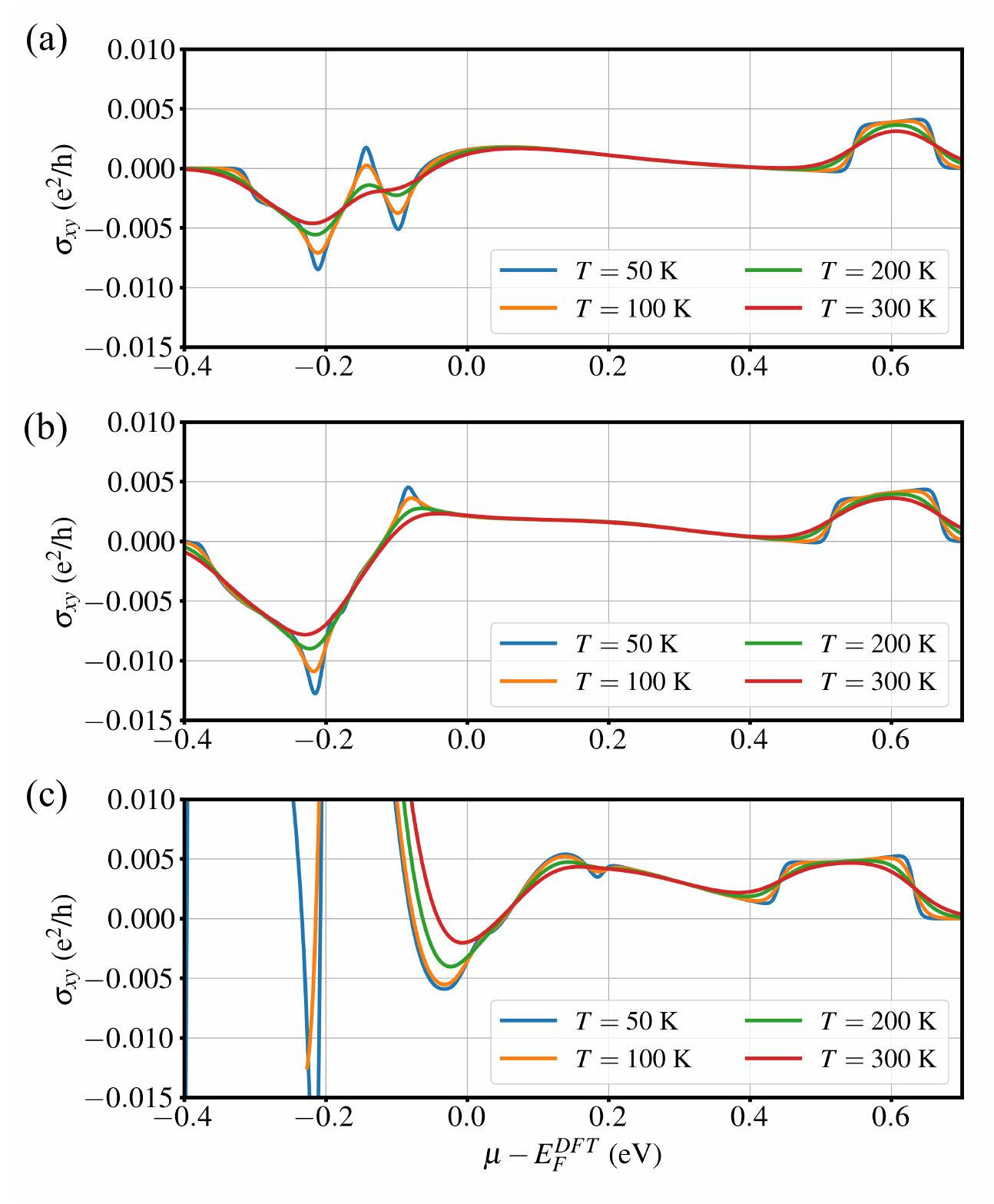}
    \caption{\textbf{Finite-temperature smearing of the anomalous Hall conductivity
    in the orthogonal two-sided configuration.}
    Sheet anomalous Hall conductivity $\sigma_{xy}^{\rm sheet}(E_{\rm F})$
    (in units of $e^2/h$) as a function of chemical potential for
    (a) NbS$_2$, (b) NbSe$_2$, and (c) NbTe$_2$ in the orthogonal two-sided
    proximity geometry, evaluated using Fermi--Dirac smoothers corresponding
    to $T=50$, $100$, $200$, and $300$~K.
    Increasing temperature mainly rounds and reduces the sharp features
    associated with interface-induced $k$-linear Rashba--exchange anticrossings near
    $\mu-E_F^{\rm DFT}\!\sim-0.2$~eV, while leaving the overall sign structure
    and high-energy plateau qualitatively unchanged in this fixed-band electronic-smearing test.}
    \label{fig:S9_T_AHC}
\end{figure}

Throughout the main text (Figs.~2 and 3), we use a Fermi--Dirac occupation corresponding to $T=50$~K. For the AHC, the occupied-band Berry curvature is weighted by $f_0(\varepsilon,T)$; for the BCD Fermi-surface form, the kernel is weighted by
\begin{equation}
-\frac{\partial f_0}{\partial \varepsilon}
= \frac{1}{4k_{\rm B}T}\,\mathrm{sech}^2
\!\left[\frac{\varepsilon-E_{\rm F}}{2k_{\rm B}T}\right].
\end{equation}
Equivalently, a finite-temperature response spectrum can be obtained by convolving its zero-temperature spectrum with this derivative. This introduces an energy broadening of order
$\Gamma_T \sim 3.5k_{\rm B}T$, which evolves from
$\Gamma_T\simeq 15$~meV at $T=50$~K to
$\Gamma_T\simeq 90$~meV at $T=300$~K.
Sharp features in $\sigma_{xy}(E_{\rm F})$ or $D_y(E_{\rm F})$ whose
intrinsic width is smaller than $\Gamma_T$ are therefore smeared, while
broader trends and sign changes are largely preserved.

Fig.~\ref{fig:S9_T_AHC} shows the resulting temperature dependence of the
sheet anomalous Hall conductivity for the orthogonal two-sided proximity
configuration in NbS$_2$ (a), NbSe$_2$ (b), and NbTe$_2$ (c).
In all three compounds, the overall structure of $\sigma_{xy}(E_{\rm F})$,
including the main sign reversals and the high-energy plateau near
$\mu-E_F^{\rm DFT}\!\sim0.5$–$0.6$~eV, is essentially unchanged as $T$ is
raised from $50$~K to $300$~K. The main effect of increasing $T$ is a
progressive reduction and broadening of the sharp peaks associated with
interface-induced $k$-linear Rashba--exchange anticrossings near $\mu-E_F^{\rm DFT}\!\sim-0.2$~eV
(visible most clearly in NbSe$_2$ and NbTe$_2$), while the zero crossings
shift only weakly.

For NbS$_2$ [Fig.~\ref{fig:S9_T_AHC}(a)], where the Berry curvature around the
$\Gamma$-centered pocket is relatively weak, $\sigma_{xy}(E_{\rm F})$ is
small at all temperatures and the modest negative minimum at
$\mu-E_F^{\rm DFT}\!\lesssim-0.2$~eV is smoothly rounded as $T$ increases.
In NbSe$_2$ [Fig.~\ref{fig:S9_T_AHC}(b)],
the pronounced negative minimum at $\mu-E_F^{\rm DFT}\!\approx-0.2$~eV is
gradually reduced in magnitude and slightly broadened, but remains clearly
visible up to $T=300$~K. In NbTe$_2$ [Fig.~\ref{fig:S9_T_AHC}(c)], the
strong SOC produces a very sharp low-temperature spike at the same energy;
finite-$T$ smearing substantially reduces its amplitude, yet a sizable
negative $\sigma_{xy}$ region remains in the $T=300$~K electronic-smearing curve, together
with the positive plateau at higher energies.

\begin{figure}[t]
    \centering
    \includegraphics[width=\columnwidth]{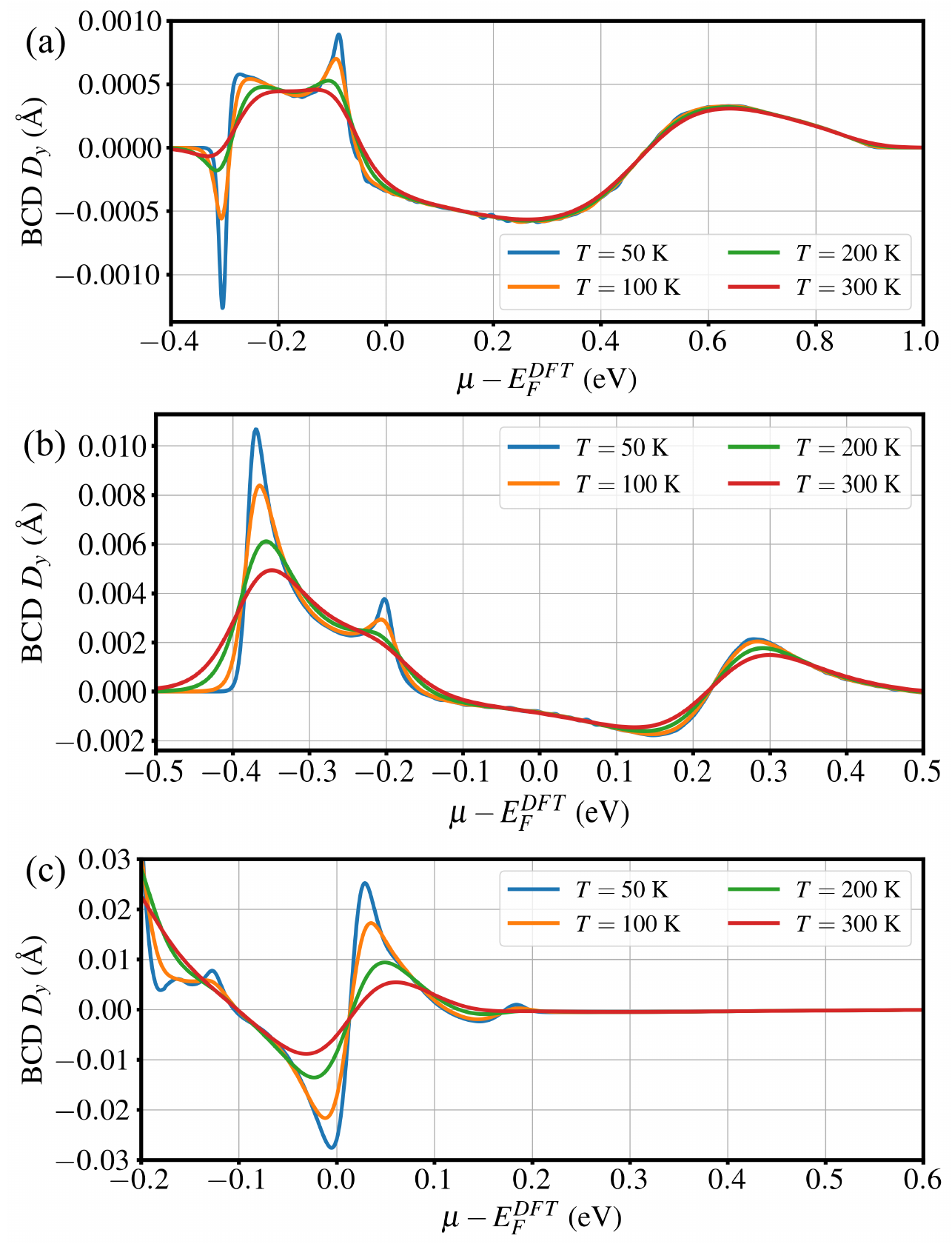}
    \caption{\textbf{Temperature dependence of the Berry-curvature dipole in
    the orthogonal two-sided configuration.}
    Berry-curvature dipole $D_y(E_{\rm F})$ as a function of chemical
    potential for (a) NbS$_2$, (b) NbSe$_2$, and (c) NbTe$_2$, evaluated
    using Fermi--Dirac smoothers corresponding to $T=50$, $100$, $200$, and
    $300$~K. Increasing temperature reduces and broadens the sharp BCD peaks
    associated with interface-induced $k$-linear Rashba--exchange anticrossings, particularly in NbTe$_2$,
    and can remove narrow sign-changing structures. Broader finite-$D_y$ windows remain in parts of the displayed spectra.}
    \label{fig:S10_T_BCD}
\end{figure}

The Berry-curvature dipole inherits the same Fermi--Dirac broadening as
$\sigma_{xy}$ but is even more sensitive to thermal smearing, since it is
essentially a weighted derivative of the anomalous velocity on the Fermi
surface. Fig.~\ref{fig:S10_T_BCD} shows the temperature evolution of
$D_y(E_{\rm F})$ in the orthogonal two-sided configuration for
NbS$_2$ (a), NbSe$_2$ (b), and NbTe$_2$ (c), complementing the
$T=50$~K reference curves in Fig.~3.

For NbS$_2$ [Fig.~\ref{fig:S10_T_BCD}(a)], the overall magnitude of $D_y$ is
small ($|D_y|\lesssim10^{-3}$~\AA) at all temperatures. The sharp negative
dip and nearby positive peak around $\mu-E_F^{\rm DFT}\!\sim-0.25$~eV at
$T=50$~K are rapidly rounded as $T$ increases, but the sign pattern and the
position of the extrema remain essentially unchanged. The broad positive
feature at higher energies ($\mu-E_F^{\rm DFT}\!\sim0.5$–$0.7$~eV) is only
weakly affected by temperature, reflecting the comparatively broad underlying
anticrossings in this compound.

In NbSe$_2$ [Fig.~\ref{fig:S10_T_BCD}(b)], the low-temperature BCD exhibits a
pronounced positive maximum near $\mu-E_F^{\rm DFT}\!\approx-0.35$~eV, followed
by a weaker negative shoulder and a smaller positive peak at
$\mu-E_F^{\rm DFT}\!\sim0.2$~eV, consistent with Fig.~3(a). Increasing $T$
progressively reduces the height of these extrema and slightly broadens the
doping window over which $D_y$ is sizable, but the sign reversals and
    some broader structures remain at the largest electronic-smearing temperature. This result does not model the temperature dependence of magnetic order, scattering time, the interface, or the NbSe$_2$ CDW, and therefore is not a room-temperature device prediction.

The strongest temperature dependence occurs in NbTe$_2$
[Fig.~\ref{fig:S10_T_BCD}(c)], where the combination of large SOC and orthogonal
proximity produces very sharp BCD features at low $T$. At $T=50$~K, $D_y$
reaches values of order $3\times10^{-2}$~\AA\ and exhibits a sequence of
sign changes as $\mu$ is swept across the interface-induced $k$-linear Rashba--exchange anticrossings.
Raising $T$ strongly suppresses the narrow positive and negative peaks near
$\mu-E_F^{\rm DFT}\!\sim0$–$0.1$~eV; by $T=300$~K most of these sharp
features are washed out, and only a broad negative lobe at small positive
doping survives with reduced amplitude. In contrast, the broader high-energy
structures in $\sigma_{xy}$ [Fig.~\ref{fig:S9_T_AHC}(c)] remain clearly visible
at the same temperatures, underscoring that the nonlinear Hall response is
more sensitive to thermal smearing than the linear AHE.

Overall, Fig.~\ref{fig:S10_T_BCD} shows that while electronic Fermi--Dirac smearing
quantitatively reduces the magnitude of the BCD—especially for the sharp,
SOC-enhanced features in NbTe$_2$—the qualitative sign structure and the
broader doping windows supporting a finite $D_y$ remain in parts of the spectra up to the $T=300$~K smearing curve. Because the band structure, proximity exchange, magnetization, CDW state, and relaxation time are held fixed, this test establishes only the sensitivity to electronic occupation broadening.

\subsection*{S5.5 Exchange-field magnitude variation effects on $\sigma_{xy}$ and $D_y$}

In experimentally realistic magnet/TMD heterostructures, the proximity exchange
strengths at the top and bottom interfaces need not be identical due to
different terminations, bonding geometries, and magnetic materials. To assess
the robustness of the \emph{orthogonal} two-sided principle against such interface
asymmetry, we fix the orthogonal exchange texture
$\boldsymbol{\Delta}_{\rm ex}^{\rm top}\perp \boldsymbol{\Delta}_{\rm ex}^{\rm bot}$
(as in the main-text orthogonal configuration) and vary only the \emph{magnitudes}
of the layer-resolved exchange fields. We compare a symmetric reference case
$\Delta_{\rm ex}^{\rm top}=\Delta_{\rm ex}^{\rm bot}=30$~meV to two increasingly
asymmetric cases with a reduced top-interface exchange,
$\Delta_{\rm ex}^{\rm top}=5$~meV and $\Delta_{\rm ex}^{\rm bot}=30$~meV or
$\Delta_{\rm ex}^{\rm bot}=50$~meV. This provides a parameter-robustness check
complementary to Sec.~S5.2 (Berry-curvature maps) and Sec.~S5.4 (finite-$T$
smearing), and connects directly to the main-text conclusion that the linear AHE
and nonlinear Hall response are controlled primarily by Berry-curvature
redistribution rather than by large Fermi-surface deformation.

\begin{figure}[t]
    \centering
    \includegraphics[width=\columnwidth]{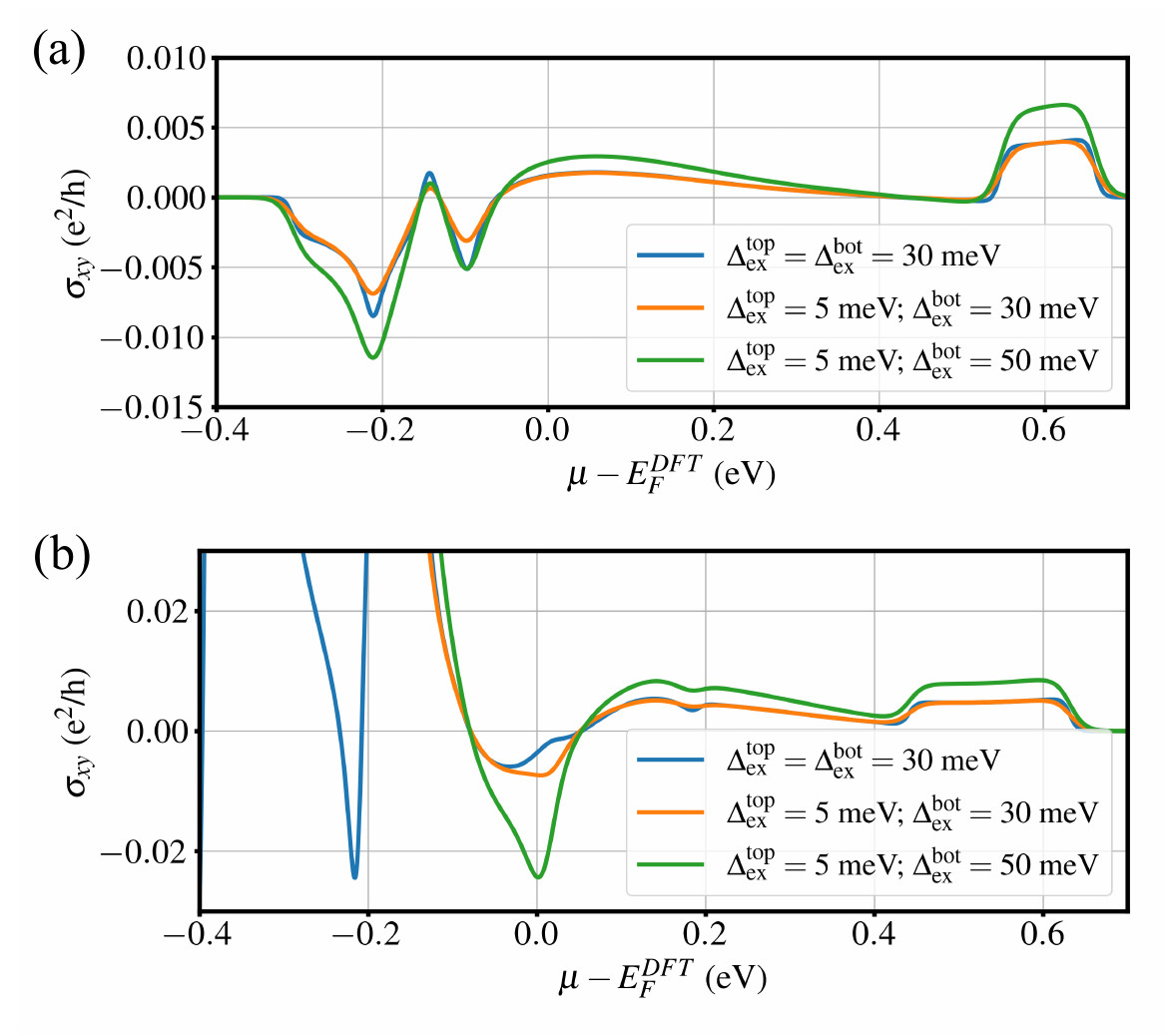}
    \caption{\textbf{Exchange-field magnitude variation and anomalous Hall conductivity.}
    Sheet anomalous Hall conductivity $\sigma_{xy}^{\rm sheet}(E_{\rm F})$ in the orthogonal
    two-sided configuration with unequal proximity strengths.
    (a) NbS$_2$ and (b) NbTe$_2$ for three choices of layer-resolved exchange magnitudes:
    $\Delta_{\rm ex}^{\rm top}=\Delta_{\rm ex}^{\rm bot}=30$~meV (symmetric),
    $\Delta_{\rm ex}^{\rm top}=5$~meV with $\Delta_{\rm ex}^{\rm bot}=30$~meV,
    and $\Delta_{\rm ex}^{\rm top}=5$~meV with $\Delta_{\rm ex}^{\rm bot}=50$~meV.}
    \label{fig:S11_exvar_AHC}
\end{figure}

\paragraph*{(i) AHC: predominantly controlled by the effective out-of-plane exchange scale.}
Figure~\ref{fig:S11_exvar_AHC} shows that the qualitative $\mu$-dependence of
$\sigma_{xy}^{\rm sheet}$ is preserved under substantial interface asymmetry,
while the magnitudes of the extrema remain tunable. This is consistent with the
symmetry analysis in the main text and the curvature-map diagnostics in
Sec.~S5.2: a finite out-of-plane exchange component generates an imbalance of
positive and negative Berry-curvature patches on the $\Gamma$-centered pocket,
leading to a nonzero Brillouin-zone integral
$\sigma_{xy}\propto\int_{\rm BZ} d^2k\,\Omega_z^{\rm occ}(\mathbf{k})$, with $\Omega_z^{\rm occ}=\sum_n f_n\Omega_{n,z}$.

For NbS$_2$ [Fig.~\ref{fig:S11_exvar_AHC}(a)], where the SOC-driven curvature around the
$\Gamma$ pocket is comparatively weak (cf.\ Sec.~S5.2(iii) and Fig.~3(c) in the main text),
$\sigma_{xy}^{\rm sheet}$ remains small for all three exchange choices. Nevertheless,
the negative feature near $\mu-E_F^{\rm DFT}\!\sim-0.2$~eV and the positive plateau at
higher doping ($\mu-E_F^{\rm DFT}\!\sim0.55$--$0.65$~eV) become more pronounced as
$\Delta_{\rm ex}^{\rm bot}$ increases, indicating that strengthening the dominant
interface enhances the net curvature imbalance without qualitatively altering the
doping windows where the AHE is active.

In NbTe$_2$ [Fig.~\ref{fig:S11_exvar_AHC}(b)], the same variation produces a much stronger
response because SOC sharpens interface-induced $k$-linear Rashba--exchange anticrossings and concentrates the
Berry curvature into narrower energy windows (cf.\ Sec.~S5.4). Increasing
$\Delta_{\rm ex}^{\rm bot}$ enhances both the low-energy extremal features and the broader
high-doping plateau, confirming that the AHE in the orthogonal geometry is robust and
tunable even when one interface couples only weakly to the TMD.

\begin{figure}[t]
    \centering
    \includegraphics[width=\columnwidth]{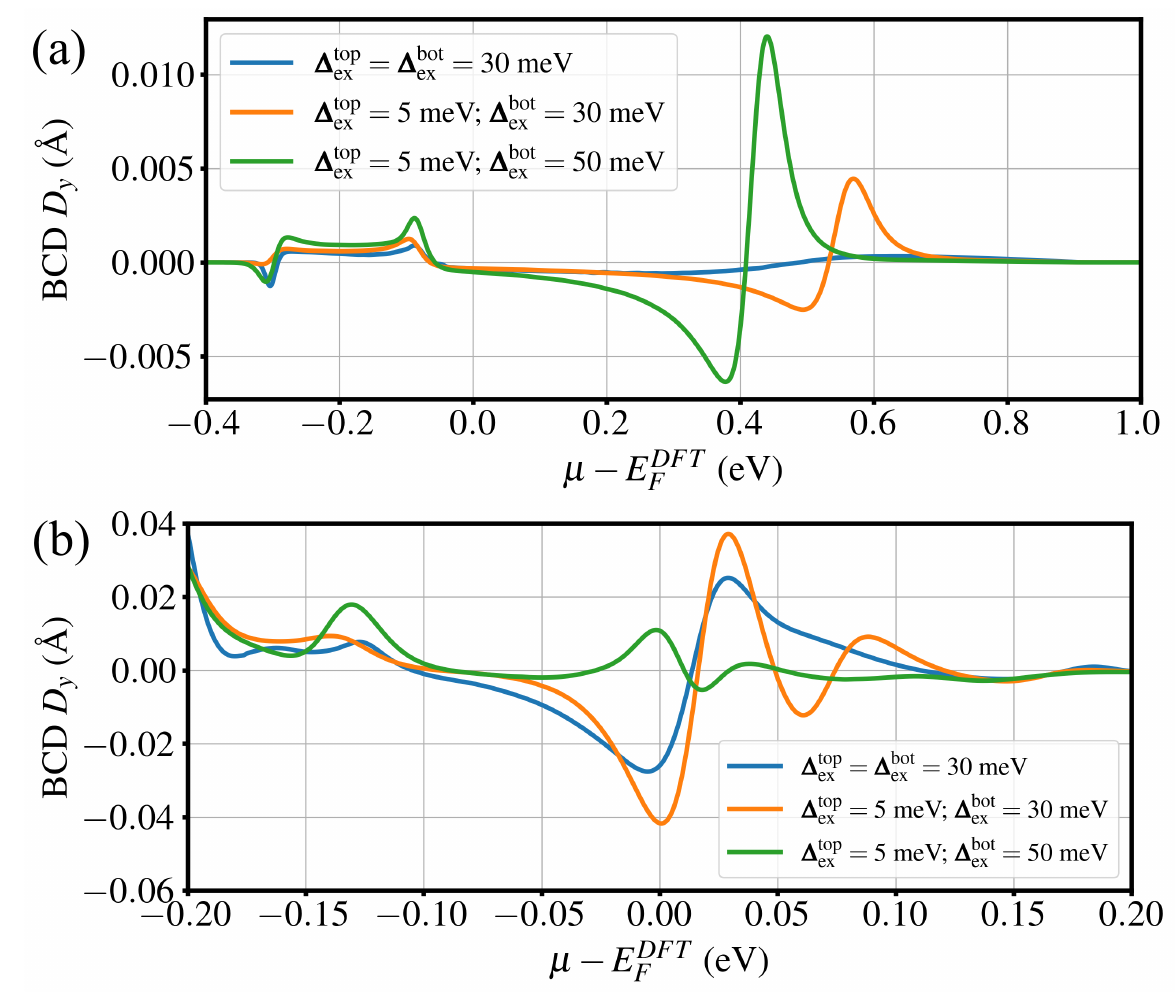}
    \caption{\textbf{Exchange-field magnitude variation and Berry-curvature dipole.}
    Berry-curvature dipole $D_y(E_{\rm F})$ for the same three choices of exchange magnitudes
    as in Fig.~\ref{fig:S11_exvar_AHC}.
    (a) NbS$_2$ and (b) NbTe$_2$.}
    \label{fig:S12_exvar_BCD}
\end{figure}

\paragraph*{(ii) BCD: more sensitive and can be non-monotonic under exchange variation.}
The Berry-curvature dipole probes a \emph{first moment} of the Berry curvature on the
thermally broadened Fermi surface, and is therefore more sensitive than $\sigma_{xy}$
to how curvature is redistributed anisotropically around the $\Gamma$-centered pocket.
In the orthogonal geometry, the simultaneous presence of in-plane and out-of-plane
exchange components breaks the remaining symmetry constraints and makes an in-plane
polar vector $D_y$ symmetry-allowed (as discussed in the main text and in Sec.~S5.3).

For NbS$_2$ [Fig.~\ref{fig:S12_exvar_BCD}(a)], the symmetric case
$\Delta_{\rm ex}^{\rm top}=\Delta_{\rm ex}^{\rm bot}=30$~meV yields only a weak dipole,
consistent with the small nonlinear Hall response across the Nb$X_2$ series in
Fig.~3(c) of the main text. Introducing interface asymmetry with
$\Delta_{\rm ex}^{\rm top}=5$~meV enhances $D_y$ and shifts the dominant structures to
higher $\mu$, indicating that in this weak-SOC compound the BCD is particularly
sensitive to how exchange-induced anticrossings move relative to the Fermi level as the
stronger interface is tuned. Increasing $\Delta_{\rm ex}^{\rm bot}$ from $30$ to $50$~meV
further amplifies the dominant positive peak, showing that even in NbS$_2$ a sizable BCD
can be obtained once the relevant avoided crossings are brought close to the active
energy window.

In NbTe$_2$ [Fig.~\ref{fig:S12_exvar_BCD}(b)], $D_y$ is large already in the symmetric case
due to SOC-enhanced curvature hot spots (cf.\ Sec.~S5.2(iii)). Reducing the top-interface
exchange to $\Delta_{\rm ex}^{\rm top}=5$~meV while keeping
$\Delta_{\rm ex}^{\rm bot}=30$~meV enhances the low-energy positive/negative lobes,
consistent with the main-text mechanism that orthogonal exchange efficiently generates
a large BCD when the exchange scale is comparable to the SOC-assisted interface-induced $k$-linear Rashba splitting.
Interestingly, increasing the stronger interface further to
$\Delta_{\rm ex}^{\rm bot}=50$~meV suppresses the BCD near $\mu\approx E_F^{\rm DFT}$ even
though the AHC increases [Fig.~\ref{fig:S11_exvar_AHC}(b)]. This non-monotonic behavior is
naturally understood as a moment effect: driving the system deeper into a strong
out-of-plane exchange regime can enhance the net Berry-curvature integral (favoring
$\sigma_{xy}$) while simultaneously making the curvature distribution around the Fermi
surface less dipolar or shifting the dominant hot spots away from the relevant chemical
potential window, thereby reducing $D_y$.

Together, Figs.~\ref{fig:S11_exvar_AHC} and \ref{fig:S12_exvar_BCD} demonstrate that the
orthogonal two-sided architecture remains functional under substantial interface
inequivalence. The linear AHE retains its characteristic sign structure and is tunable
primarily via the stronger (out-of-plane-dominant) interface, while the nonlinear Hall
response can be optimized by balancing the two exchange magnitudes so that
SOC-assisted interface-induced $k$-linear Rashba--exchange anticrossings remain close to the targeted doping window.
This corroborates the main-text conclusion that coexisting linear and nonlinear Hall
signals in Nb$X_2$ can be achieved without fine control of a canting angle, although the sharply dispersive responses can still require chemical-potential tuning, and extends the
robustness checks provided by the $k$-resolved curvature maps (Sec.~S5.2) and finite-$T$
smearing (Sec.~S5.4).


\subsection*{S5.6 Robustness check with Hubbard-$U$ correction on Nb $4d$ states}

\begin{figure}[t]
\centering
\includegraphics[width=\columnwidth]{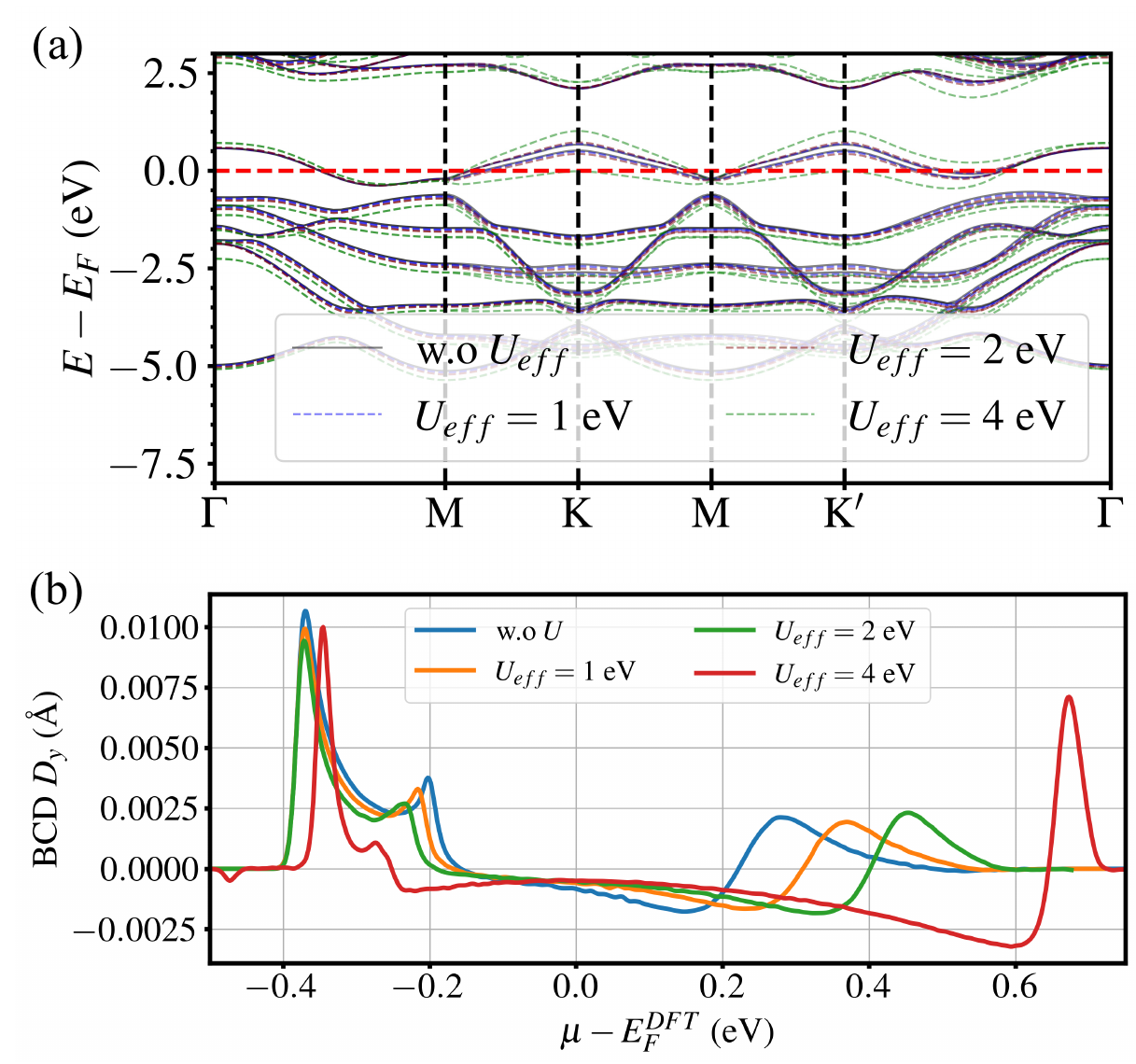}
\caption{
PBEsol+$U$ robustness check for monolayer NbSe$_2$.
(a) Band structures obtained without Hubbard correction and with $U_{\rm eff}=1$, $2$, and $4$~eV applied to the Nb $4d$ orbitals.
The Hubbard correction quantitatively shifts the Nb-derived bands but preserves the metallic low-energy band structure.
(b) Berry-curvature dipole $D_y(\mu)$ for the orthogonal two-sided proximity configuration.
The peak positions and magnitudes vary with $U_{\rm eff}$ because the Berry-curvature hot spots shift relative to the chemical potential, but a finite BCD response remains present for all tested $U_{\rm eff}$ values.
The chemical potential is measured relative to the PBEsol Fermi level $E_F^{\rm DFT}$.
In both panels, the no-$U$ curve corresponds to $U_{\rm eff}=0$.
}
\label{fig:S13_hubbardU}
\end{figure}

We also examined the effect of adding a local Hubbard correction to the Nb $4d$ orbitals.
This check was motivated by previous PBE+$U$ studies of Nb-based transition-metal dichalcogenides and by the possible role of local Coulomb corrections in Nb$X_2$~\mbox{\cite{LiuPCCP2023}}.
We performed PBEsol+$U$ calculations for monolayer NbSe$_2$ using the rotationally invariant Dudarev form~\mbox{\cite{Dudarev1998}}.
We focus on NbSe$_2$ because it is the central compound used for the detailed proximity-texture and sign-control analysis and lies between NbS$_2$ and NbTe$_2$ in the SOC-strength trend.
$U_{\rm eff}=U-J=1$, $2$, and $4$~eV were applied to the Nb $4d$ shell.
Here $U_{\rm eff}=4$~eV should be viewed as an upper-end robustness check for the metallic Nb $4d$ states, whereas $U_{\rm eff}=1$--$2$~eV represent more moderate local-correlation corrections.

Figure~\mbox{\ref{fig:S13_hubbardU}}(a) compares the corresponding band structures with the PBEsol result without Hubbard correction.
Increasing $U_{\rm eff}$ shifts the Nb-derived bands quantitatively, especially away from the Fermi level, but the system remains metallic and the qualitative low-energy band structure relevant to the Hall-response calculations is retained.
Thus, adding $U_{\rm eff}$ does not introduce a symmetry change or a qualitative reconstruction of the low-energy metallic state.

Figure~\mbox{\ref{fig:S13_hubbardU}}(b) shows the Berry-curvature dipole $D_y(\mu)$ for the orthogonal two-sided proximity configuration.
The detailed peak positions and amplitudes depend on $U_{\rm eff}$ because the Hubbard correction shifts the band energies and hence the locations of Berry-curvature hot spots relative to the chemical potential.
Nevertheless, a finite $D_y$ remains present for all $U_{\rm eff}$ values considered.
This shows that a symmetry-allowed finite BCD remains present somewhere in the displayed chemical-potential range for every tested $U_{\rm eff}$; it does not establish quantitative stability of the peak position, magnitude, or sign at a fixed operating point.
Therefore, the tested PBEsol+$U$ calculations support only the qualitative symmetry-based conclusion that the orthogonal two-sided geometry permits a BCD in NbSe$_2$. This test is restricted to the central compound and should not be interpreted as $U_{\rm eff}$-independence of the quantitative response or of the Nb$X_2$ cross-material trend.

\section*{S6. Harmonic Hall detection and order-of-magnitude estimates}

This section provides an explicit connection between the calculated linear AHE and BCD-driven nonlinear Hall
responses and a standard experimental readout based on harmonic Hall detection. Because the calculated component is $D_y\equiv D_{yz}^{\rm 2D}$, the relevant quasistatic nonlinear sheet tensor component is $\chi_{xyy}^{\rm sheet}=(e^3\tau/\hbar^2)D_y$. An ac longitudinal drive $I_y(t)$ therefore generates a first-harmonic transverse $x$-voltage dominated by the linear sheet Hall channel ($\sigma_{xy}^{\rm sheet}$), while the intrinsic nonlinear Hall channel ($\chi_{xyy}^{\rm sheet}$) produces a phase-coherent second-harmonic transverse voltage. As emphasized in the main text, the exchange-odd parts of the two harmonics
carry distinct symmetry-controlled sign reversals in the orthogonal two-sided geometry.

\subsection*{S6.1 Device geometry and magnetic switching protocol for the orthogonal two-sided configuration}

We first clarify the experimental geometry underlying Fig.~4(a,b) of the main text.
The proposed device is a dual-interface magnetic-proximity Hall bar in which a monolayer $1H$-Nb$X_2$ channel is placed between two magnetic layers.
The bottom magnetic layer is chosen to provide a predominantly out-of-plane proximity exchange component,
\begin{equation}
\boldsymbol{\Delta}^{\rm bot}\simeq s_z |\Delta_z|\hat{\mathbf z},
\qquad s_z=\pm 1,
\end{equation}
while the top magnetic layer is chosen to provide a predominantly in-plane proximity exchange component,
\begin{equation}
\boldsymbol{\Delta}^{\rm top}\simeq s_x |\Delta_x|\hat{\mathbf x},
\qquad s_x=\pm 1.
\end{equation}
Here $s_z$ and $s_x$ label the independently switchable signs of the out-of-plane and in-plane proximity components.
This realizes the orthogonal exchange texture used in the main text, with an effective field
\begin{equation}
\boldsymbol{\Delta}_{\rm eff}=
\boldsymbol{\Delta}^{\rm bot}
+
\boldsymbol{\Delta}^{\rm top}
\simeq
s_x|\Delta_x|\hat{\mathbf x}
+
s_z|\Delta_z|\hat{\mathbf z}.
\end{equation}
The out-of-plane component primarily controls the anomalous Hall conductivity, whereas the in-plane component breaks the remaining threefold constraint and shifts the Berry-curvature hot spots, thereby controlling the Berry-curvature-dipole contribution to the second-harmonic Hall response.

The calculated $D_y\equiv D_{yz}^{\rm 2D}$ is detected through $\chi_{xyy}^{\rm sheet}$; accordingly, the Hall-bar readout below uses a $y$-directed ac current and measures the transverse $x$-voltage $V_{xy}$.

For the two-sided geometry of Fig.~4(a), the bottom and top magnetic layers should ideally have different easy-axis directions or different coercive fields, so that $s_z$ and $s_x$ can be reversed independently.
Out-of-plane magnetic proximity may be supplied by Ising-like ferromagnetic insulators or magnetic van der Waals layers such as CrI$_3$-type magnets, or by EuS-type magnetic insulating substrates used in TMD proximity experiments~\cite{SeylerNanoLett18,NordenNC19}.
A predominantly in-plane exchange component may be supplied by an easy-plane magnetic insulator.
For example, atomically thin CrCl$_3$ is an insulating van der Waals magnet with easy-plane magnetic anisotropy, and monolayer CrCl$_3$ has been reported as an in-plane 2D-XY ferromagnet~\cite{CaiNanoLett2019,BedoyaPintoScience2021}.
In such a configuration, the in-plane magnetization direction can be selected or reversed by a small in-plane magnetic field, while the out-of-plane magnetic layer remains pinned by its perpendicular magnetic anisotropy.
The specific material choice is therefore device dependent; the symmetry requirement is simply that one interface supplies a switchable out-of-plane exchange component and the other supplies a switchable in-plane component.
These examples are component-wise candidates, not a validated complete stack. A practical design must also satisfy compatible magnetic ordering temperatures, separated coercive fields, electrical insulation or controlled shunting, clean interfaces, and a proximity scale appropriate to the chosen operating temperature. The present calculation does not demonstrate that one listed material combination meets all of these conditions simultaneously.

The four magnetic configurations used in Fig.~4(c,d) can be generated by a vector-field protocol.
First, an out-of-plane field $B_z$ is swept through the coercive field of the bottom magnetic layer to set $s_z=\pm1$, while the top in-plane layer is kept fixed by its in-plane anisotropy or by an in-plane bias field.
Second, at fixed $s_z$, an in-plane field $B_x$ is swept through the coercive field of the top magnetic layer to set $s_x=\pm1$, while the bottom layer remains pinned by its out-of-plane anisotropy.
Using magnetic layers with separated coercive fields, exchange bias, or different anisotropy axes allows the sequence
\begin{equation}
(s_z,s_x)=(+,+),\;(-,+),\;(+,-),\;(-,-)
\end{equation}
to be accessed reproducibly.
This protocol gives an internal sign check for the harmonic Hall measurement. At the operating chemical potential used in Fig.~4, the ideal-model exchange-odd contributions satisfy $V_{xy}^{1\omega}\propto s_z$ and $V_{xy}^{2\omega}\propto s_zs_x$ up to the global voltage convention. Thus, reversing $s_x$ at fixed $s_z$ isolates a sign change of the exchange-odd second harmonic, whereas simultaneously reversing $s_z$ and $s_x$ isolates the first-harmonic reversal. A structural $C_3$-breaking background can add a second-harmonic offset that does not obey this product law.
The purpose of Fig.~4 is therefore not to specify a unique material stack, but to show that an orthogonal dual-interface magnetic-proximity geometry provides independent experimental knobs for the linear AHE and BCD-driven nonlinear Hall response.

\subsection*{S6.2 Lock-in protocol and harmonic decomposition}

Consider a Hall-bar device with width $W$ along the transverse $x$ direction, carrying an ac current along $y$,
\begin{equation}
I_y(t)=I_0\cos\omega t,
\end{equation}
corresponding to a 2D sheet current density $j_y^{\rm sheet}(t)=I_y(t)/W$.
To leading order in the Hall angle, the transverse sheet-current response is
\begin{equation}
j_x^{\rm sheet}=\sigma_{xx}^{\rm sheet}E_x+\sigma_{xy}^{\rm sheet}E_y+\chi_{xyy}^{\rm sheet}E_y^2+\cdots,
\label{eq:S6_jy_nlhe}
\end{equation}
where $\chi_{xyy}^{\rm sheet}$ is the intrinsic nonlinear sheet conductivity discussed in the main text and Sec.~S4.4. In the open-circuit transverse geometry relevant for Hall-voltage measurements, $j_x^{\rm sheet}=0$ and the transverse electric field adjusts accordingly. Assuming $\sigma_{xx}^{\rm sheet}\simeq\sigma_{yy}^{\rm sheet}$ and $|\sigma_{xy}^{\rm sheet}|\ll\sigma_{xx}^{\rm sheet}$, one finds
\begin{equation}
E_x(t)\simeq-\frac{\sigma_{xy}^{\rm sheet}}{\sigma_{xx}^{\rm sheet}}E_y(t)
-\frac{\chi_{xyy}^{\rm sheet}}{\sigma_{xx}^{\rm sheet}}E_y^2(t),
\label{eq:S6_Ey_approx}
\end{equation}
and therefore the transverse voltage $V_{xy}(t)\equiv W E_x(t)$ contains dc, $\omega$, and $2\omega$ components. The first- and second-harmonic amplitudes are
\begin{align}
V_{xy}^{1\omega}
&\simeq-\frac{\sigma_{xy}^{\rm sheet}}{(\sigma_{xx}^{\rm sheet})^2}\,I_0,
\label{eq:S6_V1w}\\
V_{xy}^{2\omega}
&\simeq-\frac{\chi_{xyy}^{\rm sheet}}{2(\sigma_{xx}^{\rm sheet})^3}\,
\frac{I_0^2}{W},
\label{eq:S6_V2w}
\end{align}
where we used
$E_y(t)=j_y^{\rm sheet}(t)/\sigma_{yy}^{\rm sheet}\simeq I_0\cos\omega t/(W\sigma_{xx}^{\rm sheet})$
and $\cos^2\omega t=(1+\cos2\omega t)/2$.
Equations~\eqref{eq:S6_V1w}--\eqref{eq:S6_V2w} make explicit that the \emph{first-harmonic} transverse voltage probes the linear Hall channel, while the \emph{second-harmonic} transverse voltage provides a phase-sensitive probe of the nonlinear Hall channel.

For convenience one may define a second-harmonic nonlinear Hall ``resistance'' (per $I_0^2$),
\begin{equation}
R_{xy}^{2\omega}\equiv\frac{V_{xy}^{2\omega}}{I_0^2}
\simeq-\frac{1}{2W}\frac{\chi_{xyy}^{\rm sheet}}{(\sigma_{xx}^{\rm sheet})^3},
\label{eq:S6_R2w}
\end{equation}
which is directly accessible in a lock-in experiment by measuring $V_{xy}$ at $2\omega$ and normalizing by $I_0^2$.

\subsection*{S6.3 Relation to the Berry-curvature dipole and symmetry-controlled sign reversals}

Within the low-frequency semiclassical regime ($\omega\tau\ll 1$) and a constant relaxation time $\tau$, the intrinsic
nonlinear Hall sheet conductivity is controlled by the two-dimensional Berry-curvature dipole (Sec.~S4.4),
\begin{equation}
\chi_{xyy}^{\rm sheet}(E_F)\;=\;\frac{e^3\tau}{\hbar^2}\,D_y(E_F),
\label{eq:S6_chi_D}
\end{equation}
(with $D_y\equiv D_{yz}$ in the notation of the main text). Combining Eqs.~\eqref{eq:S6_R2w} and \eqref{eq:S6_chi_D} gives
\begin{equation}
R_{xy}^{2\omega}(E_F)\;\simeq\;-\frac{e^3\tau}{2\hbar^2}\,\frac{1}{W}\,\frac{D_y(E_F)}{[\sigma_{xx}^{\rm sheet}(E_F)]^3}.
\label{eq:S6_R2w_D}
\end{equation}

The sign control can be stated systematically by denoting the bottom-interface out-of-plane exchange by $\Delta_z$ and the top-interface in-plane exchange by $\Delta_x$. With the explicit convention $V_{xy}=WE_x$ adopted in Eqs.~\eqref{eq:S6_V1w} and \eqref{eq:S6_V2w}, both harmonic voltages carry the same overall minus sign relative to $\sigma_{xy}^{\rm sheet}$ and $D_y$, respectively. A global reversal of the transverse voltage leads multiplies all entries by the same sign. After absorbing this common overall sign into the Hall-voltage lead convention, the ideal-model exchange-odd contributions obey $\mathrm{sgn}\,V_{xy}^{1\omega}\sim\mathrm{sgn}\,\Delta_z$ and $\mathrm{sgn}\,V_{xy}^{2\omega}\sim\mathrm{sgn}(\Delta_z\Delta_x)$ at the selected $\mu^*$. The product term is even under simultaneous reversal $(\Delta_z,\Delta_x)\to(-\Delta_z,-\Delta_x)$ but changes sign when either component is reversed individually. The relative sign table is shown in Table~\ref{tab:S2_sign_control}.

A real dual interface can break $C_3$ structurally through inequivalent magnets, strain, twist, or moir\'e reconstruction. The measured BCD can then contain a background $D_y^{(0)}$ and other symmetry-allowed terms in addition to the product-odd contribution, schematically $D_y=D_y^{(0)}+\lambda\Delta_z\Delta_x+\cdots$. Antisymmetrizing under $\Delta_x$ reversal at fixed $\Delta_z$ isolates the part odd in $\Delta_x$, while comparing simultaneous and single-component reversals separates additional parity channels. Table~\ref{tab:S2_sign_control} therefore describes the ideal exchange-odd component, not an assertion that every experimental $2\omega$ background vanishes.
\begin{table}[tb]
\centering
\caption{\textbf{Ideal-model exchange-odd sign pattern in the orthogonal two-sided geometry.} The entries give the relative pattern at the operating chemical potential used in Fig.~4 after absorbing the common overall sign of Eqs.~\eqref{eq:S6_V1w}--\eqref{eq:S6_V2w} into the Hall-voltage lead convention. The first-harmonic pattern follows $\Delta_z$, whereas the product-odd second-harmonic component follows $\Delta_z\Delta_x$. Structural or extrinsic offsets are not included.}
\label{tab:S2_sign_control}
\small
\setlength{\tabcolsep}{8pt}
\renewcommand{\arraystretch}{1.15}
\begin{tabular}{c c c c}
\hline\hline
$\Delta_z$ & $\Delta_x$ & $V_{xy}^{1\omega}$ & $V_{xy}^{2\omega}$ \\
\hline
$+$ & $+$ & $+$ & $+$ \\
$+$ & $-$ & $+$ & $-$ \\
$-$ & $+$ & $-$ & $-$ \\
$-$ & $-$ & $-$ & $+$ \\
\hline\hline
\end{tabular}
\end{table}

The orthogonal two-sided architecture permits component-selective reversal of the ideal exchange-odd harmonics by combining two switching operations. Reversing the in-plane component at fixed out-of-plane component, $(s_z,s_x)\to(s_z,-s_x)$, leaves $V_{xy}^{1\omega}$ unchanged and reverses the part of $V_{xy}^{2\omega}$ odd in $s_x$. Simultaneously reversing both components reverses $V_{xy}^{1\omega}$ and leaves the ideal product-odd second-harmonic part unchanged. These parity combinations provide an internal symmetry check even when an experimental offset is present (see Sec.~S6.5).

\subsection*{S6.4 Order-of-magnitude estimates for $V_{xy}^{1\omega}$ and $V_{xy}^{2\omega}$}

To make the experimental estimate explicit, we assume throughout this subsection a constant relaxation time
$\tau=25$~fs, a metallic sheet conductance
$\sigma_{xx}^{\rm sheet}=1$--$5$~mS
(corresponding to a sheet resistance $R_{\square}\sim 200$--$1000~\Omega$),
a Hall-bar width $W=5~\mu$m, and an ac drive current amplitude in the range
$I_0=0.1$--$1$~mA.
For the intrinsic Hall responses, we take representative values from the calculated spectra:
$|\sigma_{xy}^{\rm sheet}|\sim 10^{-2}\,(e^2/h)$ near the AHC maxima
(as in Figs.~2 and S9),
and a conservative representative range $|D_y|\sim 10^{-3}$--$3\times10^{-2}$~\AA\ rather than the global extrema of every calculated curve (Figs.~3 and S7--S10). These AHC, BCD, and longitudinal-conductance ranges are combined only for an order-of-magnitude envelope; they are not all evaluated at one common $\mu$.

For the AHE, Eq.~\eqref{eq:S6_V1w} implies a first-harmonic Hall resistance
\begin{equation}
R_{xy}^{1\omega}\;\equiv\;\frac{V_{xy}^{1\omega}}{I_0}\;\simeq\;-\frac{\sigma_{xy}^{\rm sheet}}{(\sigma_{xx}^{\rm sheet})^2}.
\end{equation}
Using the above values gives
\begin{equation}
|R_{xy}^{1\omega}|\sim 1.5\times10^{-2}\text{--}3.9\times10^{-1}~\Omega,
\end{equation}
and therefore
\begin{equation}
|V_{xy}^{1\omega}|\sim (1.5\times10^{-2}\text{--}3.9\times10^{-1}~\Omega)\,I_0.
\end{equation}
For $I_0=0.1$--$1$~mA, this corresponds to
\begin{equation}
|V_{xy}^{1\omega}|\sim 1.5\text{--}390~\mu{\rm V},
\end{equation}
which is readily detectable with standard low-frequency lock-in techniques.

For the nonlinear Hall response, inserting Eq.~\eqref{eq:S6_chi_D} into Eq.~\eqref{eq:S6_V2w} yields
\begin{equation}
|V_{xy}^{2\omega}|\;\simeq\;\frac{e^3\tau}{2\hbar^2}\,\frac{|D_y|}{(\sigma_{xx}^{\rm sheet})^3}\,\frac{I_0^2}{W}.
\label{eq:S6_V2w_est}
\end{equation}
Using the same parameter set, one obtains
\begin{equation}
|V_{xy}^{2\omega}|\sim 10^{-10}\text{--}10^{-6}~{\rm V}
\qquad \text{for} \qquad I_0\sim 1~{\rm mA},
\label{eq:S6_V2w_numeric}
\end{equation}
over the full stated ranges of $D_y$ and $\sigma_{xx}^{\rm sheet}$. In the most active BCD windows, where $|D_y|\sim10^{-2}$ to a few$\times10^{-2}$~\AA, the expected signal is of order
\begin{equation}
|V_{xy}^{2\omega}|\sim 10^{-9}\text{--}10^{-6}~{\rm V},
\end{equation}
i.e., from a few nanovolts to approximately the microvolt scale, depending primarily on the sheet conductance and the operating chemical potential. Because $V_{xy}^{2\omega}\propto I_0^2$, reducing the drive to $I_0=0.1$~mA lowers the signal by two orders of magnitude, while increasing the drive enhances it provided Joule heating remains under control.

Expressed as a normalized coefficient, Eq.~\eqref{eq:S6_R2w} gives
\begin{equation}
|R_{xy}^{2\omega}|\sim 10^{-4}\text{--}1~{\rm V/A^2}
\end{equation}
over the full stated parameter range, with typical values in the most active BCD windows of order $10^{-3}$--$1~{\rm V/A^2}$.

Finally, the low-frequency approximation used in Eqs.~\eqref{eq:S6_V1w} and \eqref{eq:S6_V2w} is well satisfied in the intended lock-in regime:
for $\tau=25$~fs and excitation frequencies in the $10^2$--$10^3$~Hz range, one has $\omega\tau\ll 1$.
These envelope estimates indicate a readily measurable first harmonic and a much broader sub-nanovolt-to-microvolt second-harmonic range. A quantitative prediction at the proposed $\mu^*$ requires $\sigma_{xy}^{\rm sheet}$, $D_y$, $\sigma_{xx}^{\rm sheet}$, carrier density, and $\tau$ at that same operating state. The upper end of the current range is most appropriate for short, well-heat-sunk devices, and possible extrinsic $2\omega$ backgrounds are discussed separately in Sec.~S6.5.

\subsection*{S6.5 Background signals and practical separation strategies}

In harmonic Hall experiments, several extrinsic mechanisms can contribute to a measured $2\omega$ transverse voltage, including:
(i) thermoelectric voltages from Joule heating (Seebeck/Nernst-type contributions that scale as $I_0^2$),
(ii) nonlinearities in the longitudinal channel mixed into the transverse leads by geometric misalignment,
and (iii) in structures containing a conducting ferromagnet, current-induced torques that modulate the magnetization and generate
a $2\omega$ Hall response through the ordinary AHE of the ferromagnet.

The orthogonal two-sided proximity design provides two reversal handles to isolate the exchange-odd BCD contribution:
\begin{itemize}
\item \emph{Parity under exchange reversal.} Reversing the in-plane proximity component at fixed $m_z$ isolates the part of $V_{xy}^{2\omega}$ odd in $m_x$ while leaving the first-harmonic AHE unchanged. Simultaneously reversing both components provides a complementary parity channel. A structural BCD offset even in $m_x$, an anomalous-Nernst contribution tied only to $m_z$, and the ideal product-odd BCD term can therefore be separated by combining all four states rather than assuming that the full measured $2\omega$ voltage follows a single sign law.
\item \emph{Current scaling and frequency window.} The intrinsic nonlinear Hall signal scales as $V_{xy}^{2\omega}\propto I_0^2$
and is essentially frequency independent as long as $\omega\tau\ll 1$ (trivially satisfied for kHz lock-in frequencies even for
$\tau\sim10$--$100$~fs). Thermal backgrounds also scale as $I_0^2$ but typically exhibit stronger dependence on heat sinking and
geometry; systematic checks versus $I_0$, frequency, and base temperature, together with the symmetry-controlled sign reversal,
provide a stringent discrimination protocol.
\end{itemize}

Taken together, the harmonic Hall readout offers a standard route to access the Hall-valve AHE
(first harmonic) and the orthogonal-configuration nonlinear Hall response (second harmonic). The reversal parities provide internal checks, while actual signal levels require a self-consistent device stack and operating-point transport characterization.


\end{document}